\PassOptionsToPackage{unicode}{hyperref}
\PassOptionsToPackage{hyphens}{url}
\documentclass[
  final]{article}
\usepackage[x11names]{xcolor}
\usepackage{colortbl}
\usepackage{amsmath,amssymb}
\usepackage{iftex}
\ifPDFTeX
  \usepackage[T1]{fontenc}
  \usepackage[utf8]{inputenc}
  \usepackage{textcomp} 
\else 
  \usepackage{unicode-math} 
  \defaultfontfeatures{Scale=MatchLowercase}
  \defaultfontfeatures[\rmfamily]{Ligatures=TeX,Scale=1}
\fi
\usepackage{lmodern}
\ifPDFTeX\else
\fi
\IfFileExists{upquote.sty}{\usepackage{upquote}}{}
\IfFileExists{microtype.sty}{
  \usepackage[]{microtype}
  \UseMicrotypeSet[protrusion]{basicmath} 
}{}
\makeatletter
\@ifundefined{KOMAClassName}{
  \IfFileExists{parskip.sty}{%
    \usepackage{parskip}
  }{
    \setlength{\parindent}{0pt}
    \setlength{\parskip}{6pt plus 2pt minus 1pt}}
}{
  \KOMAoptions{parskip=half}}
\makeatother
\usepackage{longtable,booktabs,array}
\usepackage{multirow}
\usepackage{calc} 
\usepackage{etoolbox}
\makeatletter
\patchcmd\longtable{\par}{\if@noskipsec\mbox{}\fi\par}{}{}
\makeatother
\IfFileExists{footnotehyper.sty}{\usepackage{footnotehyper}}{\usepackage{footnote}}
\makesavenoteenv{longtable}
\usepackage{graphicx}
\makeatletter
\newsavebox\pandoc@box
\newcommand*\pandocbounded[1]{
  \sbox\pandoc@box{#1}%
  \Gscale@div\@tempa{\textheight}{\dimexpr\ht\pandoc@box+\dp\pandoc@box\relax}%
  \Gscale@div\@tempb{\linewidth}{\wd\pandoc@box}%
  \ifdim\@tempb\p@<\@tempa\p@\let\@tempa\@tempb\fi
  \ifdim\@tempa\p@<\p@\scalebox{\@tempa}{\usebox\pandoc@box}%
  \else\usebox{\pandoc@box}%
  \fi%
}
\def\fps@figure{htbp}
\makeatother
\ifLuaTeX
  \usepackage{luacolor}
  \usepackage[soul]{lua-ul}
\else
  \usepackage{soul}
\fi
\usepackage[round,authoryear]{natbib}
\usepackage{neurips_2024}
\usepackage{bookmark}
\IfFileExists{xurl.sty}{\usepackage{xurl}}{} 
\hypersetup{
  pdftitle={Verifying International Agreements on AI},
  pdfcreator={LaTeX via pandoc},
  colorlinks=true,linkcolor=black,urlcolor=RoyalBlue4,citecolor=RoyalBlue4}

\title{\protect\phantomsection\label{_5807bavuhsio}{}Verifying International Agreements on AI}
\usepackage{etoolbox}
\makeatletter
\providecommand{\subtitle}[1]{
  \apptocmd{\@title}{\par {\large #1 \par}}{}{}
}
\makeatother
\subtitle{\protect\phantomsection\label{_65se8sbjgi7k}{}Six Layers of Verification for Rules on\\
Large-Scale AI Development and Deployment}
\author{
  Mauricio Baker\thanks{Correspondence: \href{mailto:mbaker@rand.org}{mbaker@rand.org} and \href{mailto:lheim@rand.org}{lheim@rand.org}} \\
  RAND \\
  \And
  Gabriel Kulp \\
  RAND \\
  \And
  Oliver Marks \\
  University of Bristol \\
  \AND
  Miles Brundage \\
  AI Verification and Evaluation Research Institute \\
  \And
  Lennart Heim \\
  RAND
}
\date{}
\usepackage{enumitem}
\setlist[itemize,3]{label=\textbullet}
\usepackage{needspace}
\usepackage{placeins}

\usepackage{titlesec}
\titleformat{\paragraph}
  {\normalfont\small\bfseries}
  {\theparagraph}
  {0em}
  {}

\begin{document}
\maketitle

\begin{abstract}
The risks of frontier AI may require international cooperation, which in
turn may require verification: checking that all parties follow
agreed-on rules. For instance, states might need to verify that powerful
AI models are widely deployed only after their risks to international
security have been evaluated and deemed manageable. However, research on
AI verification could benefit from greater clarity and detail. To
address this, this report provides an in-depth overview of AI
verification, intended for both policy professionals and technical
researchers. We present novel conceptual frameworks, detailed
implementation options, and key R\&D challenges. These draw on existing
literature, expert interviews, and original analysis, all within the
scope of confidentially overseeing AI development and deployment that
uses thousands of high-end AI chips. We find that states could
eventually verify compliance by using six largely independent
verification approaches with substantial redundancy: (1) \emph{built-in
security features} in AI chips; (2-3) \emph{separate monitoring devices}
attached to AI chips; and (4-6) \emph{personnel-based mechanisms}, such
as whistleblower programs. While promising, these approaches require
guardrails to protect against abuse and power concentration, and many of
these technologies have yet to be built or stress-tested. To enable
states to confidently verify compliance with rules on large-scale AI
development and deployment, the R\&D challenges we list need significant
progress.\textsuperscript{\dag}
\begingroup
  \renewcommand{\thefootnote}{\dag}
  \footnotetext{As a RAND working paper, this 
paper intends to share early insights and solicit informal peer review, 
so it has not yet been formally peer reviewed or professionally edited. Review comments are welcome on \href{https://www.alphaxiv.org/abs/2507.15916}{alphaXiv}.}
\endgroup
\end{abstract}

\addtocontents{toc}{\protect\setcounter{tocdepth}{1}}

\clearpage

\phantomsection\section{Summary}\label{summary}

\phantomsection\subsection{Background}\label{background}

\textbf{The risks of frontier AI may require international cooperation, which could require compliance to be verified.} As frontier AI systems become more advanced, they could bring many benefits but also pose global risks, such as lowering barriers to developing weapons of mass destruction and losing human control over AI. Such extreme, shared risks can motivate even rival powers to cooperate, as shown by the history of arms control. Still, states may be unwilling to limit their own advanced AI development and deployment if they cannot trust that other states will do the same, as one-sided restraint could put states at a strategic disadvantage. Thus, for international agreements on AI to be made and followed, verification of compliance could be crucial (\hyperref[introduction]{\ul{Section 1}}).

\phantomsection\subsection{Contributions and Approach}\label{contributions-and-approach}

\textbf{This report overviews options for verifying compliance with rules on AI, addressing gaps in the field's clarity and detail.} To date, the field has explored many verification mechanisms, but there is limited clarity on how these mechanisms can be combined effectively, so that they reinforce weak points rather than create new vulnerabilities. The field also has yet to examine many implementation questions in any detail, hindering assessments of what is feasible and what research is needed. To address both of these issues, our overview features new conceptual frameworks and relatively detailed implementation analyses (\hyperref[contributions]{\ul{Section 1.1}}). As examples of its detail, our analysis outlines options for: detecting hardware backdoors at scale, controlling inference non-determinism, doing ``compute accounting'' when operations are duplicated, and protecting whistleblowers from state suppression. Throughout, we draw on a literature review and 18 expert interviews (\hyperref[methodology]{\ul{Section 2.4}}).

\phantomsection\subsection{Scope}\label{scope}

\textbf{We focus on verification options that oversee large-scale AI compute use, and which have the potential to be confidential and effective internationally.} Large-scale AI compute use, in this report, refers to the use of thousands of AI chips for multiple months (\hyperref[rules-on-large-scale-ai-compute]{\ul{Section 2.2}}). These can be overseen to set rules on frontier AI development and associated mass deployment, without intruding on smaller-scale activities. Confidentiality refers to protecting sensitive information---especially models, data, and code---from unauthorized access, especially theft. Finally, international effectiveness means effectiveness even against states determined to conceal non-compliance (\hyperref[introduction]{\ul{Section 1}}).

\textbf{In the above context, we cover options for verifying compliance with rules on AI models, data, and code.} These include many potential rules aimed at AI safety and security, transparency, benefit-sharing, military limits, and international norms (\hyperref[tab:examples_of_potential_rules]{\ul{Table 3}}). For example, one such rule could be that future frontier AI models must be subject to specific tests that aim to measure how much they will increase access to weapons of mass destruction. Then, if these test results are deemed to indicate unmanageable risk, the risks could be required to be addressed, as determined through further tests, before large-scale deployment may proceed. Regardless of the exact rules one might wish to verify, much of the verification problem is the same: identifying what models, data, and code were used, so one can run desired tests on them (\hyperref[rules-on-ai-models-data-and-code]{\ul{Section 2.1}}).

\textbf{Determining which technical rules actually achieve policy objectives remains challenging.} Even if one can robustly verify that, say, a model was trained with fewer than X operations, or that it scores below Y on a specific risk evaluation, we still need a separate analysis to determine whether these technical rules truly address the relevant policy concerns. This report focuses on verifying compliance with technical rules, while acknowledging that specifying effective technical rules is an important area of ongoing and future work (\hyperref[addressing-broader-challenges-for-verification]{\ul{Section 3.3}}).

\phantomsection\subsection{Key Findings}\label{key-findings}

\begin{figure}[ht]
  \noindent
  \makebox[\textwidth][c]{%
    \includegraphics[width=5.5575in,height=5.4506in]{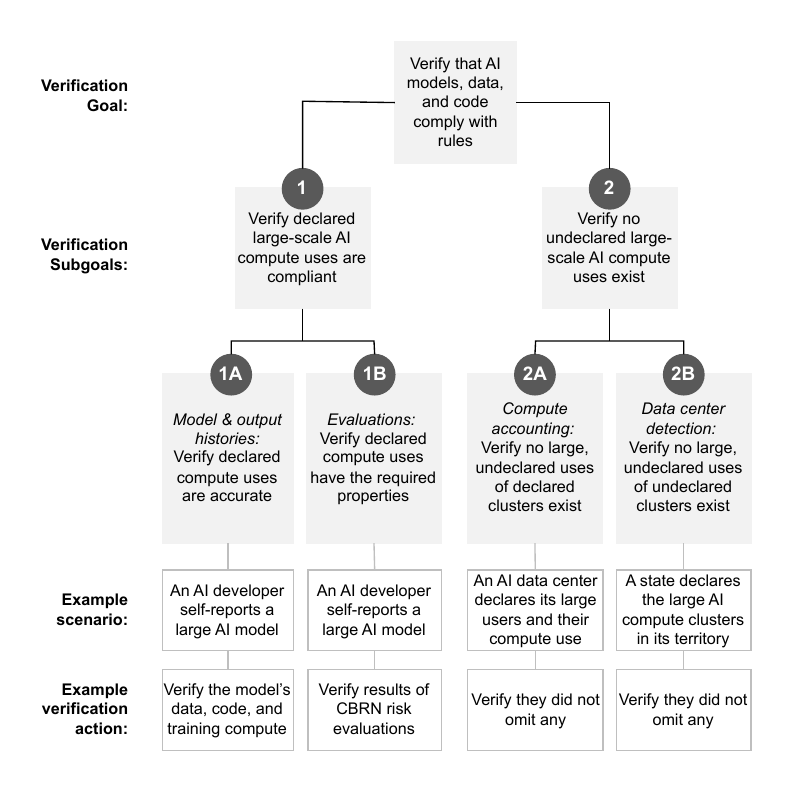}%
  }
  \caption{Framework of verification subgoals. We decompose a broad verification goal into subgoals, to clarify how verification can be achieved. Note the italicized terms such as ``Evaluations'' are only rough summaries of each subgoal; the figure's non-italicized descriptions are more precise.}
  \label{fig:framework_of_verification_subgoals}
\end{figure}

\textbf{Verifying rules on AI models, data, and code in the above context can be decomposed into subgoals} (\hyperref[verification-framework]{\ul{Section 3.2}}; \hyperref[fig:framework_of_verification_subgoals]{\ul{Figure 1}})\textbf{.} These are:

\begin{enumerate}
\def\labelenumi{\arabic{enumi}.}
\item
  Verify that \emph{declared} uses of large-scale AI compute are compliant by (A) verifying that trained AI models and their outputs were generated as claimed; and (B) verifying evaluation results, or more generally, verifying that declared models, data, and code have the required properties. ``Declared'' means self-reported, preferably via confidentiality-preserving technologies.
\item
  Verify that there are no \emph{undeclared} uses of large-scale AI compute by (A) verifying that the use of known AI data centers is accounted for; and (B) verifying that no actor has hidden AI data centers or large, decentralized collections of AI chips that can be used for violations.
\end{enumerate}

\begin{figure}[ht]
  \noindent
  \makebox[\textwidth][c]{%
    \includegraphics[width=6.5in,height=5.08333in]{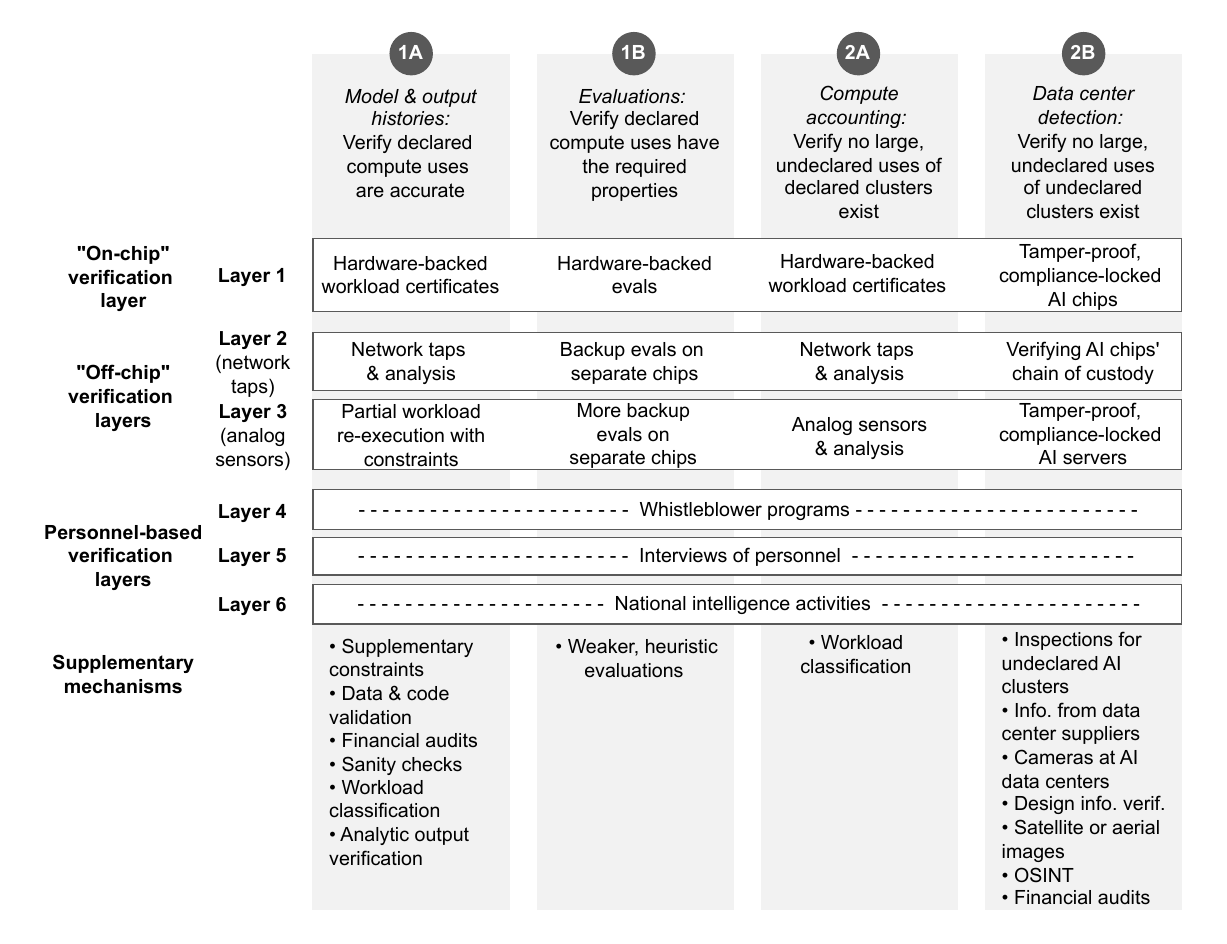}%
  }
  \caption{Verification layers consist of distinct mechanisms for each verification subgoal.}
  \label{fig:verification_layers_consist_of_distinct_mechanisms_for_each_verification_subgoal}
\end{figure}

\textbf{To complete these subgoals, states could create six layers of verification}---six largely independent assurances of compliance (\hyperref[verification-mechanisms-and-layers]{\ul{Section 4}}; \hyperref[tab:six_layers]{\ul{Table 1}}; \hyperref[fig:verification_layers_consist_of_distinct_mechanisms_for_each_verification_subgoal]{\ul{Figure 2}}). Like ``layers of defense,'' a full implementation of each layer could verify compliance on its own, and multiple layers would reinforce each other. Thus, a stack of layers is an effective combination of verification mechanisms; it completes each subgoal with redundancy. In brief, the layers are:

\begin{itemize}
\item
  \textbf{On-chip:} (1) This layer would perform verification by using functionalities \emph{built into} AI chips. Due to their cybersecurity benefits, some versions of these are already commonplace.
\item
  \textbf{Off-chip:} These layers would use \emph{external devices} to oversee AI chips. The devices could be (2) network taps, to intercept data exchanged between AI chips; and (3) analog sensors, to record measurements such as power use. These data could then be confidentially analyzed.
\item
  \textbf{Personnel-based:} Several layers would leverage the large \emph{workforce} involved in AI: (4) whistleblower programs, (5) interviews of personnel, and (6) national intelligence activities.
\end{itemize}

\FloatBarrier

\renewcommand{\arraystretch}{2}
{
  \setlength{\LTleft}{-33pt}
  \setlength{\LTright}{-26pt}
  \centering
    \Needspace*{8\baselineskip}%
    \begin{longtable}[]{
      >{\raggedright\arraybackslash}p{(\linewidth - 6\tabcolsep) * \real{0.2706} * \real{1.15}}
      >{\raggedright\arraybackslash}p{(\linewidth - 6\tabcolsep) * \real{0.3191} * \real{1.15}}
      >{\raggedright\arraybackslash}p{(\linewidth - 6\tabcolsep) * \real{0.2051} * \real{1.15}}
      >{\raggedright\arraybackslash}p{(\linewidth - 6\tabcolsep) * \real{0.2051} * \real{1.15}}}
    \toprule\noalign{}
    \begin{minipage}[t]{\linewidth}\raggedright
    \textbf{Potential verification layer}
    \end{minipage} & \begin{minipage}[t]{\linewidth}\raggedright
    \textbf{Summary of layer}
    \end{minipage} & \begin{minipage}[t]{\linewidth}\raggedright
    \textbf{Key advantages}
    \end{minipage} & \begin{minipage}[t]{\linewidth}\raggedright
    \textbf{Key disadvantages}
    \vspace{0.5em}
    \end{minipage} \\
    \endhead
    \toprule\noalign{}
    \rowcolor{black!5!white}
    \begin{minipage}[t]{\linewidth}\raggedright
    \textbf{On-chip security features}
    
    (i.e., secure boot, Confidential Computing) (\hyperref[on-chip-verification-layer]{\ul{Section 4.1}})
    
    \begin{center}
        \includegraphics[width=0.71354in,height=0.71354in]{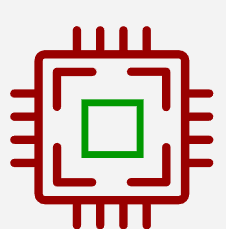}
    \end{center}
    \vspace{0.1em}
    \end{minipage} & \begin{minipage}[t]{\linewidth}\raggedright
    \cellcolor{black!5!white}
    Security features built into AI chips may enable verification, such as by ensuring that AI chips log traces of their activities for confidential analysis.
    \end{minipage} & \begin{minipage}[t]{\linewidth}\raggedright
    \cellcolor{black!5!white}
    Offers maximum transparency into AI chips' uses.
    \end{minipage} & \begin{minipage}[t]{\linewidth}\raggedright
    \cellcolor{black!5!white}
    Poses unsolved technical problems and severe security challenges (e.g., untrusted suppliers). Insecure AI chips may need to be replaced.
    \end{minipage} \\
    \begin{minipage}[t]{\linewidth}\raggedright
    \textbf{Off-chip network tap} (and analysis) (e.g. ``FlexHEGs'')
    
    (\hyperref[off-chip-verification-layers]{\ul{Section 4.2}})
    
    \begin{center}
        \includegraphics[width=1.09383in,height=0.66791in]{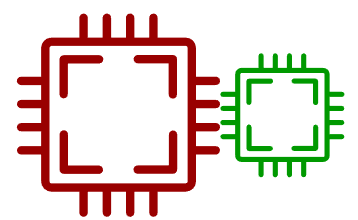}
    \end{center}
    \vspace{0.1em}
    \end{minipage} & \begin{minipage}[t]{\linewidth}\raggedright
    Mutually vetted devices could intercept data exchanged between chips, then check for discrepancies with declared uses.
    \end{minipage} & \begin{minipage}[t]{\linewidth}\raggedright
    Could be retrofitted to existing AI chips and optimized for security.
    \end{minipage} & \begin{minipage}[t]{\linewidth}\raggedright
    Poses technical, logistical, and security challenges. Strongest versions need redesigned chip-adjacent hardware.
    \end{minipage} \\
    \begin{minipage}[t]{\linewidth}\raggedright
    \cellcolor{black!5!white}
    \textbf{Off-chip analog sensors} (and analysis, e.g., proof-of-learning)
    
    (\hyperref[off-chip-verification-layers]{\ul{Section 4.2}})
    
    \begin{center}
        \includegraphics[width=0.92188in,height=0.6887in]{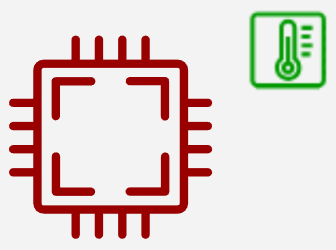}
    \end{center}
    \vspace{0.1em}
    \end{minipage} & \begin{minipage}[t]{\linewidth}\raggedright
    \cellcolor{black!5!white}
    Physically secured chips could check that (i) declared AI compute uses are accurate (e.g., reproducible) and (ii) their compute use adds up to the expected total (estimated with analog sensors, e.g., power meters, in AI data centers).
    \end{minipage} & \begin{minipage}[t]{\linewidth}\raggedright
    \cellcolor{black!5!white}
    Could be retrofitted to existing AI chips and optimized for security.
    \end{minipage} & \begin{minipage}[t]{\linewidth}\raggedright
    \cellcolor{black!5!white}
    Poses unsolved technical problems. Likely requires separate trusted clusters for analysis, and manufacturing \& installing sensors.
    \end{minipage} \\
    \begin{minipage}[t]{\linewidth}\raggedright
    \textbf{Whistleblower programs}
    
    (\hyperref[personnel-based-verification-layers]{\ul{Section 4.3}})
    
    \begin{center}
        \includegraphics[width=0.6927in,height=0.5494in]{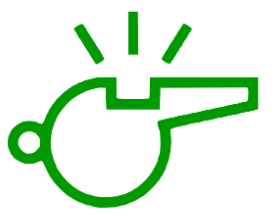}
    \end{center}
    \vspace{0.1em}
    \end{minipage} & \begin{minipage}[t]{\linewidth}\raggedright
    Programs may enable and incentivize (narrowly scoped, non-public) staff whistleblowing, for all verification subgoals.
    \end{minipage} & \begin{minipage}[t]{\linewidth}\raggedright
    Relatively simple, precedented, and implementation-
    
    ready.
    \end{minipage} & \begin{minipage}[t]{\linewidth}\raggedright
    Unclear effectiveness: depends on the number and loyalty of accomplices.
    \end{minipage} \\
    \begin{minipage}[t]{\linewidth}\raggedright
    \cellcolor{black!5!white}
    \textbf{Interviews of personnel}
    
    (\hyperref[personnel-based-verification-layers]{\ul{Section 4.3}})
    
    \begin{center}
        \includegraphics[width=0.76948in,height=0.56049in]{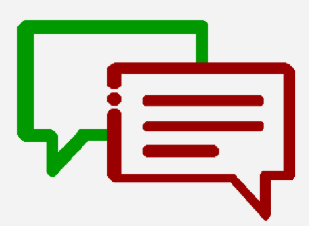}
    \end{center}
    \vspace{0.1em}
    \end{minipage} & \begin{minipage}[t]{\linewidth}\raggedright
    \cellcolor{black!5!white}
    Interviews may reveal violations at any verification subgoal, e.g., via inconsistencies or perhaps improved lie detection tech, but such tech is abusable.
    \end{minipage} & \begin{minipage}[t]{\linewidth}\raggedright
    \cellcolor{black!5!white}
    Relatively simple and precedented.
    \end{minipage} & \begin{minipage}[t]{\linewidth}\raggedright
    \cellcolor{black!5!white}
    Unclear effectiveness: depends on accomplices' ability to lie undetected.
    \end{minipage} \\
    \begin{minipage}[t]{\linewidth}\raggedright
    \textbf{National intelligence activities} (\hyperref[personnel-based-verification-layers]{\ul{Section 4.3}})
    
    \begin{center}
        \includegraphics[width=0.5625in,height=0.53856in]{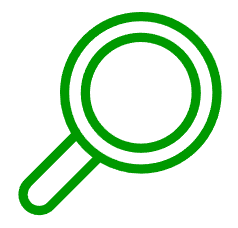}
    \end{center}
    \vspace{0.1em}
    \end{minipage} & \begin{minipage}[t]{\linewidth}\raggedright
    Intelligence agencies could collect and analyze intelligence for all verification subgoals, including via human, cyber, and signals intelligence.
    \end{minipage} & \begin{minipage}[t]{\linewidth}\raggedright
    Precedented and may be feasible unilaterally.
    \end{minipage} & \begin{minipage}[t]{\linewidth}\raggedright
    More adversarial, harder for third parties to verify, and unclear effectiveness.
    \end{minipage} \\
    
    \bottomrule\noalign{}
    \caption{Six layers for verifying the compliance of large-scale AI development and deployment. Green (red) icons are Verifier-(un)trusted.} \label{tab:six_layers}
    \endlastfoot
    \end{longtable}
}

To enhance the above layers, various less robust mechanisms could serve as supplements. These could include satellite images and open-source information.

\textbf{Each of the above verification layers has different strengths and weaknesses} (\hyperref[tab:six_layers]{\ul{Table 1}}). For example, on-chip functionalities could ultimately leave the smallest margins of error when verifying claims. Off-chip devices could be the best way to adapt existing hardware to make it verifiable, as some versions could be retrofitted onto insecure AI chips. Finally, personnel-based layers could be most readily available, as whistleblowers, for instance, do not require new or complex technology.

\textbf{Verification could also be harmful, which warrants guardrails.} Verification, if implemented without adequate restraint, could compromise sensitive information, hinder innovation, and centralize information and power. Throughout this report, we discuss potential mitigations, including: using confidentiality-preserving technologies (\hyperref[the-framework]{\ul{Section 3.2}}); limiting the scope of verification to large-scale AI compute (\hyperref[rules-on-large-scale-ai-compute]{\ul{Section 2.2}}); extensively vetting and constraining network taps and analog sensors (\hyperref[prerequisites-off-chip-devices]{\ul{Section 4.2.1.1}}); and being cautious with verification mechanisms that double as enforcement tools, e.g., by requiring multiple parties to approve any enforcement actions (\hyperref[analysis]{\ul{Section 4.1.2}}).

\textbf{Overall, states could become confident in verification of compliance with rules on large-scale AI development and deployment, but this requires progress on the R\&D and infrastructure challenges we list} (\hyperref[open-problems-in-verification]{\ul{Section 5}}; \hyperref[tab:rd_challenges]{\ul{Table 2}}). Throughout the six verification layers, we were able to outline plausible ways to address every implementation challenge we identified, including guarding against a range of potential circumvention tactics. This suggests states could create six verification layers that each offer separate, meaningful evidence of compliance. Additionally, the personnel-based layers could provide some assurances with relatively little preparation required. However, the on- and off-chip layers will likely be circumventable until there is substantial progress on the R\&D and infrastructure challenges we list. This progress could be driven by investments, pilot programs, and red teaming. Such efforts will determine whether, when states consider verification, they will see a few nascent options or many tried-and-tested technologies.

\begin{longtable}[]{
  >{\raggedright\arraybackslash}p{(\linewidth - 2\tabcolsep) * \real{0.1551}}
  >{\raggedright\arraybackslash}p{(\linewidth - 2\tabcolsep) * \real{0.8449}}}
\toprule\noalign{}
\begin{minipage}[t]{\linewidth}\raggedright
\textbf{Verification layer}
\end{minipage} & \begin{minipage}[t]{\linewidth}\raggedright
\textbf{Selected R\&D and infrastructure challenges}

(Legend: \textcolor{gray}{\rule{0.8em}{0.8em}}: hardware; \textcolor{blue}{\rule{0.8em}{0.8em}}: CS/ML; \textcolor{brown}{\rule{0.8em}{0.8em}}: organizational)
\vspace{0.5em}
\end{minipage} \\
\endhead
\toprule\noalign{}
\rowcolor{black!5!white}
\begin{minipage}[t]{\linewidth}\raggedright
1. Security features in AI chips (\hyperref[on-chip-verification-layer]{\ul{Section 4.1}})
\end{minipage} & \begin{minipage}[t]{\linewidth}\raggedright
\textcolor{blue}{\rule{0.8em}{0.8em}} \textbf{A.} \textbf{System software protocol}: Given verifiable system software (with access to e.g., kernels, memory), verify workload code, and model and data locations, despite any potential obfuscation (\hyperref[a.2-hardware-backed-workload-certificates-and-evaluations]{\ul{Appendix A.2}}).
\vspace{0.5em}

\textcolor{gray}{\rule{0.8em}{0.8em}} \textbf{B.} \textbf{Hardware design attestation}: Given a scanned hardware layout and Hardware Description Language (HDL), check if they match (\hyperref[a.1-full-stack-security-for-technical-verification-mechanisms-implementation]{\ul{Appendix A.1}}).
\vspace{0.5em}

\textcolor{gray}{\rule{0.8em}{0.8em}} \textbf{C.} \textbf{Hardware security features}: Develop highly vetted, open-source, dedicated hardware designs for secure boot, Confidential Computing, and on-chip tamper-proofing (\hyperref[a.2-hardware-backed-workload-certificates-and-evaluations]{\ul{Appendix A.2}}).
\vspace{0.5em}

\textcolor{brown}{\rule{0.8em}{0.8em}} \textbf{D.} \textbf{Design adoption}: Adopt the above designs into leading AI chips.
\vspace{0.5em}
\end{minipage} \\
\begin{minipage}[t]{\linewidth}\raggedright
2. Off-chip network taps \& analysis (\hyperref[off-chip-verification-layers]{\ul{Section 4.2}})
\end{minipage} & \begin{minipage}[t]{\linewidth}\raggedright
\textcolor{blue}{\rule{0.8em}{0.8em}} \textbf{A.} \textbf{Network tap analysis:} Given a cluster's inter-accelerator data (including kernels), verify that the cluster executed only a claimed workload (\hyperref[a.3-network-taps-analysis]{\ul{Appendix A.3}}).
\vspace{0.5em}

\textcolor{gray}{\rule{0.8em}{0.8em}} \textbf{B.} \textbf{Network taps}: Design and manufacture appropriate network taps, or identify suitable existing tech (\hyperref[a.3-network-taps-analysis]{\ul{Appendix A.3}}).
\vspace{0.5em}

\textcolor{brown}{\rule{0.8em}{0.8em}} \textbf{C.} \textbf{Trusted clusters}: Build small compute clusters that are or can be mutually physically secured (\hyperref[off-chip-verification-layers]{\ul{Section 4.2}}).
\vspace{0.5em}
\end{minipage} \\
\rowcolor{black!5!white}
\begin{minipage}[t]{\linewidth}\raggedright
3. Off-chip analog sensors \& analysis

(\hyperref[off-chip-verification-layers]{\ul{Section 4.2}})
\end{minipage} & \begin{minipage}[t]{\linewidth}\raggedright
\textcolor{blue}{\rule{0.8em}{0.8em}} \textbf{A.} \textbf{Code \& data checks}: Develop tests to detect code and data that are designed to spoof partial program re-execution with constraints (e.g. proof-of-learning) (\hyperref[a.4-partial-workload-re-execution-with-constraints]{\ul{Appendix A.4}}; \hyperref[a.5-data-and-code-validation]{\ul{Appendix A.5}}).
\vspace{0.5em}

\textcolor{blue}{\rule{0.8em}{0.8em}} \textbf{B.} \textbf{Workload modeling}: Given an AI workload and cluster specs, estimate the optimal utilization (MFU) and the associated physical signature, e.g., power (\hyperref[a.6-compute-accounting-via-analog-sensors]{\ul{Appendix A.6}}).
\vspace{0.5em}

\textcolor{gray}{\rule{0.8em}{0.8em}} \textbf{C.} \textbf{Analog sensors}: Design and manufacture appropriate analog sensors, or identify suitable existing tech (\hyperref[a.6-compute-accounting-via-analog-sensors]{\ul{Appendix A.6}}).
\vspace{0.5em}

\textcolor{gray}{\rule{0.8em}{0.8em}} \textbf{D.} \textbf{Tamper-proofing}: Design and manufacture appropriate tamper-proof server enclosures (\hyperref[off-chip-verification-layers]{\ul{Section 4.2}}).
\vspace{0.5em}
\end{minipage} \\
\begin{minipage}[t]{\linewidth}\raggedright
Cross-cutting
\end{minipage} & \begin{minipage}[t]{\linewidth}\raggedright
\textcolor{brown}{\rule{0.8em}{0.8em}} \textbf{A.} \textbf{R\&D funding} (e.g., by AISIs, philanthropists, DARPA)
\vspace{0.5em}

\textcolor{brown}{\rule{0.8em}{0.8em}} \textbf{B.} \textbf{Pilot programs} (e.g., voluntary corporate commitments, AISI collaborations)
\vspace{0.5em}

\textcolor{brown}{\rule{0.8em}{0.8em}} \textbf{C.} \textbf{Red teaming} (e.g., by companies, ICs, NIST contest) of developed proposals
\vspace{0em}
\end{minipage} \\

\bottomrule\noalign{}
\caption{Selected open challenges in verification R\&D. We discuss open problems more comprehensively (\hyperref[appendices]{\ul{Appendices}}; \hyperref[c.3-additional-rd-problems-for-verification]{\ul{Appendix C.3}}); this list is filtered (\hyperref[open-problems-in-verification]{\ul{Section 5}}). These filters and the unclassified nature of this research result in the table not listing challenges for layers (4-6).} \label{tab:rd_challenges}
\endlastfoot
\end{longtable}

\clearpage

\tableofcontents

\clearpage

\addtocontents{toc}{\protect\setcounter{tocdepth}{2}}

\phantomsection\section{1. Introduction}\label{introduction}

\textbf{Cutting-edge AI systems are increasingly capable at varied tasks, raising their potential for benefits and harms.} General-purpose AI systems have the potential to improve education, medicine, and research and prosperity more broadly, but continued progress in capabilities also could enable catastrophic misuse (e.g., in biological and cyber attacks) and accidents (e.g., via deployment of unreliable systems in high-stakes use cases, as well as loss of human control of highly intelligent systems) \citep{doi:10.1126/science.adn0117, bengio2025internationalaisafetyreport, bengio2025internationalscientificreportsafety, ngo2025alignmentproblemdeeplearning, BletchleyDeclaration2023}.

\textbf{To mitigate AI's risks while reaping its benefits, verified AI regulations and international agreements may be needed.} Rules may need to be enforced because even well-meaning actors face competitive incentives to deprioritize AI's national security risks and other externalities \citep{Armstrong2013Racing, askell2019rolecooperationresponsibleai, Ganguli_2022}, and malicious actors may have little concern for unenforced rules. To enable timely and effective enforcement, verification of compliance is essential \citep{Dai2002InformationSystems}.\footnote{A verification procedure might not directly detect a substantive violation; a non-compliant party might refuse to cooperate. Still, such uncooperativeness could be presumed to indicate a substantive violation (as in nuclear non-proliferation) \citep{rosenthal_iaea_safeguards}, and it could be defined as a procedural violation.} These benefits of verification arise in both international agreements and domestic regulation.

\textbf{The shortage of rules on AI and verification of compliance is already motivating lowered safety standards and blunt policies.} For instance, leading AI companies have in some cases fallen short of commitments to invest in safety \citep{Kahn2024OpenAI}, reportedly deprioritized ``safety culture and processes'' \citep{robison2024openairesign, roose2024openai}, suppressed whistleblowers \citep{verma2024openaiwhistle}, published safety frameworks lacking commitments \citep{Dragan2024FrontierSafety}, and rushed product releases \citep{devynck2023bingai, Gordon2024Google}. Nationally, both U.S. and Chinese lawmakers appear to be backing off from more ambitious proposals for regulating large-scale AI, as each government seeks to avoid falling behind the other in AI.\footnote{In the U.S. Congress, tech lobbyists appear to have reversed a push for regulation by ``arguing that strict safety rules would hand America's AI edge to China'' \citep{Bordelon2024}. In the months since, senior legislators from both parties have opposed prominent AI safety proposals \citep{HillValley2024Speaker, Pelosi2024SB1047}, at times explicitly highlighting tradeoffs between safety and competitiveness: ``some of us are also concerned---and I know some of you share these concerns---that {[}AI{]} could also be a very destructive tool {[}...{]} {[}but{]} if we're too aggressive in our regulations {[}...{]} we\textquotesingle ll wind up ceding ground to China.'' Similarly, the Chinese government has backpedaled from stricter AI safety proposals, amidst discussion that ``{[}f{]}ailing to develop is the greatest threat to security'' \citep{Sheehan2023ThreadX, Sheehan2024ChinaViews}.} With effective verification of risk mitigation practices globally, countries could have more confidence that caution on national security will not be exploited by corner-cutting in adversary countries. Meanwhile, the United States is controlling exports of high-end AI chips to over a dozen countries \citep{bis_semiconductor_restrictions_2023, heim2025diffusion}, affecting benign uses along with the targeted military uses. With effective verification of how exported chips are used rather than sweeping bans at a country level, American companies and other companies in the supply chain could potentially attain greater revenue, while still achieving the national security objectives that motivate sweeping bans.

\textbf{As AI capabilities continue to advance, verification will likely become increasingly important.} More capable AI systems will pose greater opportunities and risks, increasing the value of the incentives and precise enforcement ability brought by rules with verified compliance. Meanwhile, today's debated view that AI poses risks on par with nuclear war \citep{CAIS2025Risk, bengio2025internationalaisafetyreport} might become a consensus. Consider, for example, the potential scientific and political response to an AI-enabled terrorist attack, a near-miss crisis, or experimental results that provide clear evidence of severe AI threats. The historical responses to the 9/11 terrorist attacks \citep{unsc_1373_2001, cfr_911_foreign_policy} and to the Cuban Missile Crisis \citep{iaea_evolution_1998, timerbaev_npt_interview_2017} suggest threats of such magnitude can motivate deep and relatively quick international cooperation, even among geopolitical rivals like the United States and Soviet Union. Even then, states might only cooperate if they can verify each other's compliance with agreements.\footnote{With some risks, one-sided compliance could put the compliant party at a disadvantage while doing little to reduce risk. For example, suppose some AI system, if not developed securely, would make it easier for terrorists to obtain weapons of mass destruction, or would lead to an uncontrolled AI system causing global damage. If only one state takes care to ensure its AI systems do not cause these hazards, it may still face the same hazards from other states' AI systems, while having weakened its own AI industry.}

\textbf{\emph{Confidentiality-preserving} and secure verification of rules on \emph{large-scale} AI has unique potential.} Historically, it has been politically crucial for international verification methods to avoid information leaks that create serious security risks, and (to a lesser extent) to avoid leaks of trade secrets \citep{coe_arms_control_2019, baker2023nucleararmscontrolverification}.\footnote{By ``leaks,'' we mean any unauthorized access, public or not (assuming narrow access exclusively for verifying compliance with agreed-on rules is authorized).} The AI industry has an abundance of highly sensitive information, from AI model weights \citep{Nevo2024} and algorithms to training/user data. Confidentiality-preserving\footnote{We use ``confidentiality-preserving`` to refer to protecting the confidentiality of corporate intellectual property and state secrets, as well as the privacy of individuals with sensitive data contained in AI datasets (i.e., ``privacy-preserving'').} verification may be especially important and feasible for \emph{large-scale} AI development and deployment \citep{shavit2023doescatchchinchillaverifying}, such as training a future, powerful model and deploying it at scale. Such large-scale AI activities carry unique risks \citep{doi:10.1126/science.adn0117, bengio2025internationalscientificreportsafety, BletchleyDeclaration2023}. They are also industrial, billion-dollar-scale undertakings \citep{henshall_billion_2024, amazon_anthropic_2024, efrati2024openailoss}, requiring ``thousands of specialized chips'' \citep{sastry2024computingpowergovernanceartificial} and counting \citep{epoch2024trainingcomputeoffrontieraimodelsgrowsby45xperyear}. This broad trend continues to hold despite algorithmic advances, such as those of reasoning models and DeepSeek's R1 \citep{Heim2025DeepSeek}.

\textbf{Despite the growing need for them, methods to verify rules on large-scale AI development and deployment are nascent, and they face major challenges.} While some fields such as nuclear arms control have had decades to refine verification methods \citep{cfr_us_russia_arms_control, rosenthal_iaea_safeguards}, large-scale AI's growing capabilities only attained intense policy attention since 2023 \citep{henshall_ai_policy_2023, maslej2023artificialintelligenceindexreport}. In addition, AI verification is not straightforward: AI data centers might be hidden or dispersed, there is a large leap from locating a data center to determining how it is used, and simply examining public AI products would miss non-public violations (among other limitations).\footnote{Checking public AI products (via their websites or APIs) would have several limitations. (1) These checks would not reveal \emph{non-public} large-scale AI deployments, such as large AI disinformation farms, cyber attack deployments, or unsafe R\&D uses. (2) Dishonest actors might recognize when they are being tested (e.g., based on prompts or IP addresses) and spoof tests, though recognizing tests may be difficult. (3) Stakeholders may wish to know not just how an AI model behaves, but what its training history and innate properties are. For example: ``Is a model vulnerable to malicious use because its safeguards were fine-tuned away, or because the intended safeguards fall short?''} Further, verifying rules on large-scale AI will require obtaining assurances about a very complex and rapidly changing technology stack---semiconductors and AI \citep{Khan2021SemiconductorSupply}---against adversaries who may be extremely capable \citep{Nevo2024} and incentivized to circumvent verification.

\textbf{While research has broken ground on AI verification, the field could use improved conceptual clarity and more developed proposals.} From 2023 to 2025, there have been a handful of research papers explicitly studying frontier AI verification, from developing specific proposals to overviewing the landscape (\hyperref[c.5-related-work]{\ul{Appendix C.5}}). This report was motivated by the view that progress on verification would benefit from improved clarity on important conceptual questions, in particular: how can specific verification mechanisms be combined effectively? How many distinct layers of redundancy could verification achieve? In addition, existing research has limited discussion of how some proposals could be implemented. This report aims to make progress on these fronts, by proposing clarifying frameworks and outlining new implementation options (especially for how hardware security features, compute accounting, and whistleblower programs could be used for effective verification).

The remainder of this report is structured as follows. \hyperref[verification-scope-and-research-methodology]{\ul{Section 2}} describes the scope of verification considered in this report and this report's methodology. \hyperref[verification-framework]{\ul{Section 3}} explains our framework of verification subgoals. \hyperref[verification-mechanisms-and-layers]{\ul{Section 4}} describes and analyzes potential verification mechanisms and layers. \hyperref[open-problems-in-verification]{\ul{Section 5}} lists open problems. \hyperref[conclusion]{\ul{Section 6}} concludes.

\phantomsection\subsection{1.1 Contributions}\label{contributions}

Our report primarily makes the following contributions:

\begin{enumerate}
\def\labelenumi{\arabic{enumi}.}
\item
  \textbf{Verification goal} (\hyperref[verification-scope-and-research-methodology]{\ul{Section 2}}): We start by outlining a verification goal that is widely applicable for AI governance: verifying that large-scale AI development and deployment complies with rules, for rules on models, data, and code.
\item
  \textbf{Verification subgoals} (\hyperref[verification-framework]{\ul{Section 3}}): We identify four subgoals by which this goal can be achieved. A verification regime is only as robust as its weakest completion of a subgoal, so identifying subgoals helps us assess the robustness of verification proposals.\footnote{More precisely, a Prover who minimizes their probability of being caught would choose to circumvent the verification subgoal that has the lowest detection probability, so the overall detection probability would equal that of the lowest-detection-probability subgoal. However, this heuristic does not consider small violations at multiple subgoals that add up (e.g., using a mix of unaccounted-for compute in declared data centers and undeclared data centers).}
\item
  \textbf{Verification mechanisms and layers} (\hyperref[verification-mechanisms-and-layers]{\ul{Section 4}}): We identify over 20 verification mechanisms and analyze their challenges. These are the ``building blocks'' of verification regimes; each can help complete at least one of the above subgoals. We show how the verification mechanisms can be assembled into 6 distinct ``layers'' of verification, and we analyze these layers. By ``verification layer,'' we mean a set of similar mechanisms, with one mechanism for each verification subgoal.
\item
  \textbf{Open problems} (\hyperref[open-problems-in-verification]{\ul{Section 5}}): We list open R\&D problems for advancing AI verification.
\end{enumerate}

For more details, readers may be interested in the following, mostly more technical contributions within the above, where we examine how under-explored verification mechanisms could be implemented:

\begin{itemize}
\item
  We outline approaches to implementing the following verification mechanisms, analyzing potential attacks and countermeasures:

  \begin{itemize}
  \item
    Hardware-backed workload certificates, using secure boot (\hyperref[a.2-hardware-backed-workload-certificates-and-evaluations]{\ul{Appendix A.2}})
  \item
    Compute accounting via analog sensors on AI chips (\hyperref[a.6-compute-accounting-via-analog-sensors]{\ul{Appendix A.6}})
  \item
    Whistleblower programs (\hyperref[a.8-whistleblower-programs]{\ul{Appendix A.8}})
  \end{itemize}
\item
  We highlight how verification of training can be generalized to verifying large-scale inference, considering replicability (\hyperref[a.9-deterministic-replication-of-neural-network-inference]{\ul{Appendix A.9}}) and sensitive data storage (\hyperref[a.10-storing-sensitive-data-for-verification]{\ul{Appendix A.10}}), and we outline additional tests to detect spoofs (\hyperref[a.4-partial-workload-re-execution-with-constraints]{\ul{Appendix A.4}}; \hyperref[a.5-data-and-code-validation]{\ul{Appendix A.5}}).
\item
  We explore in some detail the major challenge of ensuring verification protocols are implemented securely throughout the infrastructure stack (\hyperref[a.1-full-stack-security-for-technical-verification-mechanisms-implementation]{\ul{Appendix A.1}}).
\end{itemize}

\phantomsection\subsection{1.2 Limitations}\label{limitations}

The verification mechanisms and layers we identify are not necessarily comprehensive, nor is our analysis of them. In addition, this report has various scope limitations (\hyperref[verification-scope-and-research-methodology]{\ul{Section 2}}).

\phantomsection\section{2. Verification Scope and Research Methodology}\label{verification-scope-and-research-methodology}

\phantomsection\subsection{2.1 Rules on AI Models, Data, and Code}\label{rules-on-ai-models-data-and-code}

\textbf{This report covers options for verifying compliance with rules on the \emph{AI models, data, and code} created or used in large-scale AI development and deployment.} That is, if a rule can be specified in terms of measurable properties of these AI models, data, and code,\footnote{The ``data'' here refers to training data as well as deployment data, i.e., usage prompts/inputs and associated outputs (and, if desired, intermediate outputs/activations). We use ``code'' to refer primarily to the algorithms used for training AI models (i.e., model architectures, optimization algorithms, and hyperparameter values). We specifically consider properties that affect system behavior, rather than e.g., code formatting.} \footnote{The ``created or used in large-scale AI development or deployment'' criterion excludes rules even on small-scale compute uses within large-scale AI activities. For example, verifying a rule on 90\% of a frontier AI model's training data would be in scope, but verifying that 100\% of the training data complies with the rule would not necessarily be in scope, as this rule could be violated with small-scale compute use.} the options we cover should allow verifying compliance with the rule. This covers many hypothetical rules for (near-)frontier AI (\hyperref[tab:examples_of_potential_rules]{\ul{Table 3}}).\footnote{This analysis builds on discussion by \cite{shavit2023doescatchchinchillaverifying} of ``What types of rules can we enforce\ldots''} These rules may seem quite different, but much of the challenge of verifying is the same: identifying what models, code, and data were used for large-scale AI development and deployment, so one can run desired tests on them. Still, we do not necessarily cover verifying compliance with rules on other AI-related matters, such as the general cybersecurity of AI companies or their ownership structures, though some of the mechanisms we review could help there too.\footnote{On-chip mechanisms and network taps could help verify compliance with intra-chip and intra-cluster cybersecurity rules. Personnel-based verification layers such as whistleblower programs have wide applicability. For example, they would be well-suited to revealing whether most employees are subject to cybersecurity practices that plainly impact their day-to-day work.} Note that precisely specifying rules that meaningfully achieve security or other governance objectives is a critical open challenge beyond this report's scope (\hyperref[other-scope-limitations]{\ul{Section 2.3}}).

{
  \setlength{\LTleft}{-33pt}
  \setlength{\LTright}{-26pt}
  \centering
    \begin{longtable}[]{
      >{\raggedright\arraybackslash}p{(\linewidth - 4\tabcolsep) * \real{0.1616} * \real{1.15}}
      >{\raggedright\arraybackslash}p{(\linewidth - 4\tabcolsep) * \real{0.3741} * \real{1.15}}
      >{\raggedright\arraybackslash}p{(\linewidth - 4\tabcolsep) * \real{0.4643} * \real{1.15}}}
    \toprule\noalign{}
    \begin{minipage}[t]{\linewidth}\raggedright
    \textbf{Potential governance goal}
    \end{minipage} & \begin{minipage}[t]{\linewidth}\raggedright
    \textbf{Hypothetical rule}
    
    (using ``frontier model'' as shorthand for AI models trained with large-scale AI compute (\hyperref[rules-on-large-scale-ai-compute]{\ul{Section 2.2}}))
    \vspace{0.5em}
    \end{minipage} & \begin{minipage}[t]{\linewidth}\raggedright
    \textbf{How the rule could be specified as a rule on AI models, data, and code}
    
    (illustrative, high-level)
    \end{minipage} \\
    \endhead
    \toprule\noalign{}
    \rowcolor{black!5!white}
    \multirow[t]{4}{=}{\begin{minipage}[t]{\linewidth}\raggedright
    Safety \& security
    \end{minipage}} & \begin{minipage}[t]{\linewidth}\raggedright
    \textbf{Model evaluations:} Regularly test frontier models for risks to international security.
    \end{minipage} & \begin{minipage}[t]{\linewidth}\raggedright
    Models and outputs in deployment could be evaluated for dangerous capabilities and propensities, at regular intervals of training compute or benchmark performance.
    \end{minipage} \\
    \rowcolor{black!5!white}
    & \begin{minipage}[t]{\linewidth}\raggedright
    \textbf{Deployment mitigations:} Implement risk mitigation practices for large-scale frontier AI deployment.
    \end{minipage} & \begin{minipage}[t]{\linewidth}\raggedright
    Mitigations may often be specifiable in terms of input data and output data (e.g., filter out some kinds of inputs, run oversight checks on outputs), proportionately to evaluated risks.
    \end{minipage} \\
    \rowcolor{black!5!white}
    & \begin{minipage}[t]{\linewidth}\raggedright
    \textbf{Development mitigations:} Implement risk mitigation practices in frontier AI development.
    \end{minipage} & \begin{minipage}[t]{\linewidth}\raggedright
    Mitigations could be practices for model architectures, training data, and code, proportionate to evaluated risks.
    \end{minipage} \\
    \rowcolor{black!5!white}
    & \begin{minipage}[t]{\linewidth}\raggedright
    \textbf{Halts:} Refrain from further development or large-scale deployment of a model while it would pose unmanageable security risks.
    \end{minipage} & \begin{minipage}[t]{\linewidth}\raggedright
    Further development would create a new model with higher performance and a suspect training history. Further deployment would create new output data from the risky model.
    \vspace{0.5em}
    \end{minipage} \\
    \begin{minipage}[t]{\linewidth}\raggedright
    Transparency
    \end{minipage} & \begin{minipage}[t]{\linewidth}\raggedright
    \textbf{Disclosures:} Publicly disclose all frontier models, their training compute, and/or evaluation results.
    \end{minipage} & \begin{minipage}[t]{\linewidth}\raggedright
    Training compute is a function of model architecture and training data, and the other rules here are about models.
    \vspace{0.5em}
    \end{minipage} \\
    \rowcolor{black!5!white}
    \multirow[t]{2}{=}{\begin{minipage}[t]{\linewidth}\raggedright
    Benefit-sharing
    \end{minipage}} & \begin{minipage}[t]{\linewidth}\raggedright
    \textbf{Sharing claimed models:} Share access to a specific model internationally.
    \end{minipage} & \begin{minipage}[t]{\linewidth}\raggedright
    Output data from remote deployment are in fact generated by a model with claimed specifications and performance, run on the intended input data.
    \end{minipage} \\
    \rowcolor{black!5!white}
    & \begin{minipage}[t]{\linewidth}\raggedright
    \textbf{Sharing leading models:} Share access to one's best-performing models internationally.
    \end{minipage} & \begin{minipage}[t]{\linewidth}\raggedright
    As above, output data from remote deployment are in fact generated by a model with claimed specifications and performance, run on the intended input data. Further, the model outperforms the developer's other models.
    \vspace{0.5em}
    \end{minipage} \\
    \begin{minipage}[t]{\linewidth}\raggedright
    Arms control
    \end{minipage} & \begin{minipage}[t]{\linewidth}\raggedright
    \textbf{Military limits:} Restrict frontier models specialized for military purposes
    \end{minipage} & \begin{minipage}[t]{\linewidth}\raggedright
    Restrictions may be specifiable by limits on data (e.g., use of weapons data), though small-scale AI may often suffice for narrow applications.
    \vspace{0.5em}
    \end{minipage} \\
    \rowcolor{black!5!white}
    \begin{minipage}[t]{\linewidth}\raggedright
    Usage norms
    \end{minipage} & \begin{minipage}[t]{\linewidth}\raggedright
    \textbf{International norms:} Deploy large-scale AI while meeting norms beyond safety \& security.
    \end{minipage} & \begin{minipage}[t]{\linewidth}\raggedright
    Mass-deployed models could have fine-tuning data, input filters, etc. that promote norms such as upholding international agreements, individual rights, and truthfulness.
    \vspace{0.5em}
    \end{minipage} \\
    
    \bottomrule\noalign{}
    \caption{Examples of potential rules on AI, in-scope for this report.} \label{tab:examples_of_potential_rules}
    \endlastfoot
    \end{longtable}
}

\phantomsection\subsection{2.2 Rules on Large-Scale AI Compute}\label{rules-on-large-scale-ai-compute}

\textbf{Large-scale AI---motivation.} This report focuses on options for verifying that large-scale AI development and large-scale AI deployment complies with rules. These activities involve unique risks and scale of resource use (\hyperref[introduction]{\ul{Section 1}}), so verifying their compliance could be both important and practical. In contrast, verifying small-scale compute use could require intruding more into lower-performance compute clusters and even consumer hardware. Verifying small-scale compute use would also require more precise verification methods.\footnote{Historical and proposed verification methods often rely on random sampling and approximation \citep{rosenthal_iaea_safeguards, shavit2023doescatchchinchillaverifying} (\hyperref[a.4-partial-workload-re-execution-with-constraints]{\ul{Appendix A.4}}; \hyperref[a.6-compute-accounting-via-analog-sensors]{\ul{Appendix A.6}}), so they may struggle to distinguish small violations from statistical noise.}

\textbf{Large-scale---definitions.} We define ``large-scale'' with a few corresponding terms, chosen to approximately track the compute use of near-frontier AI development over time \citep{EpochNotableModels2024} (\hyperref[addressing-broader-challenges-for-verification]{\ul{Section 3.3}}):\footnote{Thus, a much higher threshold might fail to include future, near-frontier AI activities that may pose national security risks. Meanwhile, a much lower threshold might be impractically intrusive on consumer hardware (\hyperref[c.4-scope-additional-notes]{\ul{Appendix C.4}}). Finally, we avoid defining ``large-scale'' in terms of today's leading chips, since today's industrial-grade compute may be a future year's consumer-grade compute.}

\begin{itemize}
\item
  A \emph{compute cluster} or data center is ``large-scale'' if it has the computing power of thousands of high-end AI chips (even if the chips are distributed over many locations).\footnote{We do not define any large enough set of chips as a computing cluster; they must also be controlled by a single entity. It is unconventional to refer to large, distributed inference compute as a single cluster, but doing so allows us to discuss verification more concisely (by being able to refer to ``detecting large-scale clusters'' instead of ``...clusters or comparably high-performance distributed inference compute'').}
\item
  \emph{Compute use} is ``large-scale'' if it is an amount of computation that thousands of high-end AI chips can do over multiple months.\footnote{That is, assuming hardware utilization that is common industry practice for the workload.} \footnote{Frontier AI development tends to be done on the scale of multiple months \citep{EpochNotableModels2024}.}
\item
  \emph{AI development or deployment} is ``large-scale'' if it uses thousands of high-end AI chips over multiple months.
\end{itemize}

Though this report uses the more general definition of ``thousands'' of AI chips, it may be practical to limit verification to an even higher threshold, such as hundreds of thousands of AI chips, as frontier AI development is already near that scale and growing as of early 2025 \citep{EpochNotableModels2024, Pilz2025TrendsSupercomputers}.

\textbf{Small-scale compute use.} Due to this report's focus on large-scale AI compute use, the verification options we cover do not necessarily enable verifying rules on small-scale AI compute use, i.e., the development of small (typically narrow-purpose) models or the small-scale deployment of frontier models. Other compute-centric governance approaches share this scope limitation \citep{shavit2023doescatchchinchillaverifying, sastry2024computingpowergovernanceartificial}. This could be seen as a feature, as it limits the potential for government overreach. At the same time, this gap has downsides; it is not clear that frontier AI deployment must be done at scale to be dangerous. Still, two of the verification layers we consider might enable verifying rules even on small-scale deployment of frontier AI models.\footnote{On-chip mechanisms and network taps offer full transparency into AI chips' code and data exchanges, which might allow for verifying that a frontier AI model remains highly contained (e.g., copies of model weights are encrypted and only decryptable by other AI chips with on-chip mechanisms) and that all its uses are known \citep{Scher2024Verification}. In contrast, the personnel-based layers appear unlikely to be robust here due to the few people needed for small-scale AI deployment, and analog sensors appear unlikely to be robust because small-scale deployments could hide in their margins of error.}

\phantomsection\subsection{2.3 Other Scope Limitations}\label{other-scope-limitations}

As further scope limitations, this report primarily:

\begin{itemize}
\item
  Identifies concrete R\&D challenges, leaving open the R\&D work to solve them if possible.
\item
  Focuses on verifying compliance with rules, rather than specifying rules well---another important unsolved problem (which includes improving model evaluations) (\hyperref[open-problems-in-verification]{\ul{Section 5}}).\footnote{Many hypothetical rules on AI rely on evaluations to measure risks and the extent to which risks have been mitigated. However, existing evaluations of both risks and mitigations are limited (partly because existing mitigations are limited), and improving them is an active area of research \citep{Hobbhahn2024EvalsGap, bengio2025internationalaisafetyreport}. This report focuses on protocols by which---given an evaluations suite---it could help verify that AI models, data, and code are compliant. These are quite different problems, e.g., one could have a perfect evaluations suite but be unable to run it confidentially, or be unaware of hidden models that should be tested.}
\item
  Examines verification protocols, not what organizations should carry them out.
\item
  Focuses on verification as in detecting non-compliant parties \citep{Dai2002InformationSystems}, after which one still needs to penalize or stop them (\hyperref[addressing-broader-challenges-for-verification]{\ul{Section 3.3}}).\footnote{This conceptual division matches an organizational division in nuclear nonproliferation: the International Atomic Energy Agency (IAEA) is responsible for verifying compliance, and the IAEA primarily refers cases of apparent non-compliance to states for enforcement (primarily by reporting to the UN Security Council) \citep{rosenthal_iaea_safeguards}.} We do not cover this latter step of enforcement, though a few verification mechanisms double as enforcement tools.\footnote{In particular, tamper-proof, compliance-locked hardware (\hyperref[verification-mechanisms]{\ul{Section 4.1.1.2}}; \hyperref[verification-mechanisms-1]{\ul{Section 4.2.1.2}}) would verify compliance by blocking AI hardware from being used for violations at all---which also enforces compliance. The known locations of AI compute clusters (Subgoal 2.B) could also inform enforcement actions.}
\end{itemize}

\phantomsection\subsection{2.4 Methodology}\label{methodology}

\phantomsection\subsubsection{2.4.1 Sources}\label{sources}

\textbf{Literature review.} We reviewed a range of open sources---primarily research reports from academic, industry, and think tank sources---in relevant fields.\footnote{These included areas of machine learning, AI policy, computer security \& cryptography, international relations, and arms control verification.} \footnote{To identify sources, we drew on relevant papers we were familiar with from our previous work in this field \citep{brundage2020trustworthyaidevelopmentmechanisms, baker2023nucleararmscontrolverification, Kulp2024, Heim2024Cloud}, followed citation trails, searched relevant terms (\hyperref[c.5-related-work]{\ul{Appendix C.5}}) on Google Scholar and Google Search, and considered publications mentioned by interviewees.} \hyperref[c.5-related-work]{\ul{Appendix C.5}} provides a more detailed list of areas we examined, including key search terms and major sources.

\textbf{Expert interviews} (\hyperref[c.2-methodology-for-expert-interviews-and-interview-protocol]{\ul{Appendix C.2}}). We conducted 18 semi-structured interviews with a range of subject-matter experts\footnote{Our interviewees' backgrounds spanned computer security, hardware, machine learning, AI policy, and international relations, as well as experience in academia, industry, and nonprofits.} to validate our findings. In our interviews, we asked these experts to highlight possible errors or omissions in our findings, or to answer narrower technical questions.

\phantomsection\subsubsection{2.4.2 Methodology for Analysis}\label{methodology-for-analysis}

Our analytic methodology, detailed in \hyperref[c.1-methodology-for-analysis]{\ul{Appendix C.1}}, consisted of:

\begin{enumerate}
\def\labelenumi{\arabic{enumi}.}
\item
  Developing a framework of verification subgoals: We took as a starting point a framework used by the International Atomic Energy Agency (IAEA), identified through our literature review, and modified it until it met our criteria of deductive validity, flexibility, and simplicity.
\item
  Identifying verification mechanisms: We identified candidate verification mechanisms by compiling verification mechanisms from the above sources.
\item
  Assessing, red teaming, and enhancing verification mechanisms; and identifying open problems: We evaluated each mechanism's effectiveness, rate of false alarms, confidentiality, security, setup speed, and cost, while iterating on outlined implementations. These analyses drew from the literature review and expert interviews.
\item
  Identifying and analyzing verification layers: We grouped the more positively assessed mechanisms into layers, using our above analysis and definition of verification layer.
\end{enumerate}

\phantomsection\section{3. Verification Framework}\label{verification-framework}

\phantomsection\subsection{3.1 Context for This Framework}\label{context-for-this-framework}

\begin{figure}[ht]
  \noindent
  \makebox[\textwidth][c]{%
    \includegraphics[width=6.5in,height=2.98611in]{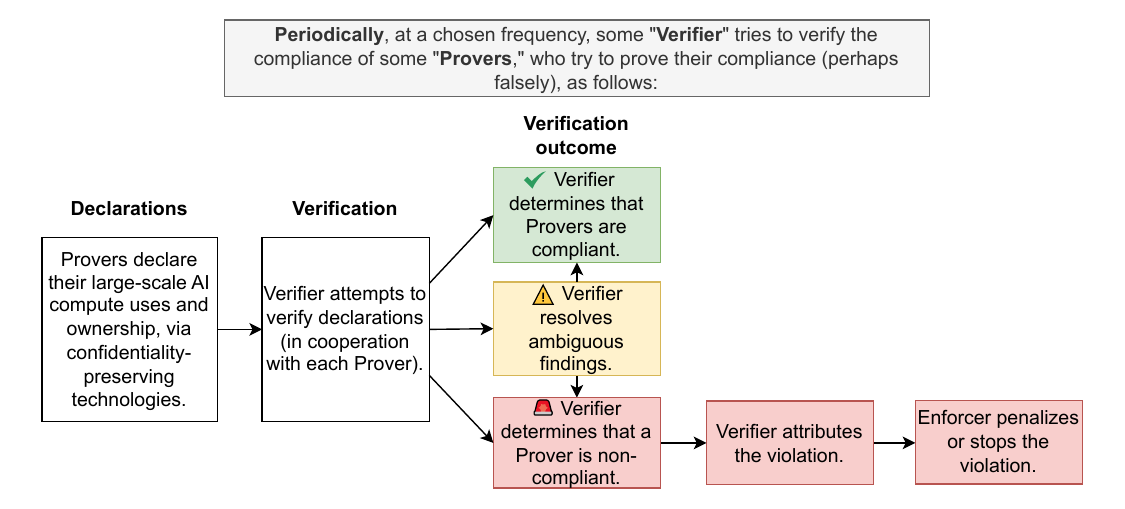}%
  }
  \caption{Assumed context for verification: verification is preceded by declarations and can lead to enforcement.}
  \label{fig:assumed_context_for_verification:_verification_is_preceded_by_declarations_and_can_lead_to_enforcement}
\end{figure}

\textbf{A Prover and a Verifier.} We consider a ``Prover'' who is trying to prove their compliance to a ``Verifier,'' drawing these terms from computer security \citep{Goldwasser1985KnowledgeComplexity}. The Prover could be a private institution or (in the case of international agreements) a government, which could constrain private companies within its territory as part of the agreement. The Verifier could be a government body or a third party.\footnote{In nuclear arms control, multilateral verification is done both by a third party---the International Atomic Energy Agency---while bilateral verification was done by national institutions---the United States and Soviet Union directly verifying each other's compliance.}

\textbf{Framework structure and assurances.} This report's framework breaks down verification into various subgoals, each of which consists of verifying some claim. By verifying all the claims in the framework, the Verifier can be confident in the Prover's compliance. This is because, by design (\hyperref[c.1-methodology-for-analysis]{\ul{Appendix C.1}}), if all the claims in the framework are true, then it will logically follow that the Prover cannot have used large-scale AI compute in a non-compliant manner.\footnote{In practice, the Verifier will not be completely certain about each subgoal they completed, so the conclusion will also not hold with complete certainty (for example, the Verifier may deem 90\% credence sufficient).}

\textbf{AI compute accounting.} The framework not only has large-scale AI compute use as its scope but also takes the approach of AI compute accounting; it seeks to verify compliance on the basis that all large-scale AI compute use is accounted for in compliant activities. One could hypothetically attempt to instead account for the use of other resources associated with large-scale AI compute use, such as electrical power or algorithms. Compared to AI compute, these other resources have broader uses (e.g., the many uses of electrical power), or they have much smaller physical footprints (e.g., the smaller spaces taken up by stored data or algorithms) (\hyperref[b.1-compute-accounting-vs.-other-kinds-of-accounting]{\ul{Appendix B.1}}). These factors suggest that AI compute use can be accounted for less intrusively than other resources.

\phantomsection\subsection{3.2 The Framework}\label{the-framework}

Our verification framework breaks down verifying the compliance of large-scale AI models, data, and code into subgoals, each of which is verifying some claim. These subgoals can be used to design or evaluate options for verifying rules on AI.

\textbf{Declarations.} The framework involves requirements for organizations to confidentially self-report, or ``declare,'' information. In this framework, organizations that own or use large-scale AI compute (e.g., major AI companies and cloud compute providers) would be required to declare facts about (i) their \emph{ownership} of large AI compute clusters (or large, decentralized quantities of AI compute),\footnote{An AI chip registry is one possible structure for reporting AI compute ownership; declaring individual AI chips is not strictly necessary but facilitates verification.} and (ii) their \emph{use} of large AI compute clusters. Verification focuses on checking that these declarations are correct and complete.\footnote{Checking completeness means checking that there are no omissions---that is, no undeclared, large-scale AI compute clusters, nor undeclared large-scale uses of AI compute clusters.} The declarations would include model weights, training and usage data, and training code, as well as other information required by verification protocols (e.g., intermediate results of computations, information about hardware utilization). We define ``large-scale'' compute use in terms of thousands of AI chips (\hyperref[rules-on-large-scale-ai-compute]{\ul{Section 2.2}}), but this is imprecise; policymakers could define more precise thresholds (and update them over time) based on future learnings about what scale of AI compute use warrants verification and is practically verifiable.

\textbf{Use of confidentiality-preserving technology.} Importantly, in our framework, declarations of AI compute use would be reported and verified via confidentiality-preserving technology---technology that enables a Prover to demonstrate their compliance without leaking their highly sensitive IP such as model weights. Such technology could include (i) a hardware security feature known as Confidential Computing (\hyperref[prerequisites-hardware-security-features]{\ul{Section 4.1.1.1}}); and (ii) compute clusters with security that both parties can confirm, so that much information can enter these devices but only a small amount of information (e.g., compliance determinations) can leave (\hyperref[prerequisites-off-chip-devices]{\ul{Section 4.2.1.1}}). Declarations of AI compute ownership are less sensitive than declarations of AI compute usage,\footnote{Declarations of compute use would contain the core IP of AI labs (namely model weights and algorithms)---ownership of which directly confers economic (and perhaps eventually, strategic)---benefits, making these declarations highly sensitive. In contrast, declarations of compute ownership (containing information about AI data centers) are not as directly useful to competitors or adversaries. Hypothetically, an adversary could abuse compute ownership declarations to set military targets, but an analogous risk did not keep the United States and Soviet Union from sharing the locations of their nuclear weapon bases with each other \citep{baker2023nucleararmscontrolverification}. Additionally, risk of targeting could be mitigated via deterrence and shielding.} but, if desired, the confidentiality-preserving technologies we discuss could also help protect these ownership declarations. As we will discuss (\hyperref[implementation-options-across-mechanisms]{\ul{Section 4.5}}), confidentiality-preserving technologies could be used to run hard-coded compliance tests, or perhaps to facilitate iterative testing by humans or AI agents, though the human option poses more confidentiality challenges.

\begin{figure}[ht]
  \noindent
  \makebox[\textwidth][c]{%
    \includegraphics[width=6.5in,height=6.375in]{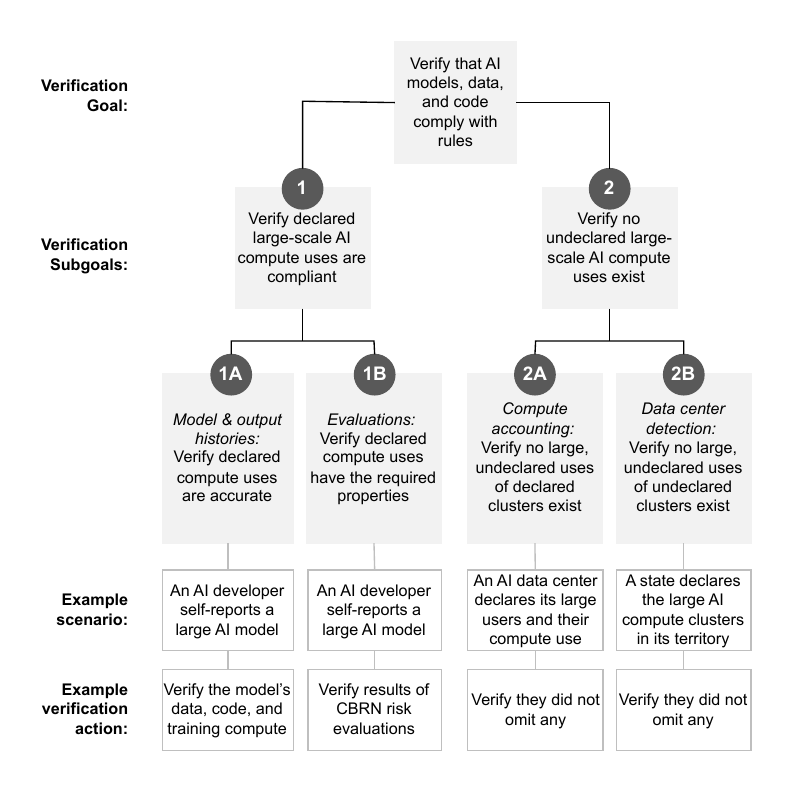}%
  }
  \caption{Framework of verification subgoals.}
  \label{fig:framework_of_verification_subgoals_duplicate}
\end{figure}

\textbf{Subgoals for verifying rules on large-scale AI:} The framework begins with a broad verification goal: verifying that AI models, data, and code comply with rules on large-scale AI development and deployment. The framework decomposes this goal into two subgoals: (1) verify that \emph{declared} uses of large-scale AI compute are compliant, and (2) verify that there are no \emph{undeclared} uses of large-scale AI compute (i.e., declarations are complete). ``Compliant'' here refers to compliance with rules on the AI models, data, or code created or used in large-scale AI development and those used in deployment (\hyperref[verification-scope-and-research-methodology]{\ul{Section 2}}). This is a valid decomposition; if declared large-scale uses are compliant and there are no undeclared large-scale uses, then \emph{all} large-scale uses of AI compute must be compliant.\footnote{However, a small-scale violation in each of the two categories could combine to constitute a large-scale violation. To address this, a buffer could be added to the thresholds defining ``large-scale'' (e.g., having the thresholds for the scope of each subgoal be half of how the threshold is defined for the larger goal). This also applies for the remaining break-downs.} These two subgoals can be decomposed further:

\begin{itemize}
\item
  \textbf{Subgoal 1:} verifying that declared uses of large-scale AI compute are compliant, faces two problems: a Prover might make a false declaration, or they might make a declaration that is honest but still non-compliant. ``Honest but non-compliant'' might mean, for example, openly declaring their AI deployment and hoping that a Verifier fails to notice that it lacked a required approval, or honestly declaring data that contains prohibited biological sequence data. Addressing each of these possible violations, we can break down Subgoal (1) into:

  \begin{itemize}
  \item
    \textbf{Subgoal 1.A:} Verify that declared uses of AI compute are declared accurately, i.e., the Prover actually did the claimed development or deployment. Equivalently,\footnote{Workload declarations are declarations of how the result of an AI workload---an AI model or output---is generated, so verifying the accuracy of a workload declaration is equivalent to verifying that AI models and outputs are generated as declared.} verify that declared AI models and outputs are generated as declared. More specifically, a Verifier could need to verify the accuracy of declared (1.A.1) AI training, (1.A.2) AI inference, or (1.A.3) non-AI uses of AI compute.
  \item
    \textbf{Subgoal 1.B:} Assuming that the declared uses are accurate (as is verified per Subgoal 1.A), verify they have the required properties. For example, the Verifier might evaluate an AI model's capabilities, or assess what kinds of data the model was run on, to check for compliance. The substance of these tests would depend on what rules are being verified (\hyperref[rules-on-ai-models-data-and-code]{\ul{Section 2.1}}), though as we will discuss, the infrastructure for running tests could be rule-agnostic.
  \end{itemize}
\item
  \textbf{Subgoal 2:} verifying that there are no undeclared uses of large-scale AI compute, also faces two problems: a Prover might try to make undeclared, large-scale uses of declared AI compute clusters, or of undeclared AI compute clusters. The Verifier can guard against both of these problems:

  \begin{itemize}
  \item
    \textbf{Subgoal 2.A:} Verify that there are no undeclared, large-scale uses of declared AI compute clusters. In other words, ensure AI compute use is accounted for, among declared AI compute clusters.
  \item
    \textbf{Subgoal 2.B:} Verify that there are no undeclared, large-scale AI compute clusters that could be used for violations. (Recall that we include large-scale, decentralized AI compute here (\hyperref[rules-on-large-scale-ai-compute]{\ul{Section 2.2}}).) This subgoal can be further broken down into verifying there are no such AI compute clusters (2.B.1) as parts of known AI data centers, nor (2.B.2) as standalone clusters.\footnote{The distinction here is that AI compute clusters are AI computing equipment, while AI data centers are the facilities that host one or more of these clusters.}
  \end{itemize}
\end{itemize}

\textbf{Verification subgoals can be completed in parallel.} Verifiers could speed up verification by completing multiple subgoals in parallel. For example, the Verifier could check for undeclared AI compute clusters and check for false usage reports at the same time.

\textbf{Verification subgoals can help design or evaluate verification regimes.} Knowing what mechanisms may complete each verification subgoal (\hyperref[fig:verification_layers_consist_of_distinct_mechanisms_for_each_verification_subgoal]{\ul{Figure 2}}), one can in principle design a robust verification regime by selecting robust mechanisms for each subgoal. Additionally, the verification subgoals can help evaluate the robustness of a proposed verification regime. One can identify what mechanisms a regime uses for completing each subgoal, and then identify the subgoal whose mechanisms are collectively least robust. This subgoal is the ``weak link'' of the regime---its robustness determines the regime's overall robustness.\footnote{That is, a verification regime's probability of detecting a violation is approximately that of the subgoal with the lowest probability of detecting a violation. This is because a Prover only needs to circumvent one verification subgoal to hide a violation, and a rational Prover seeking to hide a violation will choose to circumvent the subgoal where they are least likely to be caught. This assumes there are not practical considerations that motivate greater risk-taking from the Prover (e.g., high costs of a violation in one subgoal), and for simplicity it ignores the possibility of aggregating smaller violations across several subgoals (which may involve a lower or higher probability of detection, depending on the detection probabilities and their correlation).}

\textbf{Verification of narrower claims.} Completing all the subgoals allows for verifying fairly broad claims: claims about large-scale AI development \emph{and} deployment, even claims about what large-scale AI development or deployment was \emph{not} done. Some subgoals could be skipped or simplified if one only wished to verify narrower claims (\hyperref[b.2-verification-of-narrower-rules]{\ul{Appendix B.2}}), such as positive claims about how large-scale compute was used, or claims just about AI development.

\phantomsection\subsection{3.3 Addressing Broader Challenges for Verification}\label{addressing-broader-challenges-for-verification}

Beyond implementing specific verification mechanisms, international verification faces broader questions and challenges, including: increasingly efficient compute, enforcement, ambiguous findings, and attribution. These challenges may be manageable (\hyperref[tab:some_broader_challenges]{\ul{Table 4}}).

\clearpage

{
  \setlength{\LTleft}{-33pt}
  \setlength{\LTright}{-26pt}
  \centering
    \begin{longtable}[]{
      >{\raggedright\arraybackslash}p{(\linewidth - 2\tabcolsep) * \real{0.4119} * \real{1.15}}
      >{\raggedright\arraybackslash}p{(\linewidth - 2\tabcolsep) * \real{0.5881} * \real{1.15}}}
    \toprule\noalign{}
    \begin{minipage}[t]{\linewidth}\raggedright
    Broader challenge
    \end{minipage} & \begin{minipage}[t]{\linewidth}\raggedright
    Options (non-comprehensive)
    \end{minipage} \\
    \endhead
    \toprule\noalign{}
    \rowcolor{black!5!white}
    \begin{minipage}[t]{\linewidth}\raggedright
    \textbf{Managing increases in effective compute:} What to do about the trend that violations will become easier over time, as AI chips are produced in larger quantities, with higher performance, and used with more efficient algorithms?
    \end{minipage} & \begin{minipage}[t]{\linewidth}\raggedright
    Because of the mentioned trends, it is likely unviable to \emph{indefinitely} prevent violations with a specific amount of compute or specific AI capabilities. Instead, defensive measures will likely be needed eventually \citep{pilz2024increasedcomputeefficiencydiffusion, bernardi2025societaladaptationadvancedai}. Thus, the role of verification could be to oversee the (near-)frontier of AI capabilities---to ensure the newest AI capabilities are understood and leveraged defensively, before they proliferate.
    
    If more time were needed to prepare defenses, a costly option could be to extend verification's viability by slowing the increase in effective compute, namely through rules on AI hardware manufacturing and acquisition, and perhaps on computationally expensive AI algorithmic experiments.
    \end{minipage} \\
    \begin{minipage}[t]{\linewidth}\raggedright
    \textbf{Specifying rules to be verified:} How can AI governance goals be operationalized as technical rules on models, data, and code?
    \end{minipage} & \begin{minipage}[t]{\linewidth}\raggedright
    Existing efforts to translate high-level policy concerns into concrete technical rules should continue and accelerate (\hyperref[rules-on-ai-models-data-and-code]{\ul{Section 2.1}}; \hyperref[open-problems-in-verification]{\ul{Section 5}}). This translation poses significant challenges: technical rules like ``model trained with fewer than X FLOP'' or ``model scores Y on benchmark'' may not fully capture policy goals like ``preventing harmful capabilities.'' Rules should ideally narrowly target harmful AI development and deployment while minimizing constraints on beneficial uses. When precise targeting is impossible, conservative thresholds may be needed but should be periodically reassessed as our understanding improves. Processes involving technical experts, policymakers, and affected stakeholders could help develop rules that better align with governance goals.
    \end{minipage} \\
    \rowcolor{black!5!white}
    \begin{minipage}[t]{\linewidth}\raggedright
    \textbf{Taking enforcement actions:} In an international agreement, what should parties do if they find that another party is seriously violating the agreement?
    \end{minipage} & \begin{minipage}[t]{\linewidth}\raggedright
    Options include: reciprocating non-compliance, using non-technical enforcement actions (e.g., sanctions, export controls, withdrawal of security guarantees and of AI benefit-sharing, cyber or kinetic attacks) \citep{sastry2024computingpowergovernanceartificial, Dennis2025BenefitSharing, Hendrycks2025Superintelligence}, and (if available) using technical mechanisms for enforcement \citep{Kulp2024, Petrie2024FlexHEG}. Due to their mutual costs, these options would ideally be effective deterrents and thus never carried out.
    \end{minipage} \\
    \begin{minipage}[t]{\linewidth}\raggedright
    \textbf{Attaining relevant parties' participation:} How to attain compliance commitments from all states that host large-scale AI compute (as such states could directly misuse it or rent it to an agreement party)?
    \end{minipage} & \begin{minipage}[t]{\linewidth}\raggedright
    States could: use export controls and energy policy to keep AI hardware concentrated in cooperative states \citep{sastry2024computingpowergovernanceartificial, Pilz2025AIPower}, use international dialogues and collaborations to build consensus \citep{FARAI2024RedLines, bengio2025internationalaisafetyreport}, and use the leverage listed in the above row.
    \end{minipage} \\
    \rowcolor{black!5!white}
    \begin{minipage}[t]{\linewidth}\raggedright
    \textbf{Acting on ambiguities:} What should be done if the Verifier encounters ambiguous indicators of a potential violation?
    \end{minipage} & \begin{minipage}[t]{\linewidth}\raggedright
    As increasingly escalatory and costly options, the Verifier could: request clarifications, carry out focused investigations, demand the narrow application of more intrusive verification measures, demand temporary pauses of AI compute clusters, or impose partial penalties (\hyperref[b.3-acting-on-ambiguous-findings]{\ul{Appendix B.3}}).
    \end{minipage} \\
    \begin{minipage}[t]{\linewidth}\raggedright
    \textbf{Attributing violations:} How to determine whether a cloud provider or a compute user is to blame for a violation, and whether a corporate violation of international rules is government-backed?
    \end{minipage} & \begin{minipage}[t]{\linewidth}\raggedright
    Attribution could potentially be done via reporting requirements. With appropriate communication channels, a cloud provider or host government may be reasonably expected to detect violations on their compute/territory more quickly than an international Verifier and to report them. Thus, reporting could indicate that the reported party is complicit; failure to report could indicate that the party with reporting responsibilities is complicit.
    \end{minipage} \\
    
    \bottomrule\noalign{}
    \caption{Some broader challenges for verification and high-level options to address them.} \label{tab:some_broader_challenges}
    \endlastfoot
    \end{longtable}
}

\phantomsection\section{4. Verification Mechanisms and Layers}\label{verification-mechanisms-and-layers}

Having clarified our verification scope (\hyperref[verification-scope-and-research-methodology]{\ul{Section 2}}) and subgoals (\hyperref[verification-framework]{\ul{Section 3}}), we now turn to overviewing mechanisms and layers by which these subgoals can be achieved. This section will overview verification mechanisms and layers together, explaining six potential layers by describing the mechanisms that would make them up.

\textbf{Defining verification mechanisms and layers.} We define terms as follows:

\begin{itemize}
\item
  A verification \emph{mechanism} is a method or technology that helps complete at least one of the subgoals in our verification framework (\hyperref[the-framework]{\ul{Section 3.2}}).\footnote{The term ``mechanisms'' are discussed somewhat similarly in prior work \citep{brundage2020trustworthyaidevelopmentmechanisms}.} An example verification mechanism is inspecting AI chips to verify that they have not been sent to undeclared AI data centers; this helps complete Subgoal 2.B.
\item
  A verification \emph{layer} is a collection of similar verification mechanisms, with one mechanism for each verification subgoal. An example is a comprehensive set of ``on-chip'' verification mechanisms, as described below. In other words, a verification layer is a set of similar\footnote{``Similar'' refers to the mechanisms having similar assumptions and tradeoffs. This gives a layer as a whole distinctive assumptions and tradeoffs.} mechanisms capable of end-to-end verification (i.e., completing all subgoals) without redundancy (i.e., without having multiple mechanisms for the same subgoal). As a result, three verification layers can be stacked together to achieve three layers of redundancy, for example.\footnote{Some verification mechanisms, such as whistleblower programs, are so generally applicable that they can be used for every verification subgoal. In these cases, a whole verification layer can be made up of a single mechanism, applied to every subgoal.}
\end{itemize}

\phantomsection\subsection{4.1 On-Chip Verification Layer}\label{on-chip-verification-layer}

\begin{longtable}[]{
  >{\raggedright\arraybackslash}p{(\linewidth - 6\tabcolsep) * \real{0.2693}}
  >{\raggedright\arraybackslash}p{(\linewidth - 6\tabcolsep) * \real{0.3210}}
  >{\raggedright\arraybackslash}p{(\linewidth - 6\tabcolsep) * \real{0.2049}}
  >{\raggedright\arraybackslash}p{(\linewidth - 6\tabcolsep) * \real{0.2049}}}
\toprule\noalign{}
\begin{minipage}[t]{\linewidth}\raggedright
\textbf{Potential verification layer}
\end{minipage} & \begin{minipage}[t]{\linewidth}\raggedright
\textbf{Summary of layer}
\end{minipage} & \begin{minipage}[t]{\linewidth}\raggedright
\textbf{Key\\advantages}
\end{minipage} & \begin{minipage}[t]{\linewidth}\raggedright
\textbf{Key disadvantages}
\vspace{0.5em}
\end{minipage} \\
\endhead
\toprule\noalign{}
\rowcolor{black!5!white}
\begin{minipage}[t]{\linewidth}\raggedright
\textbf{On-chip security features}

(i.e., secure boot, Confidential Computing)

\begin{center}
    \includegraphics[width=0.71354in,height=0.71354in]{media/media/image5.png}
\end{center}
\end{minipage} & \begin{minipage}[t]{\linewidth}\raggedright
Security features built into AI chips may enable verification, such as by ensuring that AI chips log traces of their activities for confidential analysis.
\end{minipage} & \begin{minipage}[t]{\linewidth}\raggedright
Offers maximum transparency into AI chips' uses.
\end{minipage} & \begin{minipage}[t]{\linewidth}\raggedright
Poses unsolved technical problems and severe security challenges (e.g., untrusted suppliers). Insecure AI chips may need to be replaced.
\vspace{1em}
\end{minipage} \\

\bottomrule\noalign{}
\caption{Summary of the on-chip verification layer and its tradeoffs.} \label{tab:summary_of_the_on-chip_verification}
\endlastfoot
\end{longtable}

\begin{figure}[ht]
  \noindent
  \makebox[\textwidth][c]{%
    \includegraphics[width=6.5in,height=2.09722in]{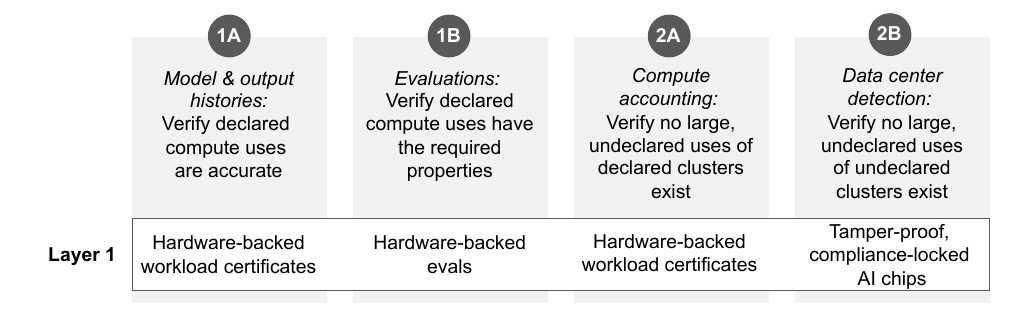}%
  }
  \caption{Summary of how the on-chip verification layer would complete each subgoal.}
  \label{fig:summary_of_how_the_on-chip_verification_layer_would_complete_each_subgoal}
\end{figure}

In the ``on-chip'' verification layer, the Prover's AI chips \emph{help} \emph{verify their own compliance} through built-in security features. In other words, this verification layer distinctively assumes (and tries to verify) that the Prover's AI chips will implement specialized behaviors that enable verification, due to features physically built into the chips during manufacturing. To achieve robust verification with hardware security features, some already common hardware security features would need to be (i) present on the Prover's chips, (ii) unusually secure, and (iii) used in particular ways. We describe these features and their uses next.

\FloatBarrier

\phantomsection\subsubsection{4.1.1 Mechanisms}\label{mechanisms}

\phantomsection\paragraph{4.1.1.1 Prerequisites: Hardware Security Features}\label{prerequisites-hardware-security-features}

The verification mechanisms we consider require a hardware security feature known as secure boot, which must be at least tamper-evident (i.e., impractical to discreetly disable or undermine). They would further benefit from Confidential Computing and tamper-proofing, though these are not required.

\textbf{Secure boot:} Secure boot aims to guarantee that a chip will only run with approved\footnote{In typical implementations, the party granting approval is the hardware vendor. If the hardware vendor is not trusted, it may be feasible to transfer approval authority to a chosen Verifier (via cryptographic key revocation).} system software (i.e., firmware and operating system). System software can constrain a chip's behavior throughout operation, so secure boot could be used to ensure that a chip will always behave in ways that facilitate verification. Secure boot implementations are especially promising for this if they: (i) are at least tamper-evident, with regards to both physical and digital tampering (so, unless they are also tamper-proof, random inspections would be needed to check for tampering);\footnote{In keeping with the ``on-chip'' approach of this verification layer, tamper-evident and tamper-proof measures could be built into AI chips. Such anti-tamper measures within AI chips would add redundancy to other verification layers, which might implement anti-tamper measures around the chips (e.g., on a server enclosure or via tamper-evident video cameras). Notably, here ``tampering'' includes simply turning off a secure boot functionality if it is intended to be optional; this could be made tamper-evident through measures involving secure storage, secure memory, compute accounting, or offline licensing (\hyperref[a.2-hardware-backed-workload-certificates-and-evaluations]{\ul{Appendix A.2}})} and (ii) include a secure private key,\footnote{That is, ``secure'' in the sense that the Prover cannot read or overwrite the key, at least not without leaving tamper evidence.} allowing the system software to digitally sign messages.

\textbf{Confidential Computing:} Confidential Computing is a hardware feature that, among other uses, can allow multiple parties to share their information only for specific purposes. This could enable a Verifier to run tests on a Prover's models, data, and code---with the Prover knowing their information will not be stolen, and with the Verifier knowing their tests will be run faithfully and will not be viewed for the sake of manipulating test results.\footnote{One might wonder: how can a Prover be assured that the Verifier's test questions will not extract more information than is appropriate, if the Prover cannot access the test questions? For one, the Prover could confirm that the Prover's tests only output a very small amount of information (e.g., just a ``pass'' or ``fail'' result), so they have little bandwidth to extract information. Additionally, the approved program could enforce constraints on the Verifier's tests, e.g., that they can run a model but not arbitrarily access individual model weights. Alternatively, the Verifier could make more tests transparent to the Prover, though this could make it easier for the Prover to ``cheat the tests'' (i.e., find ways to pass them while still committing a violation).}

Confidential Computing (or more precisely, one of its functionalities\footnote{Specifically, here we describe the multi-party computation functionality of CC, though the more commonly used functionality is for running computations on cloud compute without exposing them to the cloud provider.}) is intended to work as follows. First, multiple parties share their encrypted data, giving access only to a specific computer program on specific hardware. (They can scrutinize this program's code in advance.) Then, hardware features ensure the program is executed faithfully and without leaks. Finally, a digital signature confirms that the test results came from this approved program \citep{CCC2021TechnicalAnalysis, GoogleConfidentialSpace}.

There are existing, though likely limited, implementations and uses of Confidential Computing, which includes secure boot \citep{NVIDIA2023HopperConfidential}. AI chip designers including NVIDIA \citep{NVIDIA2023HopperConfidential}, Huawei \citep{dhar2024ascendccconfidentialcomputingheterogeneous}, Graphcore \citep{vaswani2023ccaia}, and Google \citep{Chuvakin2021ConfidentialComputingBlog, Dragan2024FrontierSafety} have in recent years implemented or announced plans for versions of Confidential Computing. The UK AI Security Institute has used some of these implementations in a pilot program to run confidential AI safety evaluations with Anthropic and OpenMined \citep{Trask2024SecureEnclaves, Beers2025PETAI}. However, as discussed below, there are reasons to expect current implementations to be hackable (\hyperref[tab:challenges_for_attaining_confidence]{\ul{Table 6}}).

\phantomsection\paragraph{4.1.1.2 Verification Mechanisms}\label{verification-mechanisms}

Secure boot and Confidential Computing could be used to implement the following verification mechanisms.

\textbf{Hardware-backed workload certificates and evaluations (\hyperref[a.2-hardware-backed-workload-certificates-and-evaluations]{\ul{Appendix A.2}}):} The Prover's chips could sign cryptographic certificates that confirm how they produced their results---confirming model histories, output histories, and evaluation results. For example, a certificate could attest that some trained AI model was produced using specific code and data. Similarly, a certificate could attest that some AI model outputs such as generated text were produced by a specific model on specific inputs. A limited version of this has already been developed and open-sourced \citep{mithril2023aicert}. Secure boot may be able to ensure that these workload certificates are not only authentic but also comprehensive: that a Prover could not simply delete a certificate or refrain from making it. We outline a potential technical implementation for such comprehensive workload certificates (\hyperref[a.2-hardware-backed-workload-certificates-and-evaluations]{\ul{Appendix A.2}}), relying on permanent secure boot. This mechanism could complete verification Subgoals 1.A, 1.B, and 2.A. In addition to the security of the hardware features, their operation---especially identifying model weights and data that may be obfuscated---poses challenges.

\textbf{Tamper-proof, compliance-locked AI chips:} AI chips could potentially be compliance-locked (i.e., refuse to run non-compliant workloads) and tamper-proof.\footnote{This is similar to some aspects of proposed Flexible Hardware-Enabled Guarantees (FlexHEGs) \citep{Petrie2024FlexHEG}, but here we are discussing an entirely on-chip option rather than compliance-locking based on external devices attached to a chip.} This would assure the Verifier that, even if the Prover has an undeclared AI compute cluster, they would be unable to use it in non-compliant ways. Thus, tamper-proof, compliance-locked AI chips would address undeclared AI compute clusters (Subgoal 2.B), a gap left by the above workload certificates. Compliance-locking could be implemented by combining the above workload certificates with offline licensing \citep{Kulp2024},\footnote{In other words, suppose AI chips throttle their own performance unless they receive periodic confirmation that they are in compliance---confirmation which the Verifier provides (as a digital signature) based on workload certificates (\hyperref[a.2-hardware-backed-workload-certificates-and-evaluations]{\ul{Appendix A.2}}). Then, any non-complaint AI data centers will soon become useless.} or more ambitiously, through entirely automated on-device checks. However, tamper-proofing's use in AI governance faces major technical challenges \citep{Kulp2024, aarne_secure_chips_2024} and serious tradeoffs (\hyperref[analysis]{\ul{Section 4.1.2}}).

\phantomsection\subsubsection{4.1.2 Analysis}\label{analysis}

\textbf{On-chip mechanisms offer maximal transparency into AI chips' computations.} By using security features built into AI chips, on-chip mechanisms directly offer transparency into AI chips' code and data, with measures to preserve confidentiality. In contrast, other verification layers only examine AI chips ``from the outside.'' That is, other layers must infer AI chips' behavior based on external measurements or personnel, which creates room for circumvention and imprecision. Those challenges are significant (\hyperref[analysis-1]{\ul{Section 4.2.2}}), and on-chip verification avoids them entirely.

\textbf{Compliance-locking offers limited verification redundancy, and it carries the pros and cons of enforcement.} One of the above mechanisms, compliance-locked AI chips, offers technical assurances only after a less technical setup phase: tracking down chip fabrication facilities and perhaps older AI compute clusters, likely by non-technical mechanisms and satellite images (\hyperref[personnel-based-verification-layers]{\ul{Section 4.3}}: \hyperref[supplementary-verification-mechanisms]{\ul{Section 4.4}}). Thus, compliance-locking is only partly redundant with other verification layers' completion of Subgoal 2.B. The more significant consequence of compliance-locking may be its built-in enforcement function, which would facilitate preventing violations but raises risks of adversarial use \citep{Kulp2024}. Still, there are proposals to mitigate these risks \citep{Petrie2024FlexHEG}, including through enforcement requiring multi-party approval. A similar analysis applies to compliance-locked server enclosures (\hyperref[off-chip-verification-layers]{\ul{Section 4.2}}). There are also options with more intermediate tradeoffs, e.g., offline licensing could be set up so that the Verifier has visibility but not veto power.\footnote{In particular, a limited and known-to-the-Verifier set of \emph{Prover} chips could be authorized to approve offline license requests, and the Verifier could monitor all communication with these chips (e.g., by network tap or physical access controls). Thus, the Prover would grant licenses but the Verifier would know about them, adding redundancy for Subgoal 2.B without the tradeoffs of Verifier enforcement power. There may still be concern over these Prover chips being sabotaged, as that could deny the Prover their licenses, so it could help for some backups (i.e., other Prover chips authorized to approve license requests) to be (verifiably) stored in highly secure locations.}

\textbf{On-chip verification faces severe, unsolved hardware security challenges} (\hyperref[tab:challenges_for_attaining_confidence]{\ul{Table 6}}). Relevant security features in the current and next generation of AI chips appear unlikely to be robust; their track records and security practices are lacking. Improving matters appears challenging, as there are major, inherent challenges in developing robust hardware security features for AI chips or other high-performance chips.

\begin{longtable}[]{
  >{\raggedright\arraybackslash}p{(\linewidth - 2\tabcolsep) * \real{0.2436}}
  >{\raggedright\arraybackslash}p{(\linewidth - 2\tabcolsep) * \real{0.7564}}}
\toprule\noalign{}
\begin{minipage}[t]{\linewidth}\raggedright
\textbf{Challenge} for attaining confidence in AI chips' hardware security features\footnote{This analysis also applies to other high-performance chips in the same cluster as the AI chips, especially CPUs.}
\vspace{0.5em}
\end{minipage} & \begin{minipage}[t]{\linewidth}\raggedright
\textbf{Description}
\end{minipage} \\
\endhead
\toprule\noalign{}
\rowcolor{black!5!white}
\begin{minipage}[t]{\linewidth}\raggedright
Mixed track records
\end{minipage} & \begin{minipage}[t]{\linewidth}\raggedright
Hardware security features have a track record that is mixed at best. There have been multiple cases of individual, low-budget researchers circumventing the hardware security features of leading companies.\footnote{One hardware security expert noted that, whenever the security community has gotten their hands on an HSM, they were able to break its security guarantees (Interview \#2, 2024). For example, a graduate student on a laptop broke Intel's Trusted Platform Module in 2020, despite its high security certification \citep{moghimi2020tpmfail}. Researchers have also found many vulnerabilities in Intel SGX, a confidential computing technology from 2015 \citep{9107096}. Beyond Intel, other companies including Apple, Android, SpaceX, and NVIDIA have also had failures in robustly implementing secure boot \citep{hildenbrand_bootloaders_2018, Crider2022, Ridley2021, Wouters2022Glitched, AppleWikiPwnage, jirku_switch_boot_2022, wololo_picofly_2023}. Intel also reports that, between its own and AMD's Confidential Computing and hardware root-of-trust technologies, 55 vulnerabilities were discovered in 2024 \citep{Intel2025SecurityReport}.} On the other hand, some track records have improved in recent years.\footnote{``Circumventing secure boot on iPhones is popularly known as ``jailbreaking'' iPhones. While jailbreaks were common in the early- to mid-2010s {[}...{]} {[}t{]}oday jailbreaking {[}using publicly known methods{]} is only possible if the phone's operating system hasn't been updated in several years'' \citep{aarne_secure_chips_2024}.}
\vspace{0.5em}
\end{minipage} \\
\begin{minipage}[t]{\linewidth}\raggedright
Lacking security practices
\end{minipage} & \begin{minipage}[t]{\linewidth}\raggedright
Most high-performance chips, including AI chips, appear to be designed contrary to important security practices, such as external red teaming by leading security experts\footnote{Hardware security academics suggested that, despite its importance, it is rare for hardware companies to engage leading security research labs to stress-test their products (Interviews \#2 and 11, 2024). In line with this, Intel reports that the 21 hardware vulnerabilities they discovered in 2024 were ``all found internally'' \citep{Intel2025SecurityReport}, though Intel has had partnerships for security reviews \citep{Aktas2023TDX}. Relatedly, NVIDIA states that it ``does not currently have a bug bounty program'' \citep{nvidia_report_vuln}. Even companies with bug bounties rarely offer more than \$100,000 rewards, even for discoveries of ``critical'' or ``exceptional'' hardware vulnerabilities \citep{GoogleBugHunters, microsoft_bug_bounty, AMD_BugBounty, intel_bug_bounty}.} and use of Hardware Security Modules (HSMs).\footnote{The security experts we interviewed often expressed that HSMs---chip regions exclusively dedicated to security features, to make them simpler to secure---would improve security (Interviews \#2, 4, 6, 9, 11, 18). However, leading AI chips such as NVIDIA's lack HSMs; NVIDIA GPUs have a broader-purpose ``GPU System Processor'' \citep{klotz2022nvidiadriver}.}
\vspace{0.5em}
\end{minipage} \\
\rowcolor{black!5!white}
\begin{minipage}[t]{\linewidth}\raggedright
Performance-security tradeoffs
\end{minipage} & \begin{minipage}[t]{\linewidth}\raggedright
There are often tradeoffs between a chip's performance and its security, incentivizing higher performance and faster releases at the expense of security.\footnote{A hardware security expert underscored a tradeoff between a chip's performance and its security, saying it is ``very difficult'' to convince hardware companies ``that security\ldots{} may {[}need to{]} cost you performance'' (Interview \#4, 2024). This is in line with prior research \citep{hastings_wac_2020} and vulnerabilities \citep{lipp2018meltdown, kocher2018spectreattacksexploitingspeculative}. In addition, fixing security vulnerabilities could delay product releases.}
\vspace{0.5em}
\end{minipage} \\
\begin{minipage}[t]{\linewidth}\raggedright
Untrusted supply chains
\end{minipage} & \begin{minipage}[t]{\linewidth}\raggedright
Chips may have designers and manufacturers whom the Verifier does not trust (especially for hypothetical U.S.-China verification). This introduces many attack vectors, perhaps addressable with extensive design disclosures and hardware verification (\hyperref[a.1-full-stack-security-for-technical-verification-mechanisms-implementation]{\ul{Appendix A.1}}).
\vspace{0.5em}
\end{minipage} \\
\rowcolor{black!5!white}
\begin{minipage}[t]{\linewidth}\raggedright
Untrusted operational environments
\end{minipage} & \begin{minipage}[t]{\linewidth}\raggedright
Chips' physical and digital environments may be controlled by untrusted actors, enabling more attack vectors, though physical and digital attacks could be countered by physical monitoring (e.g., tamper-evident security cameras (\hyperref[supplementary-verification-mechanisms]{\ul{Section 4.4}})) and secure design, respectively.
\vspace{0.5em}
\end{minipage} \\
\begin{minipage}[t]{\linewidth}\raggedright
Lack of adaptability
\end{minipage} & \begin{minipage}[t]{\linewidth}\raggedright
Hardware-level security vulnerabilities typically cannot be fixed after manufacturing, though some firmware vulnerabilities can be fixed.\footnote{Some firmware---namely that stored in read-only-memory (ROM) such as Boot ROM---cannot be digitally edited.} Replacing AI chips that have flawed security features could require replacing millions of expensive chips.\footnote{One study estimates that NVIDIA shipped nearly 4 million data center GPUs in 2023 \citep{shah2024datacenter}, and NVIDIA's revenue has been growing rapidly since then \citep{NvidiaRevenue2024}. For a sense of the potential costs of replacement, replacing ``just'' 4 million H100 GPUs costing \$30,000 each \citep{eadline2023h100} would cost \$120 billion.}
\vspace{0.5em}
\end{minipage} \\

\bottomrule\noalign{}
\caption{Challenges for attaining confidence in AI chips' hardware security features. The first two rows cover current or historical problems, and the remaining rows cover inherent challenges.} \label{tab:challenges_for_attaining_confidence}
\endlastfoot
\end{longtable}

\textbf{Hardware security challenges are especially severe with on-chip verification.} While hardware security challenges arise for any verification method that uses hardware, these challenges are uniquely severe for on-chip verification. On-chip verification uses the same chips to both run and verify AI workloads. This inherently creates tradeoffs; if one wants to tighten the security of the hardware security features, one might have to delay, replace, or change the supply chain of the whole AI chip, which goes against commercial incentives. Perhaps these challenges could be overcome (\hyperref[a.1-full-stack-security-for-technical-verification-mechanisms-implementation]{\ul{Appendix A.1}}), but absent major effort (up to replacing millions of insecure AI chips), on-chip mechanisms will likely have irreparable design flaws that allow them to be circumvented, especially for complex or dynamic mechanisms. Still, such efforts may be feasible on longer timeframes (e.g., after initially using more implementation-ready verification methods), and it is possible that existing on-chip mechanisms are secure enough despite their challenges.

To avoid the tradeoffs of on-chip verification, another approach could be to separate AI hardware from verification hardware, so that each can be specialized for its own purpose. This motivates the ``off-chip'' verification layers we consider next.

\clearpage

\phantomsection\subsection{4.2 Off-Chip Verification Layers}\label{off-chip-verification-layers}

\begin{longtable}[]{
  >{\raggedright\arraybackslash}p{(\linewidth - 6\tabcolsep) * \real{0.2658}}
  >{\raggedright\arraybackslash}p{(\linewidth - 6\tabcolsep) * \real{0.3258}}
  >{\raggedright\arraybackslash}p{(\linewidth - 6\tabcolsep) * \real{0.2042}}
  >{\raggedright\arraybackslash}p{(\linewidth - 6\tabcolsep) * \real{0.2042}}}
\toprule\noalign{}
\rowcolor{black!5!white}
\begin{minipage}[t]{\linewidth}\raggedright
\textbf{Potential verification layer}
\end{minipage} & \begin{minipage}[t]{\linewidth}\raggedright
\textbf{Summary of layer}
\end{minipage} & \begin{minipage}[t]{\linewidth}\raggedright
\textbf{Key\\advantages}
\end{minipage} & \begin{minipage}[t]{\linewidth}\raggedright
\textbf{Key disadvantages}
\vspace{0.5em}
\end{minipage} \\
\endhead
\toprule\noalign{}
\begin{minipage}[t]{\linewidth}\raggedright
\textbf{Off-chip network tap} (and analysis) (e.g. ``FlexHEGs'')

\begin{center}
    \includegraphics[width=1.09383in,height=0.66791in]{media/media/image2.png}
\end{center}
\end{minipage} & \begin{minipage}[t]{\linewidth}\raggedright
Mutually vetted devices could intercept data exchanged between chips, then check for discrepancies with declared uses.
\end{minipage} & \begin{minipage}[t]{\linewidth}\raggedright
Could be retrofitted to existing AI chips and optimized for security.
\end{minipage} & \begin{minipage}[t]{\linewidth}\raggedright
Poses technical, logistical, and security challenges. Strongest versions need redesigned chip-adjacent hardware.
\vspace{0.5em}
\end{minipage} \\

\rowcolor{black!5!white}
\begin{minipage}[t]{\linewidth}\raggedright
\textbf{Off-chip analog sensors} (and analysis, e.g., proof-of-learning)

\begin{center}
    \includegraphics[width=0.92188in,height=0.6887in]{media/media/image1.png}
\end{center}
\end{minipage} & \begin{minipage}[t]{\linewidth}\raggedright
Physically secured chips could check that (i) declared AI compute uses are accurate (e.g., reproducible) and (ii) their compute use adds up to the expected total (estimated with analog sensors, e.g., power meters, in AI data centers).
\end{minipage} & \begin{minipage}[t]{\linewidth}\raggedright
Could be retrofitted to existing AI chips and optimized for security.
\end{minipage} & \begin{minipage}[t]{\linewidth}\raggedright
Poses unsolved technical problems. Likely requires separate trusted clusters for analysis, and manufacturing \& installing sensors.
\vspace{0.5em}
\end{minipage} \\

\bottomrule\noalign{}
\caption{Summary of off-chip verification layers and their tradeoffs.} \label{tab:summary_of_off-chip_verification}
\endlastfoot
\end{longtable}

\begin{figure}[ht]
  \noindent
  \makebox[\textwidth][c]{%
    \includegraphics[width=6.5in,height=2.58333in]{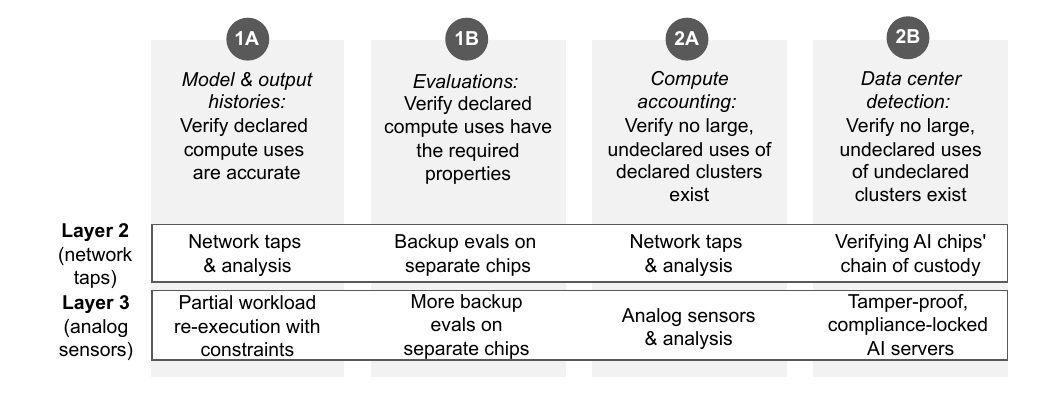}%
  }
  \caption{Summary of how off-chip verification layers would complete each subgoal. Note ``verifying AI chips' chain of custody`` and ``tamper-proof, compliance-locked AI servers`` could just as well be swapped; the point is that they are two redundant off-chip mechanisms for addressing hidden AI compute clusters. We consider them ``off-chip`` because they neither rely on features built into chips nor on leaks or disclosures from personnel. Additionally, tamper-proof enclosures and security cameras for chain-of-custody verification could be considered off-chip devices.}
  \label{fig:summary_of_how_off-chip_verification_layers_would_complete_each_subgoal}
\end{figure}

``Off-chip'' verification layers aim to avoid the security challenges of on-chip mechanisms by verifying AI chips' activities using \emph{separate devices}, rather than security features built into the AI chips. These separate devices could be (i) attached sensors to monitor the AI chips, and (ii) separate chips to analyze the sensor data and Provers' declarations. With these, the Verifier would aim to detect discrepancies between a Prover's declarations and their actual chip use, such as by detecting that chips' input data or power draw patterns tell a different story than the Prover's claims. The external devices could be mutually vetted to enable trust.

\FloatBarrier

\phantomsection\subsubsection{4.2.1 Mechanisms}\label{mechanisms-1}

\phantomsection\paragraph{4.2.1.1 Prerequisites: Off-Chip Devices}\label{prerequisites-off-chip-devices}

This section overviews, in broad terms, potential off-chip devices to collect or analyze data in AI data centers.

\textbf{Devices for data collection:} Devices could collect digital data or analog readings on AI workloads, offering redundancy and different tradeoffs (\hyperref[tab:summary_of_off-chip_verification]{\ul{Table 7}}):

\begin{itemize}
\item
  \emph{Off-chip input/output loggers} (i.e., network taps): Devices that read and log (a random sample of) digital data, such as data exchanged between AI servers or potentially between individual AI accelerators (\hyperref[a.3-network-taps-analysis]{\ul{Appendix A.3}}).
\item
  \emph{Off-chip analog sensors:} Devices that log analog measurements, such as power draw, temperature, and electromagnetic measurements (\hyperref[a.6-compute-accounting-via-analog-sensors]{\ul{Appendix A.6}}).
\end{itemize}

\textbf{Mitigating security challenges:} Without strong precautions, network taps or analog sensors could be designed with hidden functionalities for espionage or sabotage. To demonstrably prevent this, states could take measures such as the following to secure these devices, though these would be challenging:

\begin{itemize}
\item
  All parties could agree to use device designs that are: the same across parties, as simple as possible, mutually vetted, open-source, and perhaps jointly developed.
\item
  Provers could verify that these vetted devices are actually used with no additions, by physically scanning the devices (with more intensive tear-downs for a random sample) and overseeing the devices' manufacturing and assembly (\hyperref[a.1-full-stack-security-for-technical-verification-mechanisms-implementation]{\ul{Appendix A.1}}).
\item
  Provers could physically limit these devices' communication to agreed-on data quantities and data contents.\footnote{In many cases, communication limits could involve ensuring the devices are disconnected from the internet (i.e., air-gapped). This would assure Provers that (i) the devices cannot easily transmit data to the Verifier, and (ii) the Verifier cannot easily send a message to many of the devices at once (which, if possible, could presumably facilitate sabotage).}
\end{itemize}

Conversely, for Verifiers to be confident in the devices' integrity, they could rely on measures such as Verifier-trusted supply chains, mutual vetting (similar to Provers' vetting), tamper-evident enclosures (a random sample of which would be routinely inspected\footnote{Inspections introduce the problem of assuring the Prover that the inspectors do not pose threats. This could be addressed similarly to how it has been addressed over decades of nuclear arms control, where inspectors are continuously accompanied by Prover representatives, among other measures \citep{baker2023nucleararmscontrolverification}.}), and ideally tamper-proofing.

\textbf{Devices for data analysis:} Beyond devices for collecting data, collected data could be analyzed by trusted computer chips, either on-site or off-site. For on-site chips, both parties could gain confidence through measures similar to the above, though full device scanning and design disclosures would be more challenging (\hyperref[a.1-full-stack-security-for-technical-verification-mechanisms-implementation]{\ul{Appendix A.1}}). For off-site chips, which could form \emph{trusted clusters},\footnote{The term ``trusted clusters'' has also been used by related work \citep{shavit2023doescatchchinchillaverifying, Heim2024TrustedCluster}.} security could be based on both the Prover and the Verifier directly applying security measures for their respective goals. The Prover would monitor and restrict the cluster's external communications---ensuring that only the expected amount of information is extracted \citep{Greenblatt2024Exfiltration}.\footnote{Insofar as the Prover knows what tests the Verifier will run (\hyperref[implementation-options-across-mechanisms]{\ul{Section 4.5}}), the Prover may also be able to determine some of the exact information that the Verifier will extract. Then, by checking that the Verifier's limited extracted information excludes this expected information, the Prover could reduce the Verifier's room for unauthorized information extraction. Thus, the Prover could benefit from monitoring the content of the extracted information, not only its quantity.} \footnote{One potential source of information leakage is the IT equipment emitting information through non-human channels, especially wires or electromagnetic waves. By managing the ``outer room,'' the Prover-trusted actor can physically block these communication options. (Blocking electromagnetic waves may require a large Faraday cage.) By monitoring and gatekeeping personnel as they exit, the Prover-trusted actor can also ensure that operators do not take data storage devices with them (beyond data they are supposed to have). Similarly, by supervising the operators, the Prover-trusted actor can confirm that computer monitors only show what they are supposed to show. There is still some risk that these security measures may be circumvented, e.g., via bribery, but a Prover already accepts such risks when they hold sensitive information in their own data centers.} The Prover would also supervise the cluster's use. Meanwhile, the Verifier would ensure the cluster's integrity by providing, physically securing, and securely operating the chips and other IT equipment.

\phantomsection\paragraph{4.2.1.2 Verification Mechanisms}\label{verification-mechanisms-1}

Off-chip devices could be used to complete verification subgoals with redundancy as follows (\hyperref[fig:summary_of_how_off-chip_verification_layers_would_complete_each_subgoal]{\ul{Figure 6}}).

\textbf{Off-chip network taps \& analysis (\hyperref[a.3-network-taps-analysis]{\ul{Appendix A.3}}):} Mutually vetted devices could log a random sample of the data exchanged between different parts of a compute cluster, such as the data transmitted between server racks, between servers, or potentially between individual AI chips. Analysis of these logs may reveal discrepancies between actual AI chip communications and what would be expected from declared workloads, serving to verify the accuracy of a declared workload (Subgoal 1.A) and the absence of major undeclared workloads on known AI compute clusters (Subgoal 2.A). A closely related proposal for using network taps for verification is Flexible Hardware-Enabled Guarantees (FlexHEGs) \citep{petrie2025technicaloptionsflexiblehardwareenabled}, earlier proposals of which emphasized another mechanism we discuss below: tamper-proof, compliance-locked servers \citep{Petrie2024FlexHEG}.

\textbf{Partial workload re-execution with constraints (e.g., ``proof-of-learning'') (\hyperref[a.4-partial-workload-re-execution-with-constraints]{\ul{Appendix A.4}}):} Trusted, secured chips (whether on- or off-site) could run software tests to check whether the Prover's declarations are accurate (Subgoal 1.A). For Subgoal 1.A, tests could involve re-running random segments of AI training or deployment to check if claimed results can be approximately reproduced \citep{jia2021proofoflearningdefinitionspractice, shavit2023doescatchchinchillaverifying}. However, a Prover might be able to spoof results they never computed in the first place; countering these spoofs requires further tests \citep{choi2023toolsverifyingneuralmodels}. This mechanism does not inherently involve analog sensors, but it complements analog sensors well, by completing Subgoal 1.A---a gap left by analog sensors.

\textbf{(Backup) Evaluations on separate chips:} Trusted, secured chips (whether on- or off-site; \hyperref[prerequisites-off-chip-devices]{\ul{Section 4.2.1.1}}) could test whether any declared models, data, and code have the required properties (Subgoal 1.B; \hyperref[rules-on-ai-models-data-and-code]{\ul{Section 2.1}}). For example, these chips could directly run safety evaluations on declared models. These tests could be ``backup'' tests in the sense of adding redundancy to hardware-backed evaluations (\hyperref[on-chip-verification-layer]{\ul{Section 4.1}}). They could add redundancy to both the testing infrastructure (by having both off-chip and on-chip approaches to testing) and to the tests' substance (by e.g., using different test datasets).

\textbf{Off-chip analog sensors for compute accounting (\hyperref[a.6-compute-accounting-via-analog-sensors]{\ul{Appendix A.6}}):} Off-chip sensors could log measurements such as AI chips' power draw. The Verifier could then analyze these measurements with secured chips, testing for several signs of large, undeclared compute use: declared workloads would appear unnecessarily slow, AI chips would use more power than needed for the declared workloads, and/or the detailed physical signatures of chips would be different than expected. In other words, the Verifier would use analog measurements to verify the total number of operations or chip-hours done and check if approximately all of them can be ``accounted for'' by declared uses. There are various complications in implementing these checks.

\textbf{Verifying AI chips' chain of custody:} A Verifier could verify the locations and owners of random samples of AI chips from manufacturing to end-of-life destruction. This would serve to verify that large quantities of AI chips are not assembled into undeclared AI compute clusters (Subgoal 2.B). Users of small quantities could potentially be exempt. Existing AI chips would be a challenge, though perhaps many could have their locations and owners retroactively verified. Declared chains of custody could be verified with inspections, potentially supplemented by video cameras \citep{baker2023nucleararmscontrolverification} and hard-to-spoof, unique IDs.\footnote{This mechanism could loosely be considered ``off-chip verification'' in that the separate devices used are the video cameras, and the ``on-chip'' or ``personnel'' categories would fit less well. Still, it is less technically demanding or complex than the other mechanisms in this list.}

\textbf{Tamper-proof, compliance-locked AI servers:} A whole server containing AI chips could be tamper-proof and compliance-locked \citep{Petrie2024FlexHEG}. This would serve as an additional method to address undeclared AI compute clusters (Subgoal 2.B).\footnote{Technically a server could be ``compliant'' without being ``declared,'' but if this is a concern, compliance could be made to require declaration, e.g., by using offline licensing to disable undeclared data center AI servers.} This mechanism is similar to tamper-proof, compliance-locked AI chips (\hyperref[verification-mechanisms]{\ul{Section 4.1.1.2}}), but implementation would differ; here, tamper-proofing would be implemented in the server enclosure rather than on-chip, and compliance-locking would leverage the above mechanisms rather than on-chip ones to detect non-compliance.

Together, all the above mechanisms could combine to form two redundant layers of off-chip verification, one involving network taps and the other involving analog sensors (\hyperref[fig:summary_of_how_off-chip_verification_layers_would_complete_each_subgoal]{\ul{Figure 6}}).

\phantomsection\subsubsection{4.2.2 Analysis}\label{analysis-1}

\textbf{Off-chip verification avoids the worst hardware security challenges.} As discussed (\hyperref[analysis]{\ul{Section 4.1.2}}), \emph{on}-chip mechanisms face severe hardware security challenges, including that their hardware components are not retrofittable, they may have supply chains untrusted by the Verifier, and their security may be deprioritized in favor of performance. Off-chip mechanisms still need hardware security, but achieving it is not as much of an uphill battle.\footnote{To be clear, the track record of dedicated hardware security devices is also lacking \citep{moghimi2020tpmfail}. Still, as discussed (\hyperref[analysis]{\ul{Section 4.1.2}}), the challenges related to supply chains, incentives, and complexity are substantially less severe than in the on-chip case.} Off-chip sensors and chips can be retrofitted to existing chips, and their supply chains and design can be highly optimized for security.\footnote{A complication here is that the Verifier may wish to run some programs on the same kind of chips as the Prover's chips, to directly re-run Prover workloads or understand these chips' behavior under different workloads. However, such tests could be compromised by hardware backdoors; more generally, they could defeat the purpose of off-chip mechanisms: robustness to untrusted Prover chips. Fortunately, the Verifier could complete several other checks to verify the results of tests run on Prover chips. The Verifier could (i) have the Prover convert their code to run on Verifier hardware; (ii) do high-level modeling of the Prover's chips \citep{erdil2024datamovementlimitsfrontier, erdil2025inferenceeconomicslanguagemodels}, informed by physical measurements (\hyperref[a.1-full-stack-security-for-technical-verification-mechanisms-implementation]{\ul{Appendix A.1}}); and (iii) emulate the understood functionalities of the Prover's chips (though one interviewee warns this would only be practical at very small scales, as cost overheads of AI chip emulation are on the order of 10,000x; Interview \#17). Check (i) constrains any backdoored behavior to approximate regular behavior (or to be equivalent for deterministic programs), and check (ii) may detect unnecessarily altered physical signatures. Further, the Verifier could run only replicas of individual pods of Prover hardware to reduce costs.} \footnote{Even if the Verifier does run some hardware of the same kind as the Prover's hardware, the Verifier can run it more securely than if it were in the Prover's own data centers. With the Verifier in control, the Verifier can (i) use trusted hardware for some functionalities (e.g., using CPUs and network chips from a trusted supply chain, or perhaps FPGA substitutes, even if the AI chips are untrusted); (ii) implement protections against physical and digital attacks, especially known ones; (iii) check for approximate consistency across multiple kinds of hardware on compatible test cases; and (iv) perhaps manufacture a small amount of patched hardware to replace vulnerable hardware (whereas doing this for all the Prover's hardware would be more costly).} \footnote{As one potential ingredient, there are open-source HSMs \citep{nitrokey_nethsm}. One interviewee shared an anecdote of a hardware security lab that does ``90\%'' of their red teaming on open-source hardware systems, as these are much easier to study and tend to be more willing to adjust in response (Interview \#2, 2024).}

\textbf{The strongest network taps face substantial engineering challenges, requiring chip-adjacent hardware to be redesigned.} In order for network taps to tap all communication between AI chips in a compute cluster, the cutting-edge hardware that executes this communication would have to be redesigned or modified, despite being proprietary or co-packaged with AI chips (\hyperref[a.3-network-taps-analysis]{\ul{Appendix A.3}}). Alternatively, further removed taps would have visibility into a smaller portion of communications, perhaps serving as a supplement to partial workload re-execution more than as an independent guarantee.

\textbf{Off-chip verification faces substantial protocol design challenges and offers limited precision.} Off-chip devices must infer chip behavior ``from the outside.'' This involves several algorithmic problems that do not currently have complete solutions. It also introduces imprecision, e.g., in using analog measurements to infer a chip's rate of operations. Even at its best, then, off-chip verification is likely most applicable for contexts where significant error bars are acceptable, i.e., where it is not a major concern if a relatively small fraction of compute use is undeclared. With AI capabilities requiring less compute over time \citep{ho2024algorithmicprogresslanguagemodels} and the amount of compute available increasing over time, this would likely be the earlier stages of a verification regime---or longer if societal resilience improves (raising the amount of compute needed to cause harm). Separately, network taps would only be useful if the data exchanged between AI chips is unencrypted or if its decryption key is shared, which could trade off against security, though network-tap-related technologies could also support encryption and security in other ways \citep{petrie2025technicaloptionsflexiblehardwareenabled}.

\textbf{Off-chip mechanisms would require significant logistical and potentially production efforts.} To verify claims with relatively small error bars, there may need to be one off-chip sensor and/or network tap for every data center AI chip, which would require the production and installation of millions of devices (assuming there do not happen to be adequate devices available). For reference, NVIDIA was estimated to ship 1.5-2 million H100 GPUs in 2024, a significant rise from the previous year's 0.5 million H100 shipments \citep{shilov2023tripleoutput}. Building mutually secured ``trusted clusters'' would require additional work, even if concentrated into a small number of clusters. Fortunately, the sensor devices could potentially be simple sensors that do not interfere with operations, and one could use many fewer sensors (e.g., one per server rack) if one is willing to accept larger error bars.

Given the challenges of both on- and off-chip verification, it would be helpful if there were also simpler approaches to verification, or more broadly, approaches with different tradeoffs. This is where personnel-based verification comes in.

\phantomsection\subsection{4.3 Personnel-Based Verification Layers}\label{personnel-based-verification-layers}

\begin{longtable}[]{
  >{\raggedright\arraybackslash}p{(\linewidth - 6\tabcolsep) * \real{0.2658}}
  >{\raggedright\arraybackslash}p{(\linewidth - 6\tabcolsep) * \real{0.3258}}
  >{\raggedright\arraybackslash}p{(\linewidth - 6\tabcolsep) * \real{0.2042}}
  >{\raggedright\arraybackslash}p{(\linewidth - 6\tabcolsep) * \real{0.2042}}}
\toprule\noalign{}
\rowcolor{black!5!white}
\begin{minipage}[t]{\linewidth}\raggedright
\textbf{Potential verification layer}
\end{minipage} & \begin{minipage}[t]{\linewidth}\raggedright
\textbf{Summary of layer}
\end{minipage} & \begin{minipage}[t]{\linewidth}\raggedright
\textbf{Key\\advantages}
\end{minipage} & \begin{minipage}[t]{\linewidth}\raggedright
\textbf{Key disadvantages}
\vspace{0.5em}
\end{minipage} \\
\endhead
\toprule\noalign{}

\begin{minipage}[t]{\linewidth}\raggedright
\textbf{Whistleblower programs}

\begin{center}
    \includegraphics[width=0.72917in,height=0.5783in]{media/media/image9.png}
\end{center}
\vspace{0em}
\end{minipage} & \begin{minipage}[t]{\linewidth}\raggedright
Programs may enable and incentivize (narrowly scoped, non-public) staff whistleblowing, for all verification subgoals.
\end{minipage} & \begin{minipage}[t]{\linewidth}\raggedright
Relatively simple, precedented, and implementation-

ready.
\end{minipage} & \begin{minipage}[t]{\linewidth}\raggedright
Unclear effectiveness: depends on the number and loyalty of accomplices.
\end{minipage} \\

\rowcolor{black!5!white}
\begin{minipage}[t]{\linewidth}\raggedright
\textbf{Interviews of personnel}

\begin{center}
    \includegraphics[width=0.76948in,height=0.56049in]{media/media/image6.png}
\end{center}
\vspace{0em}
\end{minipage} & \begin{minipage}[t]{\linewidth}\raggedright
Interviews may reveal violations at any verification subgoal, e.g., via inconsistencies or perhaps improved lie detection tech, but such tech is abusable.
\end{minipage} & \begin{minipage}[t]{\linewidth}\raggedright
Relatively simple and precedented.
\end{minipage} & \begin{minipage}[t]{\linewidth}\raggedright
Unclear effectiveness: depends on accomplices' ability to lie undetected.
\end{minipage} \\
\begin{minipage}[t]{\linewidth}\raggedright
\textbf{National intelligence activities}

\begin{center}
    \includegraphics[width=0.5625in,height=0.53856in]{media/media/image3.png}
\end{center}
\end{minipage} & \begin{minipage}[t]{\linewidth}\raggedright
Intelligence agencies could collect and analyze intelligence for all verification subgoals, including via human, cyber, and signals intelligence.
\vspace{0.5em}
\end{minipage} & \begin{minipage}[t]{\linewidth}\raggedright
Precedented and may be feasible unilaterally.
\end{minipage} & \begin{minipage}[t]{\linewidth}\raggedright
More adversarial, harder for third parties to verify, and unclear effectiveness.
\end{minipage} \\

\bottomrule\noalign{}
\caption{Summary of personnel-based verification layers and their tradeoffs.} \label{tab:summary_of_personnel-based}
\endlastfoot
\end{longtable}

\begin{figure}[ht]
  \noindent
  \makebox[\textwidth][c]{%
    \includegraphics[width=6.5in,height=2.44444in]{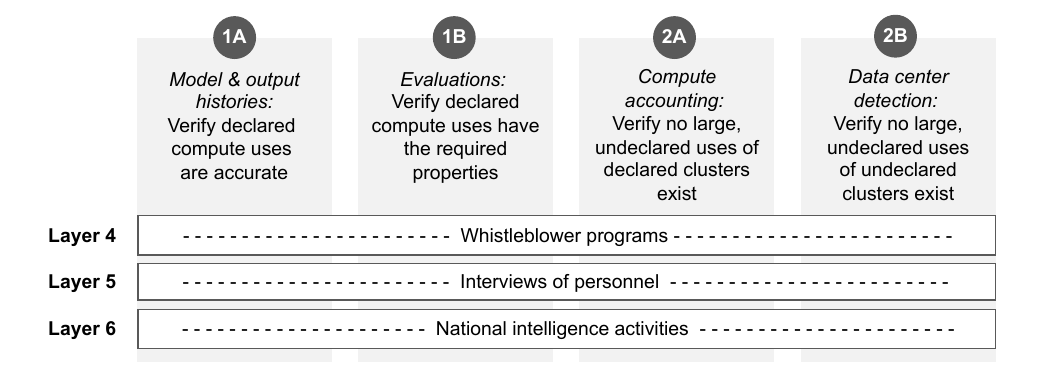}%
  }
  \caption{Summary of how personnel-based verification layers would complete each subgoal. Each of these layers simply consists of a single mechanism applied to all subgoals.}
  \label{fig:summary_of_how_personnel-based_verification_layers_would_complete_each_subgoal}
\end{figure}

In contrast to other, more technical verification mechanisms, \emph{personnel-based verification} relies on the difficulty of having large groups of people collude without disclosures or leaks. Verifiers could systematically seek disclosures or leaks through whistleblower programs, interviews of personnel, and national intelligence activities. Intelligence activities, though, might involve cyber or signals intelligence, not only direct communication with personnel.

\FloatBarrier

\phantomsection\subsubsection{4.3.1 Mechanisms}\label{mechanisms-2}

We outline three distinct mechanisms for personnel-based verification. All are well-established mechanisms for verifying compliance with regulations or international agreements.

\textbf{Whistleblower programs:} Formal, cooperative whistleblower programs could enable and encourage employees to narrowly blow the whistle on violations (\hyperref[a.8-whistleblower-programs]{\ul{Appendix A.8}}):\footnote{There could be whistleblowers without a formal, cooperative program, but then many of the below features to enable and encourage whistleblowers (e.g., regular in-person contact) would be harder to bring about.}

\begin{itemize}
\item
  To enable whistleblowing, employees could be allowed to confidentially view some of their employer's claims. They could also be given regular in-person contact with Verifiers, to counter whistleblower suppression, with measures to minimize inappropriate leaks.
\item
  To encourage whistleblowing, formal programs could, e.g., enable anonymous reports, offer financial rewards, or build pro-whistleblower norms.
\end{itemize}

\textbf{Interviews of personnel:} Employees could reveal violations unintentionally in interviews with Verifiers. This is in contrast with whistleblowers, who intentionally reveal violations. Interviewees could collude to lie, but they may struggle to do so convincingly. Another concern with interviews could be that they might reveal sensitive information; this could be mitigated by limiting interviews to questions within a narrow, agreed-on scope, similar to the questions asked to whistleblowers (\hyperref[a.8-whistleblower-programs]{\ul{Appendix A.8}}).

Hypothetically, one way interviews could be relatively robust is if they involved more reliable lie-detection technology than currently exists. However, we are not recommending the development of such technology, due to its potential for abuse.

\textbf{National intelligence activities:} States\footnote{We emphasize \emph{national} intelligence activities because states have spent decades or longer building intelligence capabilities, unlike international agencies. The International Atomic Energy Agency, for example, has often learned of violations from national agencies rather than on its own \citep{rosenthal_iaea_safeguards, baker2023nucleararmscontrolverification}. Additionally, authorizing an international agency to engage in espionage would expose states to new intelligence risks, while national intelligence activities are the status quo.} could seek evidence of violations through disciplines including human intelligence, signals intelligence, and cyber intelligence. Unlike the above mechanisms, intelligence activities may be effective without the Prover's cooperation.\footnote{Whistleblower programs could be run without the Prover's cooperation, but important practices for their effectiveness, such as regular interviews of employees, rely on Prover cooperation (\hyperref[a.8-whistleblower-programs]{\ul{Appendix A.8}}).} Intelligence-based verification may not be equally feasible for all states in multilateral contexts, but states with relatively weak intelligence capabilities could rely on stronger states' intelligence-based verification. For example, in a hypothetical agreement with the United States and China as parties, if a third party lacked confidence in their own intelligence capabilities, they might trust that the United States would verify (and enforce) China's compliance, because doing so would be in U.S. interests, and vice versa.

\phantomsection\subsubsection{4.3.2 Analysis}\label{analysis-2}

\textbf{Personnel-based verification offers simplicity, but its reliability is unclear.} Personnel-based verification mechanisms tend to offer relatively cheap, simple verification methods, without necessarily depending on complex, new technical protocols or hardware (which may be slow to develop, stress-test, and physically set up). Consequently, personnel-based mechanisms may be unusually feasible to use on short notice. These mechanisms leverage the large number of individuals involved in large-scale AI development---typically hundreds (\hyperref[tab:the_number_of_contributors]{\ul{Table 9}}). Human-based verification would also tend to strengthen other verification mechanisms, catching efforts to circumvent other mechanisms via significant collusion. However, human-based verification mechanisms may be hindered by counterintelligence, such as a violator involving only a small group of well-vetted and surveilled individuals in any violation (possibly for security reasons), though there are potential countermeasures.

\textbf{Personnel-based mechanisms would not only provide their own assurances but also strengthen technical assurances.} In addition to providing independent ways to detect violations, personnel-based mechanisms could detect efforts to circumvent on- and off-chip layers. For example, if a state tried to circumvent on-chip mechanisms, it might use personnel to: identify or design hardware vulnerabilities, physically tamper with chips, or swap a compromised chip design into a chipmaking machine.\footnote{Per one interviewee with experience in hardware manufacturing, a change to the design (i.e., the ``mask'') of an existing chip would by default be known to many people in the manufacturing company, as there are software-enforced approval requirements and physical testing for anomalies in designed chips. Moreover, the standard process for changing a design involves paper trails and several layers of signoff (Interview \#7, 2024).} Thus, a significant number of personnel may be able to disclose or leak evidence of the violation, not even counting the personnel involved in the AI aspects of the violation. In this sense, verification layers do not only provide backup in case other layers fail; they also make it less likely that other layers will fail.

\textbf{The number of human personnel needed for AI development and deployment could fall due to AI automation, reducing the effectiveness of personnel-based verification.} AI models are increasingly capable of software engineering and AI R\&D \citep{bengio2025internationalaisafetyreport, kwa2025measuringaiabilitycomplete}. Still, even if AI engineers become fully automatable, some human involvement could persist. For example, human personnel may remain as organizational leaders, as staff who construct or maintain data centers, and as overseers of AI deployments---especially if significant human oversight is required.\footnote{As an additional hurdle for violators, perhaps all automated AI engineers could be, by requirement, trained to refuse to assist with violations and instead attempt to blow the whistle on any violations (effectively replacing human whistleblowers with ``AI whistleblowers''). This could be circumvented by humans at some point removing this behavior or refraining from implementing it, but that would still introduce humans who could whistleblow.} Separately, capable AI agents could benefit on- and off-chip verification layers (\hyperref[on-chip-verification-layer]{\ul{Section 4.1}}; \hyperref[off-chip-verification-layers]{\ul{Section 4.2}}).

\begin{longtable}[]{
  >{\raggedright\arraybackslash}p{(\linewidth - 4\tabcolsep) * \real{0.3333}}
  >{\raggedright\arraybackslash}p{(\linewidth - 4\tabcolsep) * \real{0.3333}}
  >{\raggedright\arraybackslash}p{(\linewidth - 4\tabcolsep) * \real{0.3333}}}
\toprule\noalign{}
\begin{minipage}[t]{\linewidth}\raggedright
\textbf{AI model} (or model family)
\end{minipage} & \begin{minipage}[t]{\linewidth}\raggedright
\textbf{Number of ``core contributors''}\footnote{Numbers are based on counting the number of distinct names in each category of the models' papers' ``Contributions'' sections \citep{openai2024gpt4technicalreport, llamaTeam2024Llama3, geminiteam2024geminifamilyhighlycapable}, excluding listed external collaborators. OpenAI's GPT-4 paper thanks ``Microsoft Azure for supporting model training with infrastructure design and management,'' suggesting Azure staff are not named as contributors.}
\end{minipage} & \begin{minipage}[t]{\linewidth}\raggedright
\textbf{Number of ``contributors''} (including core contributors)
\vspace{0.5em}
\end{minipage} \\
\endhead
\toprule\noalign{}
\rowcolor{black!5!white}
\begin{minipage}[t]{\linewidth}\raggedright
GPT-4 (OpenAI)
\end{minipage} & \begin{minipage}[t]{\linewidth}\raggedright
85 (excl. Microsoft)
\end{minipage} & \begin{minipage}[t]{\linewidth}\raggedright
287 (excl. Microsoft)
\end{minipage} \\
\begin{minipage}[t]{\linewidth}\raggedright
Llama 3 (Meta)
\end{minipage} & \begin{minipage}[t]{\linewidth}\raggedright
222
\end{minipage} & \begin{minipage}[t]{\linewidth}\raggedright
529
\end{minipage} \\
\rowcolor{black!5!white}
\begin{minipage}[t]{\linewidth}\raggedright
Gemini (Google DeepMind)
\end{minipage} & \begin{minipage}[t]{\linewidth}\raggedright
712
\end{minipage} & \begin{minipage}[t]{\linewidth}\raggedright
1254
\end{minipage} \\

\bottomrule\noalign{}
\caption{The number of contributors and core contributors publicly listed for several prominent AI models. Note these are not defined consistently, and they likely exclude various employees who could blow the whistle on certain violations, e.g., data center construction/maintenance staff, supplier company staff, or other AI developer staff such as product staff.} \label{tab:the_number_of_contributors}
\endlastfoot
\end{longtable}

\phantomsection\subsection{4.4 Supplementary Verification Mechanisms}\label{supplementary-verification-mechanisms}

\begin{figure}[ht]
  \noindent
  \makebox[\textwidth][c]{%
    \includegraphics[width=6.5in,height=3.40278in]{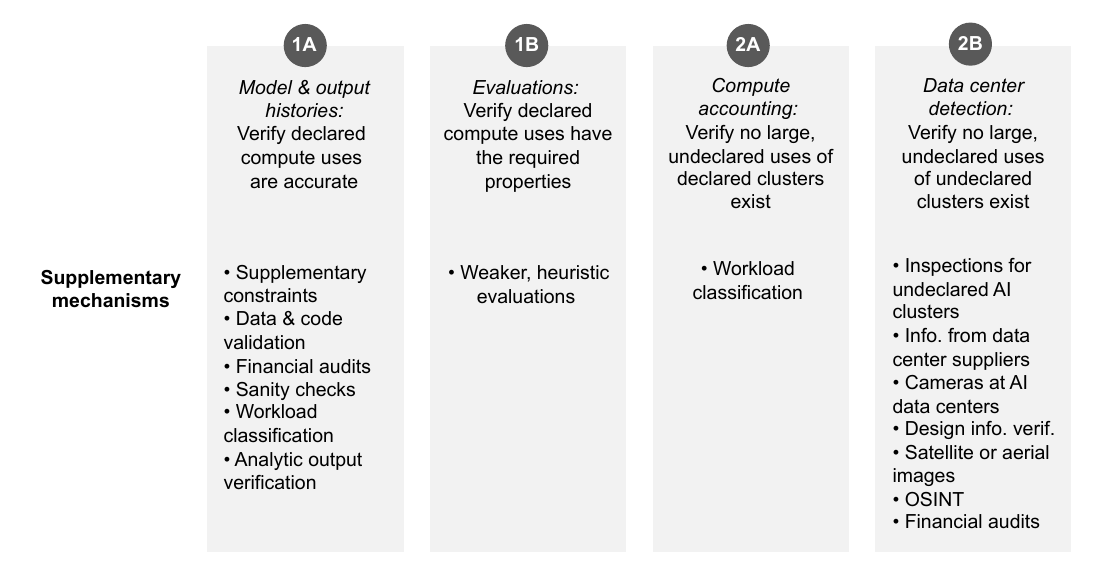}%
  }
  \caption{Supplementary verification mechanisms, each under the verification subgoal(s) it could help complete.}
  \label{fig:supplementary_verification_mechanisms,_each_under_the_verification_subgoal(s)_it_could_help_complete}
\end{figure}

We identified 15 verification mechanisms (\hyperref[fig:supplementary_verification_mechanisms,_each_under_the_verification_subgoal(s)_it_could_help_complete]{\ul{Figure 6}}) that could supplement the verification mechanisms and layers discussed above. We consider these merely supplemental because they appear highly limited in scope (i.e., unable to complete any verification subgoal) or clearly circumventable even if implemented well, whereas for the mechanisms discussed in previous sections, we were able to outline plausible defenses against all identified circumvention tactics. Still, even if these supplemental verification mechanisms can be circumvented, efforts to circumvent them may trigger other, more reliable verification mechanisms. For example, a Prover might seek to hide a data center from satellite images by building it underwater, but this may require contracting one of the few companies with experience building underwater data centers \citep{Gooding2025UnderwaterDC}, which could expose the Prover to hypothetical whistleblower programs or intelligence activities in these companies (\hyperref[personnel-based-verification-layers]{\ul{Section 4.3}}).\footnote{Foreseeing this, a Prover may instead choose a less niche way to hide their data center, such as building it underground or in what appears to be another kind of facility. Still, these options would likely entail higher costs and more accomplices---and thus more potential whistleblowers or human sources---than if the Prover could simply build a data center without worrying about satellite detection.}

The supplemental verification mechanisms we identified are:

\begin{itemize}
\item
  \textbf{Supplementary constraints:} Heuristic tests could supplement partial workload re-execution with constraints (\hyperref[a.4-partial-workload-re-execution-with-constraints]{\ul{Appendix A.4}}).
\item
  \textbf{Data \& code validation:} One could test whether declared data and code have been maliciously engineered to circumvent verification (\hyperref[a.5-data-and-code-validation]{\ul{Appendix A.5}}). This could supplement protocols that are intended to be robust to malicious data and code.
\item
  \textbf{Financial audits:} One could check documentation to see whether organizational cash flows add up. Documents could be forged, but that would be more difficult in the presence of personnel-based mechanisms like whistleblower programs.
\item
  \textbf{Sanity checks:} One could check whether claimed organizational activities are plausibly consistent given an organization's objectives and incentives. For example, a major AI company claiming to spend \$1 billion of compute on weather simulations could be suspicious.
\item
  \textbf{Workload classification:} High-level workload measurements such as power draw may allow for workload classification: distinguishing whether a workload is AI development or deployment \citep{Heim2024Cloud}.\footnote{In the context of compute accounting via analog sensors (\hyperref[a.6-compute-accounting-via-analog-sensors]{\ul{Appendix A.6}}), one implementation option involves physical signatures. This could also be considered a form of workload classification, but it would be more fine-grained---in both its measurements and classifications---than workload classification as discussed here.} Workload classification could play a valuable role (\hyperref[b.2-verification-of-narrower-rules]{\ul{Appendix B.2}}; \hyperref[a.7-verifying-non-ai-workloads-in-ai-data-centers]{\ul{Appendix A.7}}) \citep{Scher2024Verification}. At the same time, the information it gives is too coarse-grained to complete any of our verification subgoals in full; more fine-grained methods (\hyperref[a.6-compute-accounting-via-analog-sensors]{\ul{Appendix A.6}}) are needed to verify compliance with rules that prohibit specific types of development or specific types of deployment.
\item
  \textbf{Analytic output verification:} For some compute uses, most prominently cryptocurrency mining, outputs' authenticity can be verified quickly and with near-certainty by examining the outputs instead of needing to reproduce them.
\item
  \textbf{Weaker, heuristic evaluations:} More uncertain evaluations could supplement higher-confidence ones.
\item
  \textbf{Inspections for undeclared AI clusters:} Inspections could help identify hidden AI compute clusters. To enable inspections, suspect sites could first be identified by other mechanisms in this list, e.g., open-source information and satellite images. However, these methods would struggle to find data centers that are underwater \citep{Gooding2025UnderwaterDC} or otherwise hidden and have strong operational security. They are also challenged by the growing success of decentralized AI development---AI training that is spread out over many smaller compute clusters. In contrast, the other verification mechanisms we highlight in prior sections, such as whistleblower programs and chain-of-custody verification, could be effective even if AI development were split over relatively small compute clusters.\footnote{In scenarios where extremely decentralized training is highly effective, e.g., if violations could be done through extremely decentralized networks of thousands of consumer electronics, this would undermine the on-chip and off-chip verification layers, which focus on AI chips. Still, the personnel-based verification layers could remain effective; extremely decentralized networks would likely involve many accomplices.}
\item
  \textbf{Information from data center suppliers:} Data center suppliers' public statements and privately disclosed documentation could help reveal hidden AI data centers. Data centers have many specialized components, not just AI chips \citep{pilz2023computescalebroadinvestigation}.
\item
  \textbf{Cameras at AI data centers:} Security cameras could help ensure that AI chips are not tampered with or diverted, and that any other verification equipment is also not tampered with. These cameras, like those of the International Atomic Energy Agency \citep{IAEA2016Surveying, rosenthal_iaea_safeguards}, could be tamper-evident and checked on-site. Security cameras are already standard in data centers \citep{DatacenterDynamics2024Field, GluckMazzoli2024DatacenterSecurity}
\item
  \textbf{Design information verification:} Design information verification is a practice of the International Atomic Energy Agency \citep{rosenthal_iaea_safeguards}. It consists of verifying information about a facility's design, including through on-site inspections during construction, to ensure the facility does not have hidden rooms or piping. Applied to AI, analogous checks could help ensure that AI data centers are not built with hidden rooms that could store undeclared AI compute clusters.
\item
  \textbf{Satellite or aerial images:} Satellite or aerial images, including infrared images, could help identify hidden AI data centers, especially in combination with inspections of suspect sites (``Inspections for undeclared AI clusters'' above).
\item
  \textbf{Open-source intelligence (OSINT):} Open-source information, such as social media posts and news reports, could reveal information such as hidden data center constructions, especially in combination with inspections of suspect sites (``Inspections for undeclared AI clusters'' above). Open-source information could also find signs of non-compliant AI deployment, e.g., through its economic, scientific, or military impacts. However, these impacts might not be evident before the deployment causes harm or yields an unfair advantage.\footnote{Broader societal impacts of AI deployment could be delayed if a violator focuses on internal deployment within one organization, such as using AI for AI R\&D. In addition, perhaps it would be challenging to quickly determine whether economic, scientific, or military advances reflect ordinary processes, compliant AI, or non-compliant AI.}
\end{itemize}

\phantomsection\subsection{4.5 Implementation Options Across Mechanisms}\label{implementation-options-across-mechanisms}

Across different verification mechanisms, some common implementation questions arise. Here, we briefly discuss some of these questions, to highlight options and their tradeoffs.

\textbf{Technical tests---hard-coded tests, human auditors, or AI auditors?} On-chip and off-chip verification layers rely on technical tests to check Provers' claims. These tests could involve:

\begin{itemize}
\item
  \emph{Hard-coded tests}: software with all analysis specified precisely in advance. Such software can require a challenging degree of foresight and generality, though constraints on the Prover's formatting could help. This report's \hyperref[appendices]{\ul{Appendices}} explore implementation in the context of hard-coded tests, conservatively assuming that human auditors and AI auditors may be unavailable. Still, protocols designed to be hard-coded tests could also be run by human or AI auditors.
\item
  \emph{Human auditors}: human inspectors who do iterative testing, designing new tests based on intermediate findings.\footnote{Note the use of auditors is a spectrum, which can be varied by a technical parameter: the amount of information that auditors receive as intermediate results. This spectrum ranges from (i) Verifier auditors only receive a single yes/no result about compliance, to (ii) Verifier auditors receive arbitrarily large amounts of intermediate results. There are intermediate options, where auditors receive a fixed amount of information beyond a yes/no compliance determination (e.g., identification of specific aspects of a Prover's declarations that warrant further scrutiny).} The transmission of intermediate results would undermine confidentiality-preservation, but this downside could be limited by the intermediate results being low-bandwidth \citep{Greenblatt2024Exfiltration}, perhaps up to physically isolating the voluntary human auditors \citep{harack_verification_2025}.
\item
  \emph{AI auditors:} AI agents that do iterative testing, designing new tests based on intermediate findings. This may become feasible, as AI systems have been rapidly improving in their ability to complete complex, long-term software engineering and research tasks \citep{bengio2025internationalaisafetyreport, kwa2025measuringaiabilitycomplete}. AI agents are dynamic, unlike hard-coded tests, and unlike humans, their memory can be wiped of confidential information. However, it is unclear when AI agents will have the needed capabilities and reliability, and deployment poses logistical and cybersecurity challenges.
\end{itemize}

\textbf{Security---through transparency or through obscurity?} The technical details of verification mechanisms could be disclosed openly so they can be stress-tested and mutually trusted. However, this openness could help Provers find ways to spoof the verification mechanisms, especially if there is not enough time to make them airtight. Perhaps the ideal approach is for some details to be publicly vetted while others are kept confidential, so that both known and unknown mechanisms can deter violations.

\textbf{Verifier---bilateral or third-party?} Though this report's focus is not on institution design, it is worth noting that Provers may more easily trust that a third-party Verifier will act in good faith, rather than seeking to steal information or compromise equipment. On the other hand, third-party Verifiers may take longer to set up and may not be as effective. In particular, third parties would lack major states' intelligence capabilities, though intelligence sharing may help. These tradeoffs might be eased through intermediate options, such as a national verification office that hires citizens of third-party countries to conduct interviews.

\phantomsection\section{5. Open Problems in Verification}\label{open-problems-in-verification}

Throughout our analysis of verification mechanisms (\hyperref[verification-mechanisms-and-layers]{\ul{Section 4}}; \hyperref[appendices]{\ul{Appendices}}), we identified implementation challenges. Here, we compile a list of selected open problems for technical research and collaboration (\hyperref[tab:rd_challenges_duplicate]{\ul{Table 10}}), as well as brief context for funders and researchers considering work in this area.

\textbf{Selected R\&D and infrastructure challenges.} The following list is filtered for R\&D problems that are relatively challenging, have been subject to relatively little work compared to their difficulty, could play major roles in verification, and would not create highly abusable tech (e.g., lie detection). We separately discuss R\&D challenges more comprehensively and in more detail (\hyperref[appendices]{\ul{Appendices}}; \hyperref[c.3-additional-rd-problems-for-verification]{\ul{Appendix c.3}}).

\begin{longtable}[]{
  >{\raggedright\arraybackslash}p{(\linewidth - 2\tabcolsep) * \real{0.1551}}
  >{\raggedright\arraybackslash}p{(\linewidth - 2\tabcolsep) * \real{0.8449}}}
\toprule\noalign{}
\begin{minipage}[t]{\linewidth}\raggedright
\textbf{Verification layer}
\end{minipage} & \begin{minipage}[t]{\linewidth}\raggedright
\textbf{Selected R\&D and infrastructure challenges}

(Legend: \textcolor{gray}{\rule{0.8em}{0.8em}}: hardware; \textcolor{blue}{\rule{0.8em}{0.8em}}: CS/ML; \textcolor{brown}{\rule{0.8em}{0.8em}}: organizational)
\vspace{0.5em}
\end{minipage} \\
\endhead
\toprule\noalign{}
\rowcolor{black!5!white}
\begin{minipage}[t]{\linewidth}\raggedright
1. Security features in AI chips (\hyperref[on-chip-verification-layer]{\ul{Section 4.1}})
\end{minipage} & \begin{minipage}[t]{\linewidth}\raggedright
\textcolor{blue}{\rule{0.8em}{0.8em}} \textbf{A.} \textbf{System software protocol}: Given verifiable system software (with access to e.g., kernels, memory), verify workload code, and model and data locations, despite any potential obfuscation (\hyperref[a.2-hardware-backed-workload-certificates-and-evaluations]{\ul{Appendix A.2}}).
\vspace{0.5em}

\textcolor{gray}{\rule{0.8em}{0.8em}} \textbf{B.} \textbf{Hardware design attestation}: Given a scanned hardware layout and Hardware Description Language (HDL), check if they match (\hyperref[a.1-full-stack-security-for-technical-verification-mechanisms-implementation]{\ul{Appendix A.1}}).
\vspace{0.5em}

\textcolor{gray}{\rule{0.8em}{0.8em}} \textbf{C.} \textbf{Hardware security features}: Develop highly vetted, open-source, dedicated hardware designs for secure boot, Confidential Computing, and on-chip tamper-proofing (\hyperref[a.2-hardware-backed-workload-certificates-and-evaluations]{\ul{Appendix A.2}}).
\vspace{0.5em}

\textcolor{brown}{\rule{0.8em}{0.8em}} \textbf{D.} \textbf{Design adoption}: Adopt the above designs into leading AI chips.
\vspace{0.5em}
\end{minipage} \\
\begin{minipage}[t]{\linewidth}\raggedright
2. Off-chip network taps \& analysis (\hyperref[off-chip-verification-layers]{\ul{Section 4.2}})
\end{minipage} & \begin{minipage}[t]{\linewidth}\raggedright
\textcolor{blue}{\rule{0.8em}{0.8em}} \textbf{A.} \textbf{Network tap analysis:} Given a cluster's inter-accelerator data (including kernels), verify that the cluster executed only a claimed workload (\hyperref[a.3-network-taps-analysis]{\ul{Appendix A.3}}).
\vspace{0.5em}

\textcolor{gray}{\rule{0.8em}{0.8em}} \textbf{B.} \textbf{Network taps}: Design and manufacture appropriate network taps, or identify suitable existing tech (\hyperref[a.3-network-taps-analysis]{\ul{Appendix A.3}}).
\vspace{0.5em}

\textcolor{brown}{\rule{0.8em}{0.8em}} \textbf{C.} \textbf{Trusted clusters}: Build small compute clusters that are or can be mutually physically secured (\hyperref[off-chip-verification-layers]{\ul{Section 4.2}}).
\vspace{0.5em}
\end{minipage} \\
\rowcolor{black!5!white}
\begin{minipage}[t]{\linewidth}\raggedright
3. Off-chip analog sensors \& analysis

(\hyperref[off-chip-verification-layers]{\ul{Section 4.2}})
\end{minipage} & \begin{minipage}[t]{\linewidth}\raggedright
\textcolor{blue}{\rule{0.8em}{0.8em}} \textbf{A.} \textbf{Code \& data checks}: Develop tests to detect code and data that are designed to spoof partial program re-execution with constraints (e.g. proof-of-learning) (\hyperref[a.4-partial-workload-re-execution-with-constraints]{\ul{Appendix A.4}}; \hyperref[a.5-data-and-code-validation]{\ul{Appendix A.5}}).
\vspace{0.5em}

\textcolor{blue}{\rule{0.8em}{0.8em}} \textbf{B.} \textbf{Workload modeling}: Given an AI workload and cluster specs, estimate the optimal utilization (MFU) and the associated physical signature, e.g., power (\hyperref[a.6-compute-accounting-via-analog-sensors]{\ul{Appendix A.6}}).
\vspace{0.5em}

\textcolor{gray}{\rule{0.8em}{0.8em}} \textbf{C.} \textbf{Analog sensors}: Design and manufacture appropriate analog sensors, or identify suitable existing tech (\hyperref[a.6-compute-accounting-via-analog-sensors]{\ul{Appendix A.6}}).
\vspace{0.5em}

\textcolor{gray}{\rule{0.8em}{0.8em}} \textbf{D.} \textbf{Tamper-proofing}: Design and manufacture appropriate tamper-proof server enclosures (\hyperref[off-chip-verification-layers]{\ul{Section 4.2}}).
\vspace{0.5em}
\end{minipage} \\
\begin{minipage}[t]{\linewidth}\raggedright
Cross-cutting
\end{minipage} & \begin{minipage}[t]{\linewidth}\raggedright
\textcolor{brown}{\rule{0.8em}{0.8em}} \textbf{A.} \textbf{R\&D funding} (e.g., by AISIs, philanthropists, DARPA)
\vspace{0.5em}

\textcolor{brown}{\rule{0.8em}{0.8em}} \textbf{B.} \textbf{Pilot programs} (e.g., voluntary corporate commitments, AISI collaborations)
\vspace{0.5em}

\textcolor{brown}{\rule{0.8em}{0.8em}} \textbf{C.} \textbf{Red teaming} (e.g., by companies, ICs, NIST contest) of developed proposals
\vspace{0em}
\end{minipage} \\

\bottomrule\noalign{}
\caption{Selected open challenges in verification R\&D. We discuss open problems more comprehensively (\hyperref[appendices]{\ul{Appendices}}; \hyperref[c.3-additional-rd-problems-for-verification]{\ul{Appendix C.3}}). Note that work on the listed ``CS/ML'' challenges would often draw on expertise on low-level implementation.} \label{tab:rd_challenges_duplicate}
\endlastfoot
\end{longtable}

\textbf{AI evaluations and standards are complementary to verification R\&D.} Beyond verification R\&D, verification will only be useful if the rules being verified (\hyperref[rules-on-ai-models-data-and-code]{\ul{Section 2.1}}) are well-specified: if compliance is actually sufficient to achieve governance goals such as security. Operationalizing AI governance goals is a challenging and active area of research, which includes work on AI evaluations and standards \citep{shenk2024evaluating, frontier_model_forum_publications, uk_ai_security_institute_work}, and it is a necessary complement to verification R\&D (\hyperref[addressing-broader-challenges-for-verification]{\ul{Section 3.3}}).

\textbf{Context for R\&D funders and researchers.} R\&D funders and researchers may wish to consider the following brief context on the landscape of technical research in AI verification:

\begin{itemize}
\item
  \emph{Relevant fields and areas of expertise:} Computer security research experience will be especially useful for AI verification R\&D.\footnote{Technical AI verification R\&D is a challenge of developing adversarially robust software and/or hardware, which is precisely what computer security focuses on. In addition, some specific topics in AI verification, such as cryptography and secure boot, are primarily studied in computer security.} Broader expertise in software, ML, and computing hardware will often be very applicable as well, including research experience with large language models, distributed computing, and GPUs / AI accelerators.
\item
  \emph{Ongoing, relevant work:} While there is much adjacent research in academia and sometimes industry and government, the specific challenges we list mostly receive little to no work, as of early 2025.\footnote{Multiple interviewees---a large fraction of the computer security experts we interviewed from academia and industry---highlighted ways in which the scenarios we consider differ from their usual threat models (Interviews \#1, 5, and 11, 2024). Still, hardware security features are active areas of R\&D, though arguably not at the needed level of robustness (\hyperref[analysis]{\ul{Section 4.1.2}}). Some workload modeling is presumably also already done by AI companies, e.g., to optimize hardware utilization.} This is because these challenges target unusual scenarios, where hardware designers, manufacturers, cloud providers, and users may all collude against a separate Verifier due to a coordinated government effort. This rules out many typical ways of gaining assurances by trusting some of these actors.\footnote{For example, hardware design attestation (as we define it) is unnecessary if you trust the hardware designer to check that the layout matches their Hardware Description Language, off-chip analog sensors for inferring compute use are unnecessary if you trust on-chip performance counters, and malicious code is less of a concern if you are the user writing the code.} The scenarios we consider also open up unusual options, such as on-site inspections. As partial exceptions, in 2024 there were \$4.1 million granted for research on Flexible Hardware-Enabled Guarantees (FlexHEGs) \citep{SFF2024flexHEGs}, and a further expected \$2-10 million for hardware-enabled mechanisms (including but not only for verification) were announced in 2025 \citep{Longview2025HEMRFP}.
\item
  \emph{Sequencing:} (1) Initial development and pilot programs, (2) world-class red teaming (e.g., an international NIST competition or red teaming by intelligence agencies), and (3) manufacturing/adoption could be done in that order, to efficiently use organizations' limited availability for the latter of these actions.
\end{itemize}

\phantomsection\section{6. Conclusion}\label{conclusion}

In managing AI's impacts, verification of rules on AI development and deployment may play a crucial role: assuring actors that all parties are upholding an agreement, so compliance does not put them at a disadvantage. Confidentiality-preserving verification of rules on large-scale AI compute use may be especially valuable. There are many options for such verification, and it could ultimately feature substantial redundancy. In a framework of several verification subgoals, we saw how all subgoals may be completed by plausible verification mechanisms, forming up to six redundant layers of verification. Personnel-based verification layers, such as whistleblower programs, would offer some assurances with limited preparation required. However, before states can have confidence in technical layers of verification, the R\&D and infrastructure challenges we list require progress, which can be promoted by R\&D funding, pilot programs, and red teaming. With such efforts, we can avoid leaving future policymakers with just a few untried verification options, and instead equip them with an arsenal of stress-tested tools.

\clearpage

\addtocontents{toc}{\protect\setcounter{tocdepth}{1}}

\phantomsection\section{About This Working Paper}\label{about-this-working-paper}

\phantomsection\subsection{Contributions}\label{contributions-1}

\textbf{Mauricio Baker} designed the research methodology, formulated the verification framework, contributed to the analysis of verification mechanisms, conducted the expert interviews, and wrote the majority of the report.

\textbf{Gabriel Kulp} contributed to the analysis of verification mechanisms and confidentiality-preserving technologies, especially on hardware mechanisms and secure implementation of technical mechanisms, and contributed to the structure of the report.

\textbf{Oliver Marks} contributed to the analysis of verification mechanisms and confidentiality-preserving technologies, especially on partial workload re-execution and compute accounting via analog measurement of AI chips, and contributed to the structure of the report.

\textbf{Miles Brundage} gave detailed feedback on the report\textquotesingle s overall argument and structure and contributed to writing.

\textbf{Lennart Heim} contributed strategic guidance, research management, editing, and feedback on the report.

Each author contributed ideas and/or writing to the report and reviewed the report, though authorship does not necessarily imply agreement with every claim made in the report.

\phantomsection\subsection{Technology and Security Policy Center}\label{technology-and-security-policy-center}

RAND Global and Emerging Risks is a division of RAND that delivers rigorous and objective public policy research on the most consequential challenges to civilization and global security. This work was undertaken by the division's Technology and Security Policy Center, which explores how high-consequence, dual-use technologies change the global competition and threat environment, then develops policy and technology options to advance the security of the United States, its allies and partners, and the world. For more information, contact tasp@rand.org.

\phantomsection\subsection{Funding}\label{funding}

This research was independently initiated and conducted within the Technology and Security Policy Center using income from operations and gifts from philanthropic supporters, which have been made or recommended by DALHAP Investments Ltd., Effektiv Spenden, Ergo Impact, Founders Pledge, Charlottes och Fredriks Stiftelse, Good Ventures, Jaan Tallinn, Longview, Open Philanthropy, and Waking Up Foundation. A complete list of donors and funders is available at www.rand.org/TASP. RAND donors and grantors have no influence over research findings or recommendations.

\phantomsection\subsection{Acknowledgements}\label{acknowledgements}

We thank our interviewees for sharing their expertise. We do not list their names to preserve confidentiality. We also thank Ben Harack for his generous feedback. We additionally thank Aaron Scher, Aidan O'Gara, Aris Richardson, Asher Brass, Ben Chang, Bria Persaud, Brodi Kotila, Casey Dugan, Casey Mahoney, Chris Byrd, Daniel Reuter, David Dalrymple, David Glickstein, David Schneider-Joseph, Everett Smith, Felipe Calero Forero, Girish Sastry, Jack Miller, Jair Aguirre, James Bradbury, Joel Predd, Jonathan Happel, Joshua Clymer, Kendrea Beers, Konstantin Pilz, Lisa Thiergart, Michael Aird, Michael Byun, Morgan Livingston, Naci Cankaya, Nora Ammann, Oliver Guest, Onni Aarne, Ying Yi Dang, and Yonadav Shavit for their helpful feedback or discussion. As usual, this report does not necessarily represent the views of acknowledged individuals, and any remaining mistakes are our own.

\clearpage

\addtocontents{toc}{\protect\setcounter{tocdepth}{3}}

\appendix

\phantomsection\section{Appendices}\label{appendices}

Our appendices are split into three sections:

\begin{enumerate}
\def\labelenumi{\Alph{enumi}.}
\item
  \textbf{Implementation analyses}, making up the majority of the appendices, outline how various, mostly technical, implementation challenges could be addressed. We list highlights from these analyses in \hyperref[contributions]{\ul{Section 1.1}}.
\item
  \textbf{Broader regime design} sections discuss: rationales for compute accounting (\hyperref[b.1-compute-accounting-vs.-other-kinds-of-accounting]{\ul{Appendix B.1}}), narrower governance goals that may be easier to verify (\hyperref[b.2-verification-of-narrower-rules]{\ul{Appendix B.2}}), and how verification may deal with ambiguous findings (\hyperref[b.3-acting-on-ambiguous-findings]{\ul{Appendix B.3}}).
\item
  \textbf{Methodology details} elaborate on methodology-related aspects, including overviews of additional R\&D problems (\hyperref[c.3-additional-rd-problems-for-verification]{\ul{Appendix C.3}}) and related work (\hyperref[c.5-related-work]{\ul{Appendix C.5}}).
\end{enumerate}

\phantomsection\subsection{A. Implementation Analyses}\label{a.-implementation-analyses}

\phantomsection\subsubsection{A.1 Full-stack Security for Technical Verification Mechanisms' Implementation}\label{a.1-full-stack-security-for-technical-verification-mechanisms-implementation}

\textbf{Full-stack security overview.} To be adversarially robust, all technical verification mechanisms must not only have secure high-level protocols; these protocols must also be implemented securely. Implementation must be secure despite the large attack surface (namely the whole infrastructure stack, from the Verifier's software to the hardware) and the numerous untrusted actors (including various software vendors and potentially the hardware design, manufacturing, and assembly companies). In requiring such secure infrastructure, many verification mechanisms face shared challenges, which may be amenable to shared solutions. Across the infrastructure stack, (relevant parts of) designs may need to be (1) disclosed, (2) verified as authentic, (3) comprehensively tested for vulnerabilities, and (4) patched or addressed as needed. These actions may leverage technologies including open-source components, secure boot, delayering, and formal verification (\hyperref[tab:overview_of_some_options_for_comprehensively_securing]{\ul{Table 11}}). However, such security may be slow, technically challenging, and expensive (especially if much hardware needs to be replaced), and it may require significant IP disclosures (though these may be less significant than they initially appear).

\textbf{Hardware security for Prover- vs Verifier-owned hardware.} Significant hardware security measures (\hyperref[tab:overview_of_some_options_for_comprehensively_securing]{\ul{Table 11}}) are most needed when relying on hardware that is (i) acquired from untrusted suppliers and (ii) operated by the Prover. The Prover-owned hardware used for on-chip verification (\hyperref[on-chip-verification-layer]{\ul{Section 4.1}}) would likely meet these criteria. In contrast, for mechanisms that rely more on the Verifier's hardware, such as off-chip mechanisms (\hyperref[off-chip-verification-layers]{\ul{Section 4.2}}), the Verifier may have simpler options to secure their hardware.\footnote{For example, the Verifier may be able to avoid intentional backdoors by acquiring hardware from Verifier-trusted suppliers, and the Verifier could potentially run computations on multiple devices from different suppliers to detect uncoordinated backdoor activation. Additionally, for Verifier-operated hardware, the Verifier has additional defenses available against efforts to trigger a Hardware Trojan or leverage side-channel attacks.}

\clearpage

\begin{longtable}[]{
  |>{\raggedright\arraybackslash}p{(\linewidth - 8\tabcolsep) * \real{0.2000}}|
  >{\raggedright\arraybackslash}p{(\linewidth - 8\tabcolsep) * \real{0.2000}}|
  >{\raggedright\arraybackslash}p{(\linewidth - 8\tabcolsep) * \real{0.2000}}|
  >{\raggedright\arraybackslash}p{(\linewidth - 8\tabcolsep) * \real{0.2000}}|
  >{\raggedright\arraybackslash}p{(\linewidth - 8\tabcolsep) * \real{0.2000}}|}
\toprule\noalign{}
\rowcolor{black!5!white}
\begin{minipage}[t]{\linewidth}\raggedright
Component of infrastructure stack
\end{minipage} & \begin{minipage}[t]{\linewidth}\raggedright
1. How designs could be disclosed
\end{minipage} & \begin{minipage}[t]{\linewidth}\raggedright
2. How disclosed designs could be verified authentic
\end{minipage} & \begin{minipage}[t]{\linewidth}\raggedright
3. How disclosed designs could be comprehensively tested for vulnerabilities
\vspace{0.5em}
\end{minipage} & \begin{minipage}[t]{\linewidth}\raggedright
4. How identified vulnerabilities could be mitigated
\end{minipage} \\
\endhead
\toprule\noalign{}
\begin{minipage}[t]{\linewidth}\raggedright
\cellcolor{black!5!white}
Application (of the Verifier\footnote{This refers to the \emph{Verifier's} software being implemented correctly; verification does not necessarily require the Prover's application-layer software to be implemented correctly.})
\vspace{0.5em}
\end{minipage} & \multirow[t]{5}{=}{\begin{minipage}[t]{\linewidth}\raggedright
Using open-source components,\footnote{As prior verification work has noted \citep{shavit2023doescatchchinchillaverifying}, there are existing initiatives on open-source hardware roots-of-trust \citep{Larabel2022Caliptra}. However, the Verifier would need to confirm that a root-of-trust is functionally incorporated into the rest of the chip, rather than e.g., being a decoy.} open-sourcing design IP,\footnote{Stakeholders may be at least partly compensated by improved confidence in product security, AI tools such as code models may facilitate vulnerability detection \citep{fang2024teamsllmagentsexploit}, and much leading AI chip IP has reportedly already been stolen (which may limit the downsides of intentional disclosures) \citep{clark2022nvidialeak, tyson2024googleleak}.} using a Verifier-trusted designer and foundry,\footnote{The U.S. Department of Defense has a Trusted Foundry Program that may achieve this \citep{dod_tapo_trusted}.} or making confidentiality-preserving IP disclosures.\footnote{Confidentiality-preserving technologies would only be secure on an infrastructure stack that the Verifier has other reasons to trust (e.g., having applied the other listed methods to the confidentiality-preserving technology).}
\end{minipage}} & \multirow[t]{3}{=}{\begin{minipage}[t]{\linewidth}\raggedright
Secure boot,\footnote{This includes remote firmware attestation \citep{Attestation2020}. We do not mark this as technically ambitious to reflect that, \emph{if} lower layers (i.e., firmware and hardware) are secure, then firmware attestation is relatively straightforward.} secure compiler.
\end{minipage}} & \multirow[t]{5}{=}{\begin{minipage}[t]{\linewidth}\raggedright
Formal verification of narrow design (with improved provers) *,\footnote{Formal verification is an active area of work in both software \citep{hasan2015formal, Souyris2009Formal} and hardware \citep{10.1145/307988.307989, nunes2019vrased}. Formal verification is labor-intensive and challenging to apply to complex software or hardware. Progress in AI has the potential to help reduce its costs and expand its applicability \citep{Lin2025Toolchain}.} red teaming *, and standard design validation techniques like test cases *.
\end{minipage}} & \multirow[t]{3}{=}{\begin{minipage}[t]{\linewidth}\raggedright
Software/firmware patches, or key revocation where applicable.
\vspace{0.5em}
\end{minipage}} \\
\cline{1-1}
\begin{minipage}[t]{\linewidth}\raggedright
\cellcolor{black!5!white}
Operating system
\vspace{0.5em}
\end{minipage} & & & & \\
\cline{1-1}
\begin{minipage}[t]{\linewidth}\raggedright
\cellcolor{black!5!white}
Hypervisor
\vspace{0.5em}
\end{minipage} & & & & \\
\cline{1-1}\cline{3-3}\cline{5-5}
\begin{minipage}[t]{\linewidth}\raggedright
\cellcolor{black!5!white}
Firmware
\end{minipage} & & \begin{minipage}[t]{\linewidth}\raggedright
The above and below *.\footnote{In secure boot, firmware attests to itself, so malicious firmware might hide its own malicious properties, analogously to compilers or malware that hide their own Trojans \citep{thompson_trusting_trust_1984, MITRE_HideArtifacts}. To address this, delayering and scanning of memory that stores firmware may be needed, ideally limited in scope (e.g., only the first-stage bootloader). Still, firmware attestation should at least detect less sophisticated firmware Trojans.}
\end{minipage} & & \begin{minipage}[t]{\linewidth}\raggedright
The above or below *.\footnote{Often, boot ROM (read-only memory) firmware is burned onto a chip during manufacturing and is not electronically updatable \citep{intel2019arria10, NVIDIA_BootROM}. This can lead to electronically unpatchable vulnerabilities, as have been found in Apple's Boot ROM \citep{kirk_bootrom_2019}.}
\vspace{0.5em}
\end{minipage} \\
\cline{1-1}\cline{3-3}\cline{5-5}
\multirow[t]{2}{=}{\begin{minipage}[t]{\linewidth}\raggedright
\cellcolor{black!5!white}
Hardware
\end{minipage}} & & \begin{minipage}[t]{\linewidth}\raggedright
Delayering *, scanning *, and circuit verification * for a random sample\footnote{Hardware designs could be verified by (i) verifying that a manufactured layout precisely matches a claimed layout, and (ii) verifying that the claimed design (i.e., netlist) corresponds to the claimed layout (with no analog additions). Fortunately, the semiconductor industry has established methods for reverse-engineering chips (e.g., for failure analysis or IP litigation). These include, for (i), ``delayering'' a chip---removing layers one by one---and scanning each layer with a sophisticated microscope, especially a scanning electron microscope \citep{Lumenci2022Delayering, thermofisher2025delayering, 7546493}; and, for (ii), methods for ``circuit extraction.'' Prior work suggests that even analog hardware Trojans may be detectable by scoped delayering \citep{7546493}. Notably, circuit extraction or reverse-engineering is not the only possible approach; one could also re-run programs that map from Hardware Description Language (HDL) to a netlist and then to a layout. However, re-running programs may be nondeterministic (for efficiency) and computationally expensive; ``{[}m{]}odern EDA {[}Electronic Design Automation{]} workloads utilize 20,000 to 100,000 parallel compute cores on a single design'' \citep{aws2023tr4958}.}, and/or oversight of manufacturing *.
\end{minipage} & & \begin{minipage}[t]{\linewidth}\raggedright
Using alternative (non-hardware-based) verification methods *, refraining from use of vulnerable hardware, or key revocation where applicable.
\vspace{0.5em}
\end{minipage} \\
\cline{2-5}
\cellcolor{black!5!white}
& \multicolumn{4}{>{\raggedright\arraybackslash}p{(\linewidth - 8\tabcolsep) * \real{0.8000} + 6\tabcolsep}@{}}{%
\begin{minipage}[t]{\linewidth}\raggedright
\emph{*: Technically challenging/ambitious (if one wishes to use this for strong, affordable, and timely security assurances).}
\vspace{0.5em}
\end{minipage}} \vline \\

\bottomrule\noalign{}
\caption{Overview of some options for comprehensively securing a technology stack for AI verification.} \label{tab:overview_of_some_options_for_comprehensively_securing}
\endlastfoot
\end{longtable}

Next, we non-comprehensively discuss some notable challenges posed by the above security measures, and how these challenges may be addressed.

\textbf{Securing a hypervisor and operating system.} Acquiring a secure hypervisor and operating system (OS) may appear to be a daunting task. The Linux OS, for instance, contains millions of lines of code \citep{Larabel2020Kernel}, and---even though it has been scrutinized for decades---over 1,000 security vulnerabilities were found just in recent years \citep{Day2023NetworkRisks, CVEDetailsLinuxKernel}. Still, hypervisor and OS security may be more feasible for AI verification than it first appears, for several reasons:

\begin{itemize}
\item
  \emph{Hypervisor security:} Hypervisors are designed to isolate virtual machines, offering security even against compromised OSs. Fortunately, hypervisors are well-studied because of their importance for cloud computing. Moreover, there are already open-source hypervisors with relatively ``tiny'' attack surfaces: tens of thousands, not millions, of lines of code. A security expert we interviewed expressed confidence that such a hypervisor, like KVM \citep{KVM}, would be suitable even for high-stakes applications, at least after extensive auditing. (Interview \#6, 2024.)
\item
  \emph{Narrow function:} An OS specialized for verifying rules on AI could have much narrower functions than a general-purpose OS, so the former could potentially consist of much less code (e.g., omitting many drivers, graphical user interface functionalities, and sophisticated support for resource allocation).
\item
  \emph{Opportunity for formal verification:} A lightweight hypervisor and OS could potentially be formally verified. There has been significant work in this direction \citep{lf_projects_sel4_microkernel, RedoxOS}, though secure OS development has so far received limited funding,\footnote{For instance, seL4, the ``first ever industrial-strength, general-purpose operating system with formally proved implementation correctness,'' was only published in 2009 rather than decades earlier \citep{ACMSoftwareSystemAward}} at least partly due to lacking incentives.\footnote{Companies often face relatively low penalties for weak cybersecurity \citep{Dean2015CybersecurityIncentives}.} Reliable formal verification may benefit from more reliable automated provers.
\item
  \emph{Other benefits:} If not already available, a secure hypervisor and OS for running AI workloads would likely more broadly benefit the security of AI model weights, a widely shared policy goal \citep{FrontierAISafety2024, Biden2023EOAI, EU_AIAct, VanceAIActionSummit}, and perhaps computer security more broadly, presumably increasing the feasibility of acquiring funding. However, security-performance tradeoffs may be challenging, and improved AI security might trade off with the feasibility of using cyber intelligence for verification.
\end{itemize}

\textbf{Verification with an untrusted hardware designer.} Hardware security research often \citep{9401143} but not always \citep{6856140} assumes a hardware \emph{manufacturer} or assembly/test company is untrusted. A threat model where the hardware \emph{designer} is also untrusted poses further challenges, including the following, though they may be resolvable:

\begin{itemize}
\item
  \emph{Key management:} Untrusted hardware design and manufacturing introduces challenges in cryptographic key management (for hardware security modules' private keys), including secure key generation and ensuring the designer does not abuse their usual authority to sign firmware updates. For the latter challenge, secure key revocation appears needed. Fortunately, key revocation is a common existing functionality of cryptographic systems \citep{NIST_Key_Revocation}, so that any stolen keys do not create permanent vulnerabilities.
\item
  \emph{Hardware design verification:} With an untrusted hardware design, security requires verifying not only that manufactured hardware (or at least the hardware security module) matches its claimed design, but also that the claimed design is free of backdoors.
\item
  \emph{Reference measures for firmware attestation:} Firmware attestation involves comparing measurements of an attesting device (e.g., iteratively computed hashes of firmware and of executable code) to reference measurements \citep{Attestation2020}. To obtain reliable reference measurements when the hardware designer's measurements are untrusted, without gaining access to private information such as the Prover's code, one option may be for the Verifier and Prover to jointly create reference measurements by emulating a compliant bootup on a confidentiality-preserving technology (e.g., Confidential Computing).
\item
  \emph{Hardware specs:} An untrusted hardware designer might state false hardware performance specifications, posing an obstacle to methods such as operations accounting (\hyperref[a.6-compute-accounting-via-analog-sensors]{\ul{Appendix A.6}}). Specifications could be (i) verified approximately by comparing new devices to known ones, or (ii) verified precisely by (delayering and) scanning chips for bottom-up verification of their operations or bandwidth per clock cycle (and otherwise verifying clock speed, such as through electromagnetic measurement).
\end{itemize}

\textbf{Low-latency chip delayering and scanning.} One might worry that delayering and scanning a random sample of chips would be too slow for timely detection of Hardware Trojans (or false specs), with modern chips having ``up to 100 layers'' \citep{asml_microchips} and nanometer-scale features. To address this, a Verifier could in principle do much faster delayering via three kinds of parallelization (at added expense)\footnote{One could simultaneously: parallelize across layers (i.e., at the limit, for a chip with N layers, delayer N separate copies of the chip to reach a different target layer in each, and then simultaneously scan all layers), parallelize across sampled chips (i.e., scan multiple samples of the same layer at once, as multiple samples are needed to remove noise), and parallelize within each layer of each chip (via multiple electron beams or perhaps slicing each chip perpendicular to the surface---there are photolithography mask writers with 260,000 beams each \citep{nuflare_eb2002}). However, these parallelization methods would come at significant expense, due to the need for more equipment, more sophisticated equipment, and destructive analysis of more chips.} and by only scanning certain components (e.g., a known hardware security module).

\textbf{Attesting to the continuous presence of firmware.} On-chip verification (\hyperref[on-chip-verification-layer]{\ul{Section 4.1}}) requires certain firmware to be continuously present, rather than only present for occasional checks. To attest to this, approaches include: (i) a firmware update that makes secure boot infeasible to disable,\footnote{Hardware often offers the option to disable secure boot \citep{NVIDIA2023SecureBoot, microsoft2021disabling}, but firmware could likely be updated to remove this option. Hardware vendors have previously attempted this, though with exploitable vulnerabilities \citep{hildenbrand_bootloaders_2018, AppleWikiPwnage, jirku_switch_boot_2022, wololo_picofly_2023}.} (ii) a hardware-level secure counter that detects firmware being swapped out,\footnote{For example, there could be an increment-only counter which increments every clock cycle that the device is run without appropriately signed firmware (with an overflow-resistant maximum size).} (iii) randomly timed queries requiring low-latency firmware attestation,\footnote{However, this may be incompatible with a security air gap.} (iv) off-chip sensors such as electricity meters attesting to continuous operation, and (v) more (\hyperref[a.2-hardware-backed-workload-certificates-and-evaluations]{\ul{Appendix A.2}}).

\textbf{Secure operation.} Beyond secure protocol design, the Verifier must also run their protocols securely. This includes personnel security, physical security for Verifier hardware, and a secure IT system for tracking findings.

\phantomsection\subsubsection{A.2 Hardware-Backed Workload Certificates and Evaluations}\label{a.2-hardware-backed-workload-certificates-and-evaluations}

Hardware security features could enable on-chip verification mechanisms including hardware-backed workload certificates and hardware-backed evaluations (\hyperref[a.2-hardware-backed-workload-certificates-and-evaluations]{\ul{Appendix A.2}}). This appendix outlines how these mechanisms could be implemented, assuming tamper-evident secure boot and optionally Confidential Computing are supported in AI compute clusters' AI chips and CPUs. If supported, secure boot could ensure the presence of system software that enforces the following behavior,\footnote{As one class of attacks, a Prover could form a cluster that consists of both compliant chips and malicious chips (i.e., chips that lack the hardware-backed workload certificates protocol). The malicious chips could then engage in data communication without logging it, or a malicious CPU could provide unlogged instructions to a compliant AI chip. One way to address these attacks is for compliant chips to check that fellow chips in their cluster are also running the required protocol, demonstrated by presenting a separate certificate signed by the Verifier. Alternatively, the master CPU could keep track of the identities of the chips in its cluster and include these in the certificate; unauthorized chip additions or omissions could then be apparent. For ensuring that a connection, once established, is maintained, the chips could use encryption, or perhaps occasional challenges, electronic measurements, or external tamper-evident security cameras.} with Confidential Computing optionally used where specified below:

\begin{enumerate}
\def\labelenumi{\arabic{enumi}.}
\item
  \emph{Declarations:} When a Prover loads an AI workload onto an AI compute cluster, the Prover includes explicit, specially formatted information about whether the workload is AI training or inference, and what memory locations and/or data packets will hold the model weights, training data, and usage data. This constitutes the Prover's declaration of their models and data. (The Prover's declaration of code is implicit in the code they load to CPUs.)

  \begin{enumerate}
  \def\labelenumii{\alph{enumii}.}
  \item
    To address \emph{non-AI} workloads that may be done on AI compute clusters, there are various options (\hyperref[a.7-verifying-non-ai-workloads-in-ai-data-centers]{\ul{Appendix A.7}}).
  \end{enumerate}
\item
  \emph{Verification Subgoal 1.A:} The system software checks whether the Prover's above claims are accurate---whether the models and data are as claimed---by analyzing the code and/or runtime data.\footnote{The system software cannot simply assume that e.g., in a training workload, a cluster's inputs from external devices will be the training data, and the cluster's outputs to external devices will be the trained model weights. This assumption would not be robust; the ``real'' models or data---that which a cluster internally uses as models or data---could be encoded in benign-looking cluster inputs or outputs. Moreover, synthetic data could be generated within a cluster rather than being supplied by external devices.} \footnote{Or the system software does not check accuracy directly, but signs a data log that allows the Verifier to later check.} (Note one cannot simply assume that e.g., trained model weights are directly visible as a cluster's outputs; non-compliant model weights might be encoded in a benign-looking cluster output. This motivates collecting and analyzing data on a cluster's internal activities.)

  \begin{enumerate}
  \def\labelenumii{\alph{enumii}.}
  \item
    Checking declarations' accuracy could involve checking that the claimed code, models, and data have various expected properties (\hyperref[a.5-data-and-code-validation]{\ul{Appendix A.5}}), from how the code is factored to the data movement patterns.\footnote{For example, model weights are involved in all-reduce operations during training and are the primary output from the cluster at the end of training.}
  \item
    Procedurally, options for carrying out these checks include the following:

    \begin{enumerate}
    \def\labelenumiii{\roman{enumiii}.}
    \item
      The systems software directly executes the checks on the cluster.
    \item
      The Verifier executes the checks via Confidential Computing (allowing for the checks to not be fully disclosed, though the result could be limited to few bits to reduce the potential for information leakage).
    \item
      The systems software outputs a certificate attesting to the logged data, which the Verifier later tests via Confidential Computing or a trusted cluster (\hyperref[off-chip-verification-layers]{\ul{Section 4.2}}).
    \end{enumerate}
  \end{enumerate}
\item
  \emph{Verification Subgoal 1.B:} The system software checks that the AI models, data, and code have the required properties, e.g., verifies the results of safety evaluations. As above, the system software could do this by:

  \begin{enumerate}
  \def\labelenumii{\alph{enumii}.}
  \item
    Directly executing these checks (or certifying that the Prover's code does so and certifying its result)
  \item
    Confidential Computing
  \item
    Certifying the logged data to allow for later checks (via Confidential Computing or a trusted cluster)
  \end{enumerate}
\item
  \emph{Verification Subgoal 2.A:} To verify that no large-scale uses of AI compute are undeclared (e.g., no certificates are discreetly deleted or hidden), options include the following,\footnote{The secure storage and offline licensing options are similar to prior proposals \citep{shavit2023doescatchchinchillaverifying}.} which could be applied jointly for extra assurances:

  \begin{enumerate}
  \def\labelenumii{\alph{enumii}.}
  \item
    \emph{Secure storage}: Chips could securely store a count of the number of certificates they have generated (or are in the process of producing),\footnote{Secure storage could potentially be implemented through e.g., fuse burning or other increment-only on-chip non-volatile storage (especially for storing a simple count), with a large enough range to prevent overflows. A further security layer could be only allowing the system software to edit this storage.} or securely store the certificates. Then, the Verifier could confirm this information by receiving it with a signature from the trusted system software (with a nonce to prevent the use of old signatures).
  \item
    \emph{Secure memory}: As a variant of the above, relevant chips could be continuously powered and thus able to store certificates in memory. Continuous power could be verified by e.g., the system software securely logging its own time since boot in memory, or storing in memory a password received from the Verifier at boot (with the password being encrypted when received and later returned).
  \item
    \emph{Compute accounting}: The system software could associate certificates with the amount of computation (e.g., in GPU-hours) used to generate the certificates.\footnote{Though logging clock time would require the ability to detect or counter attacks like overclocking.} Then, assuming the Verifier can separately verify the Prover's total compute use (e.g., they know the number of chips by Subgoal 2.B and the Prover claims to have had them all operational nearly continuously), the Prover will only be able to account for all their compute use by disclosing all their certificates.

    \begin{enumerate}
    \def\labelenumiii{\roman{enumiii}.}
    \item
      This approach is broadly similar to compute accounting in the context of off-chip mechanisms (\hyperref[a.6-compute-accounting-via-analog-sensors]{\ul{Appendix A.6}}), but the mechanisms for determining a workload's compute use differ.
    \end{enumerate}
  \item
    \emph{Offline licensing} \citep{Kulp2024}: AI chips could periodically share certificates and have limited functionalities if they do not receive a confirmation of receipt within an expected amount of time.
  \end{enumerate}
\end{enumerate}

\phantomsection\subsubsection{A.3 Network Taps \& Analysis}\label{a.3-network-taps-analysis}

Network taps, in principle, have various options for where they could operate \citep{petrie2025technicaloptionsflexiblehardwareenabled}, but all face substantial challenges:

\begin{itemize}
\item
  \textbf{Chip-level:} network taps could attempt to tap the communications of individual AI chips, by tapping AI chips' connections to memory (i.e., HBM/interposer) or to other AI chips (e.g., NVLink/NVSwitch). However, introducing network taps here would be very difficult. One would have to replace or modify these cutting-edge, TB/s-scale technologies, a task made even more difficult by the proprietary nature of NVLink and the co-packaging of AI chips with memory.
\item
  \textbf{Server-level or higher:} network taps could tap communications between AI servers, server racks, pods, etc., by tapping devices such as Network Interface Cards, InfiniBand connections, or network switches. With bandwidths on the order of 100 GB/s rather than TB/s, more open specifications, and more modular designs, replacement or modification to incorporate network taps does not face quite as steep challenges as chip-level network taps. On the other hand, the lower visibility of these further-out network taps may limit their value for verification.
\end{itemize}

As another challenge, whether dealing with communications on the scale of 1 TB/s or 100 GB/s per tap, storing all tapped data would fill a 1 TB hard drive every \textasciitilde0.1-10 seconds per tap, suggesting that substantial random sampling and/or real-time processing would be needed.

A sketch of a potential protocol (for Subgoals 1.A and 2.A) using network taps is as follows:

\begin{enumerate}
\def\labelenumi{\arabic{enumi}.}
\item
  A Verifier ensures that tamper-evident network taps are installed to tap the communications of every AI accelerator in a compute cluster (or only at a higher level of organization, such as one per server). The Verifier and Prover would both take various steps to verify these network taps lack malicious functionalities but still work as intended (\hyperref[prerequisites-off-chip-devices]{\ul{Section 4.2.1.1}}).
\item
  The network taps log random samples of all data being inputted to or outputted from the attached AI accelerators. This includes GPU kernels.

  \begin{enumerate}
  \def\labelenumii{\alph{enumii}.}
  \item
    This requires verifying limits on significant data exchanges via side-channels. Physical inspections, security cameras, physical barriers, and sensors may help here (though sensors to detect side-channel communication are conceptually distinct from sensors to oversee AI chips' activity (\hyperref[a.6-compute-accounting-via-analog-sensors]{\ul{Appendix A.6}})).
  \item
    However, workload data could be encrypted when exchanged between accelerators, especially if Confidential Computing is being used. This could be addressed through limited disclosure of private keys (confined to Prover-overseen chips) if the hardware allows for this, or otherwise not using Confidential Computing during regular operation.\footnote{For compatibility with on-chip verification that uses Confidential Computing, one could run occasional tests on a small fraction of hardware with Confidential Computing, but not use Confidential Computing for regular operation.} That said, network taps need not worsen encryption on net; they could also potentially be used to strengthen server-level encryption \citep{Petrie2024FlexHEG}.
  \end{enumerate}
\item
  In mutually secured chips (\hyperref[prerequisites-off-chip-devices]{\ul{Section 4.2.1.1}}), the Verifier:

  \begin{enumerate}
  \def\labelenumii{\alph{enumii}.}
  \item
    Re-runs a random sample of kernels (or instructions at a higher level of organization) to verify that the instruction set architecture functions as declared (i.e., that instructions in the kernel mean what the Verifier thinks they mean). This makes the Verifier's analysis of kernels more informative. Re-running kernels is analogous to proof-of-learning \citep{jia2021proofoflearningdefinitionspractice, shavit2023doescatchchinchillaverifying}, but (i) for an individual kernel (or other cluster unit) instead of a whole segment of training,\footnote{Thanks to Jonathan Happel for this idea.} and with guaranteed (probabilistic) constraints on what data was or was not communicated.
  \item
    Analyzes the kernels, other logged data, and Prover declarations to verify that the compute cluster executed (only) the declared high-level code. This is an unsolved problem.

    \begin{enumerate}
    \def\labelenumiii{\roman{enumiii}.}
    \item
      Approaches could include:

      \begin{enumerate}
      \def\labelenumiv{\arabic{enumiv}.}
      \item
        Top-down: Derive information about what kernels and inputs/outputs should contain given the declared high-level code (potentially assuming a mutually trusted compiler), and check that this matches the collected data.
      \item
        Bottom-up: Reconstruct the high-level code from the kernels and inputs/outputs \citep{Petrie2024FlexHEG}.
      \end{enumerate}
    \item
      To facilitate the above or add constraints to partial workload re-execution (\hyperref[a.4-partial-workload-re-execution-with-constraints]{\ul{Appendix A.4}}), a protocol could require the Prover to send specially formatted information to network taps over the course of a workload.
    \item
      As an additional check, network taps (if also given channels for uploading data to clusters) may be able to verify that, at a given time, the majority of a cluster's memory consists of certain data (e.g., random data or certain model weights). They could potentially do this by uploading large amounts of data to a cluster---ideally enough to fill its memory---and checking that the cluster can then return that data.
    \end{enumerate}
  \end{enumerate}
\item
  To address \emph{non-AI} workloads that may be done on AI compute clusters, there are various options (\hyperref[a.7-verifying-non-ai-workloads-in-ai-data-centers]{\ul{Appendix A.7}}).
\end{enumerate}

\phantomsection\subsubsection{A.4 Partial Workload Re-Execution With Constraints}\label{a.4-partial-workload-re-execution-with-constraints}

\textbf{Background:} A Verifier may wish to verify that a declared workload was actually run (Subgoal 1.A). For example, a Prover may make claims about what training code, data, and intermediate results (e.g., model weight checkpoints) were involved in training some model weights, and the Verifier may wish to verify these claims \citep{jia2021proofoflearningdefinitionspractice, choi2023toolsverifyingneuralmodels}.\footnote{More precisely, the problem here (in the context of verifying training, i.e., Subgoal 1.A.1) is close to the proof-of-training-data problem as defined in prior work \citep{choi2023toolsverifyingneuralmodels}. However, there is an important difference; here, the Verifier may be able to assume that the amount of compute the Prover can use to create a spoof is at most the amount of compute the declared workload would take, on the basis of compute accounting (\hyperref[a.6-compute-accounting-via-analog-sensors]{\ul{Appendix A.6}}). In this respect, the problem here may more closely resemble proof-of-learning \citep{jia2021proofoflearningdefinitionspractice}.}

\begin{itemize}
\item
  The Verifier can do this by verifying \emph{faithfulness}, i.e., that running the declared workload in fact produces the claimed results, and \emph{uniqueness}, i.e., that a faithful declaration is infeasible to produce in practice except by actually running the declared workload \citep{choi2023toolsverifyingneuralmodels}.
\item
  The Verifier can verify faithfulness and uniqueness, respectively, via (1) \emph{partial workload re-execution}, i.e., re-running (randomly sampled parts of) the Prover's program to check if the declared results are approximately reproducible, and (2) \emph{constraints}, i.e., checking that the declaration meets constraints which rule out spoofed declarations.\footnote{Further, constraints should rule out ``glue-ing'' attacks, where segments of different training workloads are presented as consecutive portions of one workload. These attacks would create discrepancies only in rare transition points, so random sampling would be unlikely to detect them \citep{jia2021proofoflearningdefinitionspractice}. Instead, cheap tests applied to all training segments would be a more viable countermeasure \citep{choi2023toolsverifyingneuralmodels}.}

  \begin{itemize}
  \item
    This approach generalizes a proposal for proof-of-training-data \citep{choi2023toolsverifyingneuralmodels}.
  \item
    These checks could be done on confidentiality-preserving technology (\hyperref[the-framework]{\ul{Section 3.2}}), to avoid leaking the Prover's sensitive information.
  \item
    Various constraints could be applied (\hyperref[tab:possible_constraints_to_rule_out_spoofs]{\ul{Table 12}}), though it is unclear which constraints will be sufficient to rule out spoofs.
  \end{itemize}
\end{itemize}

{
  \setlength{\LTleft}{-33pt}
  \setlength{\LTright}{-26pt}
  \centering
    \begin{longtable}[]{
      >{\raggedright\arraybackslash}p{(\linewidth - 8\tabcolsep) * \real{0.1806}}
      >{\raggedright\arraybackslash}p{(\linewidth - 8\tabcolsep) * \real{0.4048} * \real{1.37}}
      >{\raggedright\arraybackslash}p{(\linewidth - 8\tabcolsep) * \real{0.1382}}
      >{\raggedright\arraybackslash}p{(\linewidth - 8\tabcolsep) * \real{0.1382}}
      >{\raggedright\arraybackslash}p{(\linewidth - 8\tabcolsep) * \real{0.1382}}}
    \toprule\noalign{}
    \begin{minipage}[t]{\linewidth}\raggedright
    Constraint type
    \end{minipage} & \begin{minipage}[t]{\linewidth}\raggedright
    Constraint
    \end{minipage} & \begin{minipage}[t]{\linewidth}\centering
    Applicable to AI training?
    \end{minipage} & \begin{minipage}[t]{\linewidth}\centering
    Applicable to AI inference?
    \end{minipage} & \begin{minipage}[t]{\linewidth}\centering
    Applicable to non-AI workloads?
    \vspace{0.5em}
    \end{minipage} \\
    \endhead
    \toprule\noalign{}
    \rowcolor{black!5!white}
    \multirow[t]{2}{=}{\begin{minipage}[t]{\linewidth}\raggedright
    Randomness constraints
    \end{minipage}} & \begin{minipage}[t]{\linewidth}\raggedright
    Initialization is verifiably random \citep{choi2023toolsverifyingneuralmodels}
    \end{minipage} & \begin{minipage}[c]{\linewidth}\centering
    \textcolor{green!50!black}{\checkmark}
    \end{minipage} & \begin{minipage}[c]{\linewidth}\centering
    \textcolor{red}{X}
    \end{minipage} & \begin{minipage}[c]{\linewidth}\centering
    \textcolor{yellow!95!black}{\rule{0.8em}{0.8em}}
    \end{minipage} \\
    \rowcolor{black!5!white}
    & \begin{minipage}[t]{\linewidth}\raggedright
    Training data order is verifiably random \citep{choi2023toolsverifyingneuralmodels}\footnote{This is not applicable to training in which the data is generated iteratively (e.g., reinforcement learning with online learning). It is only feasible to a limited extent for training if the data order is fixed for curriculum learning.}
    \end{minipage} & \begin{minipage}[c]{\linewidth}\centering
    \textcolor{yellow!95!black}{\rule{0.8em}{0.8em}}
    \end{minipage} & \begin{minipage}[c]{\linewidth}\centering
    \textcolor{red}{X}
    \end{minipage} & \begin{minipage}[c]{\linewidth}\centering
    \textcolor{yellow!95!black}{\rule{0.8em}{0.8em}}
    \end{minipage} \\
    \begin{minipage}[t]{\linewidth}\raggedright
    Compute constraint
    \end{minipage} & \begin{minipage}[t]{\linewidth}\raggedright
    Compute accounting limits the compute usable to generate spoofs\footnote{If the Verifier can verify that the Prover does no undeclared, large-scale workloads (Subgoal 2) and that none of the declared workloads consist of generating a spoofed (i.e., non-unique) declaration, this will constrain the amount of compute the Prover can spend on creating a spoof.}
    \end{minipage} & \begin{minipage}[c]{\linewidth}\centering
    \textcolor{green!50!black}{\checkmark}
    \end{minipage} & \begin{minipage}[c]{\linewidth}\centering
    \textcolor{green!50!black}{\checkmark}
    \end{minipage} & \begin{minipage}[c]{\linewidth}\centering
    \textcolor{green!50!black}{\checkmark}
    \end{minipage} \\
    \rowcolor{black!5!white}
    \multirow[t]{2}{=}{\begin{minipage}[t]{\linewidth}\raggedright
    Requiring more intermediate states to be replicable
    \end{minipage}} & \begin{minipage}[t]{\linewidth}\raggedright
    Intermediate results, e.g., activations, are replicable \citep{sun_svip_2024}
    \end{minipage} & \begin{minipage}[c]{\linewidth}\centering
    \vspace{0.5em}
    \textcolor{green!50!black}{\checkmark}
    \end{minipage} & \begin{minipage}[c]{\linewidth}\centering
    \vspace{0.5em}
    \textcolor{green!50!black}{\checkmark}
    \end{minipage} & \begin{minipage}[c]{\linewidth}\centering
    \vspace{0.5em}
    \textcolor{yellow!95!black}{\rule{0.8em}{0.8em}}
    \end{minipage} \\
    \rowcolor{black!5!white}
    & \begin{minipage}[t]{\linewidth}\raggedright
    Optimizer states, e.g., gradients, are replicable
    \vspace{0.5em}
    \end{minipage} & \begin{minipage}[c]{\linewidth}\centering
    \textcolor{green!50!black}{\checkmark}
    \end{minipage} & \begin{minipage}[c]{\linewidth}\centering
    \textcolor{red}{X}
    \end{minipage} & \begin{minipage}[c]{\linewidth}\centering
    \textcolor{yellow!95!black}{\rule{0.8em}{0.8em}}
    \end{minipage} \\
    \begin{minipage}[t]{\linewidth}\raggedright
    Requiring replication of specific logged values
    \vspace{0.3em}
    \end{minipage} & \begin{minipage}[t]{\linewidth}\raggedright
    Values securely logged by on-chip mechanisms \citep{shavit2023doescatchchinchillaverifying} (\hyperref[on-chip-verification-layer]{\ul{Section 4.1}}) or network taps (\hyperref[a.3-network-taps-analysis]{\ul{Appendix A.3}}) are replicable
    \end{minipage} & \begin{minipage}[c]{\linewidth}\centering
    \vspace{2.15em}
    \textcolor{green!50!black}{\checkmark}
    \end{minipage} & \begin{minipage}[c]{\linewidth}\centering
    \vspace{2.15em}
    \textcolor{green!50!black}{\checkmark}
    \end{minipage} & \begin{minipage}[c]{\linewidth}\centering
    \vspace{2.15em}
    \textcolor{green!50!black}{\checkmark}
    \end{minipage} \\
    \rowcolor{black!5!white}
    \begin{minipage}[t]{\linewidth}\raggedright
    Time or order constraints
    \end{minipage} & \begin{minipage}[t]{\linewidth}\raggedright
    Values were produced in specific time intervals or orders\footnote{For example, a workload could potentially be modified (e.g., by perturbing activations or changing the sampling seed) so its outputs are more dependent on previous outputs, or so they are dependent on a Verifier-provided value (thus verifying they were produced after the Verifier provided that value), or to be more frequently shared with the Verifier (thus verifying intermediate outputs existed by some point). Perhaps verifiable delay functions \citep{cryptoeprint:2018/601} could also contribute to timing constraints. These constraints could make it more challenging for a Prover to retroactively modify values to create spoofs. Thanks to Daniel Reuter for a version of this idea.}
    \vspace{0.3em}
    \end{minipage} & \begin{minipage}[c]{\linewidth}\centering
    \vspace{0.2em}
    \textcolor{green!50!black}{\checkmark}
    \end{minipage} & \begin{minipage}[c]{\linewidth}\centering
    \vspace{0.2em}
    \textcolor{green!50!black}{\checkmark}
    \end{minipage} & \begin{minipage}[c]{\linewidth}\centering
    \vspace{0.2em}
    \textcolor{green!50!black}{\checkmark}
    \end{minipage} \\
    \multirow[t]{4}{=}{\begin{minipage}[t]{\linewidth}\raggedright
    Requiring consistency with empirical trends
    \end{minipage}} & \begin{minipage}[t]{\linewidth}\raggedright
    The norm of the weight changes lacks abrupt jumps \citep{choi2023toolsverifyingneuralmodels}
    \end{minipage} & \begin{minipage}[c]{\linewidth}\centering
    \vspace{1em}
    \textcolor{green!50!black}{\checkmark}
    \end{minipage} & \begin{minipage}[c]{\linewidth}\centering
    \vspace{1em}
    \textcolor{red}{X}
    \end{minipage} & \begin{minipage}[c]{\linewidth}\centering
    \vspace{1em}
    \textcolor{yellow!95!black}{\rule{0.8em}{0.8em}}
    \end{minipage} \\
    & \begin{minipage}[t]{\linewidth}\raggedright
    Models memorize the data most recently trained on \citep{choi2023toolsverifyingneuralmodels}
    \end{minipage} & \begin{minipage}[c]{\linewidth}\centering
    \vspace{1em}
    \textcolor{green!50!black}{\checkmark}
    \end{minipage} & \begin{minipage}[c]{\linewidth}\centering
    \vspace{1em}
    \textcolor{red}{X}
    \end{minipage} & \begin{minipage}[c]{\linewidth}\centering
    \vspace{1em}
    \textcolor{yellow!95!black}{\rule{0.8em}{0.8em}}
    \end{minipage} \\
    & \begin{minipage}[t]{\linewidth}\raggedright
    The performance of model checkpoints fits the functional form of AI scaling laws \citep{kaplan2020scalinglawsneurallanguage, hoffmann2022trainingcomputeoptimallargelanguage, epoch2023scalinglawsliteraturereview}
    \end{minipage} & \begin{minipage}[c]{\linewidth}\centering
    \vspace{2em}
    \textcolor{green!50!black}{\checkmark}
    \end{minipage} & \begin{minipage}[c]{\linewidth}\centering
    \vspace{2em}
    \textcolor{red}{X}
    \end{minipage} & \begin{minipage}[c]{\linewidth}\centering
    \vspace{2em}
    \textcolor{red}{X}
    \end{minipage} \\
    & \begin{minipage}[t]{\linewidth}\raggedright
    Most gradients generated in training are not highly similar to each other\footnote{Choi, et al. \citep{choi2023toolsverifyingneuralmodels} mention that randomizing the data order ``does not address hypothetical methods for constructing synthetic data points that would induce a particular gradient with respect to any weight vector that would be encountered across many possible training runs with high probability,'' though no such methods are known. In case such methods were found, one could detect their use by checking for whether most gradients in a sample of gradients are highly similar to each other. This check may also detect a hypothetical attack involving a loss function that optimizes a model to be closer to a target model (with weight checkpoints derived by interpolation); other constraints like checking for data memorization may also address that. Perhaps an attacker could still induce a particular gradient \emph{on average} (after the gradient is scaled by the learning rate or otherwise adjusted by the optimization algorithm), but one could check sample gradients for that property as well.}
    \end{minipage} & \begin{minipage}[c]{\linewidth}\centering
    \textcolor{green!50!black}{\checkmark}
    \end{minipage} & \begin{minipage}[c]{\linewidth}\centering
    \textcolor{red}{X}
    \end{minipage} & \begin{minipage}[c]{\linewidth}\centering
    \textcolor{yellow!95!black}{\rule{0.8em}{0.8em}}
    \end{minipage} \\
    \rowcolor{black!5!white}
    \begin{minipage}[t]{\linewidth}\raggedright
    Sensitivity and cross-system analyses
    \end{minipage} & \begin{minipage}[t]{\linewidth}\raggedright
    Perturbing or replacing a component (e.g., weights, data, training algorithm) has the expected effects
    \end{minipage} & \begin{minipage}[c]{\linewidth}\centering
    \vspace{1.25em}
    \textcolor{green!50!black}{\checkmark}
    \end{minipage} & \begin{minipage}[c]{\linewidth}\centering
    \vspace{1.25em}
    \textcolor{green!50!black}{\checkmark}
    \end{minipage} & \begin{minipage}[c]{\linewidth}\centering
    \vspace{1.25em}
    \textcolor{green!50!black}{\checkmark}
    \end{minipage} \\
    \begin{minipage}[t]{\linewidth}\raggedright
    Origin constraint
    \end{minipage} & \begin{minipage}[t]{\linewidth}\raggedright
    A model used for large-scale inference has a declared training history
    \end{minipage} & \begin{minipage}[c]{\linewidth}\centering
    \textcolor{red}{X}
    \end{minipage} & \begin{minipage}[c]{\linewidth}\centering
    \textcolor{green!50!black}{\checkmark}
    \end{minipage} & \begin{minipage}[c]{\linewidth}\centering
    \textcolor{yellow!95!black}{\rule{0.8em}{0.8em}}
    \end{minipage} \\
    \rowcolor{black!5!white}
    \begin{minipage}[t]{\linewidth}\raggedright
    Data and code constraints
    \end{minipage} & \begin{minipage}[t]{\linewidth}\raggedright
    For example, code for AI training has the structure of gradient descent, in terms of code factorization.\footnote{This somewhat overlaps with cross-system analyses, as a gradient calculation can be verified by running a different function for computing a gradient.} (\hyperref[a.5-data-and-code-validation]{\ul{Appendix A.5}}.)
    \vspace{0.5em}
    \end{minipage} & \begin{minipage}[c]{\linewidth}\centering
    \vspace{1.7em}
    \textcolor{green!50!black}{\checkmark}
    \end{minipage} & \begin{minipage}[c]{\linewidth}\centering
    \vspace{1.7em}
    \textcolor{green!50!black}{\checkmark}
    \end{minipage} & \begin{minipage}[c]{\linewidth}\centering
    \vspace{1.7em}
    \textcolor{green!50!black}{\checkmark}
    \end{minipage} \\
    
    \bottomrule\noalign{}
    \caption{Possible constraints to rule out spoofs in partial workload re-execution. This compilation is non-exhaustive. A yellow square denotes partial applicability, i.e., applicability to some forms of the workload type.} \label{tab:possible_constraints_to_rule_out_spoofs}
    \endlastfoot
    \end{longtable}
}

\textbf{Applicability to different subgoals in our verification framework:}

\begin{itemize}
\item
  \textbf{Subgoal 1.A.1:} For verifying the correctness of declared AI training, a Verifier could re-execute training between randomly sampled, adjacent pairs of model checkpoints (snapshots of an AI model partway through training) \citep{jia2021proofoflearningdefinitionspractice}.
\item
  \textbf{Subgoal 1.A.2:} For verifying the correctness of declared AI inference, a Verifier could re-execute inference on a random sample of inputs (i.e., prompts).
\item
  \textbf{Subgoal 1.A.3:} For verifying the correctness of many non-AI programs, such as physics simulations, a Verifier could re-execute inference between randomly sampled, adjacent pairs of checkpoints of the simulation state, or (for programs that analyze many items, such as many possible drug designs) re-execute analysis on a random sample of items.
\end{itemize}

\textbf{Challenges and mitigations:}

\begin{itemize}
\item
  \textbf{Attacks considered in the proof-of-training-data literature:} Prior research on proof-of-training data protocols \citep{choi2023toolsverifyingneuralmodels} proposes constraints to counter various attacks. These attacks include ``glue-ing'' attacks, where records of different workloads are ``glued'' together; attacks that use synthetic data, data reordering, or adversarially chosen initializations to spoof training declarations; and data addition or data subtraction attacks. To counter these attacks, Choi, et al. highlight and provide experimental support for some constraints: verifiably random initializations and data orders, and checking whether model checkpoints have ``memorized'' their most recently seen data as expected. These are powerful defenses; randomizing the data order breaks any attack involving a carefully ordered sequence of data. Supplemental constraints (\hyperref[tab:possible_constraints_to_rule_out_spoofs]{\ul{Table 12}}) could further limit an attacker's options. However, there have only been limited experimental tests of these methods \citep{choi2023toolsverifyingneuralmodels}, and online learning (where data is iteratively generated using a model as it is trained) is incompatible with a randomized data order, though it may be addressable by verifying that online learning data is not generated in a manner conducive to relevant attacks (\hyperref[a.5-data-and-code-validation]{\ul{Appendix A.5}}).
\item
  \textbf{Prover code attacks:} The above attacks leverage adversarially chosen data, initializations, or model checkpoints, but not adversarial code. However, a Prover might also attempt an attack that leverages non-standard behavior of the Prover's declared code. For example, a Prover might submit AI ``training'' code that updates weights by overwriting them with the expected pre-computed values (encoded in the ``data''), instead of executing a gradient update. As another attack, a Prover might submit code (e.g., forward pass code) that swaps in a safer model to respond to safety-related queries.\footnote{These attacks should not be confused with the Prover's code exploiting a vulnerability in the \emph{Verifier's} code to corrupt the results of a verification program; we address that separately (\hyperref[a.1-full-stack-security-for-technical-verification-mechanisms-implementation]{\ul{Appendix A.1}}).} Constraints discussed above like verifiably random initializations and data orders as well as memorization checks might defeat this attack, but further constraints (\hyperref[tab:possible_constraints_to_rule_out_spoofs]{\ul{Table 12}}) and data and code validation (\hyperref[a.5-data-and-code-validation]{\ul{Appendix A.5}}) may be needed. Code validation could leverage backdoor detection methods \citep{Thomas2018Backdoors}; the problem here is easier---perhaps much easier---than the fully general problem of backdoor detection.\footnote{(1) In the fully general problem of backdoor detection, one wants to ensure that a program will never activate a backdoor, on any input. Here, the Verifier only needs to ensure that backdoors do not trigger too often on a given set of inputs. Thus, here there is no need to detect latent backdoors; it suffices to sample from the given set of inputs and detect activated backdoor behavior. (2) In the more general problem of detecting AI models with learned backdoors, the standard or non-backdoored behavior might not be precisely known, and it might only depend on the model outputs. When replicating a claimed workload, one might demand low-error replication (e.g., demand exact replication of inference done with integer operations), as well as replication of intermediate results such as intermediate activations. These demands pose additional constraints for an attacker, perhaps making subtle backdoors infeasible.}
\item
  \textbf{Novel attacks:} The above approaches are heuristic-based, leaving open the possibility of novel attacks \citep{choi2023toolsverifyingneuralmodels}, though any novel attack would face many constraints.\footnote{As discussed, Choi, et al. do not explicitly address malicious code attacks, but their defenses may nonetheless generalize to these attacks. This is somewhat suggestive of the generalizability of current defenses to novel attacks.} Identifying more attacks and defenses would help, and a theoretically backed approach would be ideal.
\item
  \textbf{Inflated compute use attacks:} The Verifier might wish to verify the accuracy of declared workloads, not only in terms of the workloads' input-output behavior, but also in terms of their compute use. This is especially important if the Verifier wishes to use partial workload re-execution along with compute accounting via analog measurement of AI chips (\hyperref[a.6-compute-accounting-via-analog-sensors]{\ul{Appendix A.6}}). A Prover may have various ways of inflating their compute use, including: (a) claiming to have executed the same workload or approximately the same workload more times than they really did, (b) submitting code with effectively unused computations (but not actually running those), and (c) declaring an inference workload whose outputs may be predicted with less compute than actually running the model (especially if the model is engineered to facilitate this).\footnote{Another attack could be over-stating their ratio of hardware operations to model operations, but we address this in our discussion of compute accounting via analog measurement of AI chips (\hyperref[a.6-compute-accounting-via-analog-sensors]{\ul{Appendix A.6}}).} Methods to counter these attacks may include, respectively, (a) checking for (approximately) duplicate submissions and requiring the frequency of these to be low (i.e., requiring reasonable use of caching), (b) using code validation (\hyperref[a.5-data-and-code-validation]{\ul{Appendix A.5}}) (such as having a compiler and a code model check for these unused computations) and scaling law checks, and (c) requiring intermediate activations to be replicated.\footnote{This defense (c) may be especially effective if paired with a code factorization constraint (\hyperref[a.5-data-and-code-validation]{\ul{Appendix A.5}}) that hinders efforts to manipulate intermediate activations.} Of these, (b) appears most challenging.
\item
  \textbf{Full-stack secure implementation, including hardware:} The Prover could make their program available to the Verifier as a function that reads from and writes to specified files, allowing the Verifier to check whether the program's input-output behavior is as claimed. The program would need to be run with strong full-stack security, from the Verifier's software to the hardware (\hyperref[a.1-full-stack-security-for-technical-verification-mechanisms-implementation]{\ul{Appendix A.1}}).
\item
  \textbf{Confidentiality-preserving technology:} A secure, confidentiality-preserving technology such as a trusted cluster or Confidential Computing would be needed for confidential data analyses (\hyperref[the-framework]{\ul{Section 3.2}}).
\item
  \textbf{Replicability:} A Verifier may struggle to replicate a Prover's program even if the Prover is honest, since some programs---including much large-scale AI training and inference---depend on both (i) the specific choice of hardware products, and (ii) the hardware's random behavior. To address this, each source of variation could either be controlled (eliminating the variation, potentially at significant cost) or statistically modeled (enabling random variation to be statistically distinguished from signs of non-compliance, as long as the former does not swamp the latter). Fortunately, hardware randomness appears controllable at a low cost for inference (\hyperref[a.9-deterministic-replication-of-neural-network-inference]{\ul{Appendix A.9}}), and at least the second source of variation (ii) appears small relative to the signs of non-compliance in some AI training experiments \citep{choi2023toolsverifyingneuralmodels}.
\item
  \textbf{Data storage:} The storage of data could pose confidentiality concerns, especially storing user data for verifying inference, but there are several options for mitigating these concerns, on top of using confidentiality-preserving technology. Latency and cost overheads from data storage are small. (\hyperref[a.9-deterministic-replication-of-neural-network-inference]{\ul{Appendix A.9}}.)
\item
  \textbf{Imprecision of empirical trends:} Empirical trends of machine learning systems (and other workloads), such as AI scaling laws, are imprecise and may only apply in a limited domain, potentially hindering efforts to check for consistency with empirical trends as a constraint. Broader experimentation may clarify the scope and exactness with which these trends may be expected, though the scope of potential workload setups is large.
\end{itemize}

\phantomsection\subsubsection{A.5 Data and Code Validation}\label{a.5-data-and-code-validation}

\textbf{Background:} For several of the other verification mechanisms we consider in this report, a Prover could attempt to circumvent them through maliciously designed code or data:

\begin{itemize}
\item
  Malicious code or data may spoof efforts to re-execute portions of programs (\hyperref[a.4-partial-workload-re-execution-with-constraints]{\ul{Appendix A.4}}), or break the assurance given by workload certificates (\hyperref[a.2-hardware-backed-workload-certificates-and-evaluations]{\ul{Appendix A.2}}), by e.g., loading pre-computed results encoded in data rather than computing results normally.
\item
  Code with many unnecessary (i.e., padded) instructions could undermine compute accounting via analog measurement of AI chips (\hyperref[a.6-compute-accounting-via-analog-sensors]{\ul{Appendix A.6}}).
\item
  Malicious data may create learned backdoors that undermine safety evaluations.
\item
  Malicious data may obfuscate the tasks a model is being trained on or used for, such as by obfuscating biological sequence data.
\end{itemize}

To address this, in addition to the measures discussed in these other mechanisms' sections, a Verifier could use \emph{data and code validation}: checking if data and code have expected properties and lack specific malicious features of concern (such as backdoors in code). Validation could be conducted over confidentiality-preserving technologies.

\textbf{Applicability to different subgoals in our verification framework:}

\begin{itemize}
\item
  \textbf{Subgoals 1 and 2.A:} Data and code validation cannot complete any of our verification framework's subgoals on its own, but per the above it can strengthen other mechanisms for subgoals 1 and 2.A.
\end{itemize}

\textbf{Challenges and mitigations:}

\begin{itemize}
\item
  \textbf{Reliable methods for data and code validation:} Data and code validation could be done with various methods, each of which has its own challenges and limitations:\footnote{Programs could potentially be securely saved for later analysis when data and code validation methods have further improved, potentially deterring violations that would occur over a long time period (even if the Prover knows initial validation techniques are unreliable).}

  \begin{itemize}
  \item
    \emph{Options for auditing data include:}

    \begin{itemize}
    \item
      \emph{Frequency analysis of data:} Drawing from cryptanalytic frequency analysis, one could check that statistical data properties are as expected (e.g., the distribution of tokens), with expectations being based on the usual\footnote{The expected properties of data could be determined based on the fact that, currently, much of the data used to pre-train foundation models tends to be of a few types---natural language text, code, and (for multimodal models) photographs and potentially audio recordings \citep{brown2020languagemodelsfewshotlearners, epoch2022trendsintrainingdatasetsizes, touvron2023llamaopenefficientfoundation, geminiteam2024geminifamilyhighlycapable}---though synthetic data may become increasingly prominent \citep{Trinh2024AlphaGeometry}. As another caveat, when considering what data ``is expected to look like,'' it would be important to account for how linguistic and regional variation, as well as innovation (including in synthetic data) and unique use-cases, may lead to unexpected datasets.} or claimed data types.
    \item
      \emph{Data tracing:} A Prover could demonstrate the legitimate origin of (a random sample of a large portion of) their data, such as by stating the data's position in a large public dataset, having third parties that they provided the data, or (for synthetic data) producing the program that generated the data. However, it may be challenging to ensure these origins of data are not compromised.\footnote{A message could be encoded in a public dataset (either by the creator of the dataset, or by a creator of content scraped for the dataset). This may be simplest to counter by requiring that public datasets used be sufficiently old (such that a violation leveraging them would be impractically slow), but that would pose performance tradeoffs. Third parties providing data could be compromised; a whistleblower system may address this if the number of parties involved is large enough. Synthetic data could also be synthesized maliciously; one may need a further level of code and/or data auditing to detect this.}
    \item
      \emph{Checking a pre-trained model's accuracy on data:} A pre-trained model's performance on some data could serve as a measure of how similar the data is to that used to train the reference model.
    \item
      \emph{Inspection of data by a trained model:} A capable AI model (e.g., a code model) could examine data to assess whether it looks as expected and to identify malicious features of concern.\footnote{This raises the question of how to protect the weights of the ``inspector'' AI model from being stolen. This problem is an instance of wishing to protect private details of the Verifier's tests, which would be addressed by successful implementations of confidentiality-preserving technologies such as trusted clusters or Confidential Computing.}
    \end{itemize}
  \item
    \emph{Options for auditing code include:}

    \begin{itemize}
    \item
      \emph{Code factorization constraints:} The Verifier could check that the Prover's code is factored in the expected manner for the workload, such as having the structure of gradient descent (or a variant thereof).\footnote{As a slightly simplified example, for AI training code, one could check that code is factored into functions including: (i) a forward pass and loss function, which together access a data batch and the model weights to produce a loss value (potentially with a regularization term); (ii) a backward pass function, which accesses a loss value, a data batch, and the model weights to compute the gradient; and (iii) an update rule, which updates weights by adding a (simple) function of the current and potentially previous gradients (note this accommodates SGD, momentum, Adam, Shampoo, and close variants, which are also used in RL algorithms such as DQL and PPO) \citep{HuggingFaceDeepQ, schulman2017proximalpolicyoptimizationalgorithms}. Mathematically, the constraints are:
        
        $W_{i+1}=W_i+u(g_1, \dots, g_i)$.

        $g_i=\nabla_WL(f_{W_i}(X_i))$.

        ($u$ could be further constrained to simplicity, e.g., being fully encodable in < 10,000 characters in Python, including calls to standard libraries and saving variables for use in future updates.)

        When replicating portions of the Prover's training run, the Verifier could enforce these structural constraints by running the pipeline that connects the Verifier-provided functions and data files, and ensuring each function is otherwise sandboxed. Additionally, the Verifier could check that the gradient computation in (ii) is mathematically correct, by substituting a different backward pass function known to be valid. These checks would mean that, if a Prover wanted to, for example, overwrite weights with pre-computed weights encoded in training data, they could not directly do so; instead, they would need to implement this within the functional form of gradient descent or its variants. Additionally, all functions other than the loss and forward pass functions would be heavily constrained, allowing the Verifier to focus other checks (e.g., analysis by a code model) on the forward pass function.} \footnote{Provers may need to submit two versions of their code: one factored to facilitate code validation, and another with the (claimed) original speed for operations accounting. (The Verifier could check that the two codes' outputs are sufficiently similar.)}
    \item
      \emph{Formal verification of code:} Systematic analysis could prove that code behaves in certain ways, though this may come with severe formatting constraints and thus engineering slow-downs and limitations. Formal verification of software is an active \citep{hasan2015formal, Souyris2009Formal} but challenging research field.
    \item
      \emph{Inspection of code by a trained model:} Analogous to the above ``Inspection of data by a trained model.''
    \end{itemize}
  \end{itemize}
\item
  \textbf{Data storage:} The storage of data, especially usage prompts, could pose confidentiality concerns, but there are several options for mitigating these, on top of using confidentiality-preserving technology (\hyperref[the-framework]{\ul{Section 3.2}}) for disclosing and analyzing the stored data. In addition, latency and cost overheads from data storage are small. (\hyperref[a.10-storing-sensitive-data-for-verification]{\ul{Appendix A.10}}.)
\item
  \textbf{Full-stack secure implementation, including hardware:} The Verifier would need uncompromised software and hardware to securely run their checks (\hyperref[a.1-full-stack-security-for-technical-verification-mechanisms-implementation]{\ul{Appendix A.1}}).
\item
  \textbf{Confidentiality-preserving technology:} A secure, confidentiality-preserving technology such as a trusted cluster or Confidential Computing would be needed for confidential data analyses (\hyperref[the-framework]{\ul{Section 3.2}}).
\end{itemize}

\phantomsection\subsubsection{A.6 Compute Accounting via Analog Sensors}\label{a.6-compute-accounting-via-analog-sensors}

\textbf{Background:} In \emph{compute accounting}, one verifies the amount of AI compute used by a Prover, and verifies that a high fraction of this compute use can be accounted for by declared uses.\footnote{``Compute accounting'' has previously been defined as ``measurements and techniques to produce an estimation of the amount of compute consumed by a customer running one or more workloads on a specific compute cluster,'' \citep{Heim2024Cloud}, consistently with the term's earlier use \citep{brundage2020trustworthyaidevelopmentmechanisms, Jackson2022ComputeAccounting}. Our use of the term is a little different; here, the goal is not only to produce one estimate, but to compare estimates of ``sources of compute'' with ``uses of compute'' to ensure the former is legitimately accounted for.} Ideally, the declared AI compute use would add up to 100\% of the AI compute use. If a sufficiently high fraction of compute use can be accounted for, this implies the Prover cannot have done large-scale, undeclared use of AI compute, among the computing clusters being accounted for. Off-chip analog sensors could enable three partly compatible\footnote{They are all compatible except for options (A) and (B) specifically in the context of AI training. In AI training (but not inference), (A)'s conservative assumption that MFU is optimized is inconsistent (in the context of AI training) with (B)'s conservative assumption that \emph{MFU = HFU} \citep{chowdhery2022palmscalinglanguagemodeling}.} approaches to compute accounting (\hyperref[tab:three_implementation_options_for_compute_accounting]{\ul{Table 13}}), if combined with other mechanisms (such as partial workload re-execution, \hyperref[a.4-partial-workload-re-execution-with-constraints]{\ul{Appendix A.4}}) for verifying declared uses and ensuring the integrity of analog sensors (\hyperref[prerequisites-off-chip-devices]{\ul{Section 4.2.1.1}}).

A core problem here is that there is no simple way to deduce an AI chip's rate of computation, even with analog measurements. An AI chip's utilization can vary from below 1\% to around 90\% depending on workload types, hardware, and implementations \citep{pope2022efficientlyscalingtransformerinference, erdil2024dtsim}. Utilization also has a complex relationship to analog measurements, in part because of the distinction between ``model FLOP utilization (MFU)'' (which only counts unique operations) and ``hardware FLOP utilization (HFU)'' (which also counts recomputed operations).\footnote{An important concept for compute accounting via external sensors is the distinction between model FLOP and hardware FLOP, and relatedly between Model FLOP Utilization (MFU) and Hardware FLOP Utilization (HFU) \citep{chowdhery2022palmscalinglanguagemodeling}. In a given AI training workload, model FLOP are \emph{unique} operations, while hardware FLOP may be re-computed to avoid accessing a cached result (which can be inefficient due to memory bandwidth bottlenecks). This is typically done in backward passes in AI training.}

\begin{longtable}[]{
  >{\raggedright\arraybackslash}p{(\linewidth - 4\tabcolsep) * \real{0.2436}}
  >{\raggedright\arraybackslash}p{(\linewidth - 4\tabcolsep) * \real{0.2596}}
  >{\raggedright\arraybackslash}p{(\linewidth - 4\tabcolsep) * \real{0.4968}}}
\toprule\noalign{}
\begin{minipage}[t]{\linewidth}\raggedright
Implementation option
\end{minipage} & \begin{minipage}[t]{\linewidth}\raggedright
How the Verifier estimates \emph{total} AI compute use
\vspace{0.5em}
\end{minipage} & \begin{minipage}[t]{\linewidth}\raggedright
How the Verifier estimates AI compute use \emph{accounted for} by declared uses
\end{minipage} \\
\endhead
\toprule\noalign{}
\rowcolor{black!5!white}
\begin{minipage}[t]{\linewidth}\raggedright
\emph{Option A:} Assume declared workloads were executed efficiently (i.e., optimizing MFU).
\end{minipage} & \begin{minipage}[t]{\linewidth}\raggedright
Count the active AI \emph{chip-hours} (e.g., GPU-hours, TPU-hours).
\end{minipage} & \begin{minipage}[t]{\linewidth}\raggedright
$\text{Chip-hours accounted for }=$

\begin{center}
    $\frac{\text{Declared model OP}}{\text{Theoretical\ OP/hr}\  \cdot \ \text{MFU}}$, \\
\end{center}
\vspace{0.5em}
where \emph{Declared Model OP} are verified via Subgoal 1.A, and \emph{Theoretical OP/hr} and \emph{MFU} are estimated conservatively (i.e., high).\footnote{One might wish to estimate a chip's theoretical performance rather than taking the vendor's word for it because a hardware vendor could hypothetically be colluding with the Prover, especially if both are of the same country.}
\vspace{0.5em}
\end{minipage} \\
\begin{minipage}[t]{\linewidth}\raggedright
\emph{Option B:} Estimate hardware FLOP from measurements, and cap the allowed hardware FLOP per model FLOP.
\end{minipage} & \begin{minipage}[t]{\linewidth}\raggedright
Conservatively estimate \emph{hardware OP} based on analog measurements.
\end{minipage} & \begin{minipage}[t]{\linewidth}\raggedright
$\text{Hardware-OP accounted for }=$

\begin{center}
    $\text{Declared\ model\ OP} \cdot \frac{\text{HFU}}{\text{MFU}}$, \\
\end{center}
\vspace{0.5em}
where \emph{Declared Model OP} are verified via Subgoal 1.A, and \(\frac{HFU}{MFU}\) is estimated conservatively (i.e., low, an estimate of 1 being most conservative).
\vspace{0.5em}
\end{minipage} \\
\rowcolor{black!5!white}
\begin{minipage}[t]{\linewidth}\raggedright
\emph{Option C:} Check that workloads' physical signatures are accounted for (i.e., do fine-grained workload classification).
\end{minipage} & \begin{minipage}[t]{\linewidth}\raggedright
Count the active AI \emph{chip-hours} (e.g., GPU-hours, TPU-hours).
\end{minipage} & \begin{minipage}[t]{\linewidth}\raggedright
$\text{Chip-hours accounted for }= \text{Chip-hours with an expected physical signature}$,
\vspace{0.5em}

where the expected physical signature corresponds to a workload verified in Subgoal 1.A, is measured with external sensors, and may consist of a power draw pattern over time, input/output bandwidth pattern over time, etc.
\vspace{0.5em}
\end{minipage} \\

\bottomrule\noalign{}
\caption{Three implementation options for compute accounting via off-chip analog sensors.} \label{tab:three_implementation_options_for_compute_accounting}
\endlastfoot
\end{longtable}

\textbf{Applicability to different subgoals in our verification framework:}

\begin{itemize}
\item
  \textbf{Subgoal 2.A:} Operations accounting would verify the absence of undeclared, large-scale uses of known AI data centers (i.e., large-scale, declared AI compute clusters).
\end{itemize}

\textbf{Challenges and mitigations:} The above options raise many implementation challenges, including estimating the mentioned variables.

\begin{itemize}
\item
  \textbf{Challenges specific to option A:}

  \begin{itemize}
  \item
    \textbf{Estimating optimal Model FLOP Utilization (MFU):}

    \begin{itemize}
    \item
      MFU can vary greatly across workload types, hardware, and implementations \citep{pope2022efficientlyscalingtransformerinference, erdil2024dtsim}. For a given workload type and cluster, optimal MFU could be coarsely approximated as the MFU known to have been achieved on similar workloads and clusters, on which there is private and public data \citep{EpochNotableModels2024}. Optimal MFU could be more precisely estimated with theoretical modeling \citep{erdil2024datamovementlimitsfrontier, erdil2025inferenceeconomicslanguagemodels}. Still, significant error bars may remain. These could be mitigated by complementing MFU estimates with monitoring of physical signatures (Option C); if a Prover finds an algorithmic change that can improve their MFU beyond the conservatively estimated value, this algorithmic change could be reflected in an unexpected physical signature.
    \end{itemize}
  \item
    \textbf{Estimating AI chips' theoretical peak performance:}

    \begin{itemize}
    \item
      Peak performance could be estimated by comparing chips to known reference chips, or more precisely by delayering and scanning chip features (\hyperref[a.1-full-stack-security-for-technical-verification-mechanisms-implementation]{\ul{Appendix A.1}}).
    \end{itemize}
  \end{itemize}
\item
  \textbf{Challenges specific to option B:}

  \begin{itemize}
  \item
    \textbf{Estimating hardware operations from measurements:}

    \begin{itemize}
    \item
      To verify the \emph{number of total operations} done in given AI data centers in some period (from \(t_{a}\) to \(t_{b}\)), one approach could be to measure AI chips' power draw (P) over time (Watts per second), and then multiply it by the chips' energy efficiency (OP per Watt) (determined based on AI hardware measurements (M) such as active GPU-hours, power draw and temperature at the time, available cooling solutions, reference measurements, perhaps magnetometer measurements \citep{matyunin2019magneticspyexploitingmagnetometermobile}, and the extent of hardware wearing out):\\
      
      ${\text{Num. total\ operations}}_{(t_{a},\ t_{b})}\  = \ \int_{t_{a}}^{t_{b}}P(t) \cdot \text{OP\_per\_W}(M(t))\ dt$.
    \item
      \emph{Inferring energy efficiency (OP/W) from hardware measurements:} AI chips' energy efficiency has a complex relationship with the chips' power consumption and other properties; that is, determining $\text{OP\_per\_W}(M(t))$ and an appropriate set of measurements is not straightforward.\footnote{A chip's power consumption consists of the power used to (i) execute operations and (ii) communicate with memory or I/O, as well as (iii) the static power use or power leakage. (i) is affected by factors including the type of operation (e.g., bit-width) and (to a lesser extent) the distribution of numbers \citep{he2024matrix}, (ii) is affected by factors including the workload's arithmetic intensity, and (iii) is affected by factors including the availability and use of clock gating and power gating to reduce the power consumption of unused modules of a chip \citep{8728370}. Additionally, as a chip's clock speed increases (at least, if the memory clock increases equally), OP/s rise linearly, while power usage rises non-linearly \citep{FatahalianOlukotun2023Parallelism}. The maximum clock speed (before excess wear) is affected by power throttling and thermal throttling, which depend on the cooling technology used and even environmental temperature; when training Llama 3, Meta ``noted a diurnal 1-2\% throughput variation based on time-of-day'' \citep{llamaTeam2024Llama3}. Power consumption also varies non-linearly with process technology \citep{micron2018metamorphosis}. Equipment beyond AI chips adds further complexity, though as mentioned this complexity could be avoided by only measuring AI chips' power draw.} Still, it may be feasible to model this relationship with small error bars, for any given workload.\footnote{A given workload might be definable with a specification such as ``transformer inference with INT8 operations,'' or perhaps a greater level of granularity would be needed. One approach to modeling an AI chip's energy efficiency (for a given chip) could be to assume the user seeks to maximize model OP/s for the workload subject to the constraints of a maximum power draw and a maximum temperature, which are determined by available cooling and electrical infrastructure and the hardware's resilience to high temperatures and power (and perhaps ambient temperature). One could further assume that, without (significantly) lowering model OP/s from this constrained maximum, a user would then take nearly all opportunities to reduce power leakage. Empirical tests could determine the parameters of these relationships (ideally with precautions against backdoors: \hyperref[a.1-full-stack-security-for-technical-verification-mechanisms-implementation]{\ul{Appendix A.1}}), though industry standards could be a rough substitute).

        Some actors might require technical assistance to meet the assumption of approximately maximizing model OP/s under constraints; this could be provided, potentially for compensation (as better energy efficiency means lower power bills). Still, variation in energy efficiency introduced through engineering introduces uncertainty.

        With uncertainty, the mentioned assumptions, experimental results, and live measurements (perhaps with some refinements) would collectively determine all the variables previously mentioned to relate the power draw and OP/s (e.g., assuming a workload determines the type of operation and the arithmetic intensity, assumptions and measurements determine the clock speed(s) and power leakage).

        However, a Prover may have legitimate reasons not to maximize OP/s under power and temperature constraints; certain uses may demand sacrificing throughput to lower latency. Hopefully such use-cases would either not significantly affect energy efficiency, be identifiable from certain measurements, or be otherwise verifiable (e.g., consumer-end verification of low latency).} Notably, one might only be able to infer a ``menu of options'' that a chip may have been used for (e.g., N operations if used in some training configuration, N' operations if used in some inference configuration); the Prover could then account for the chip's use by showing that they have the results of one of those options.\footnote{As one wrinkle (to illustrate using simplified numbers), suppose one server could do 100 FLOP for workload A or 200 FLOP for workload B, while a second server could do 200 FLOP for workload A or 100 FLOP for workload B. A Prover could then claim to have chosen the 100-FLOP option with each server while really choosing each 200-FLOP option. The Prover would then be able to account for the two 100-FLOP workloads, while having done just as many operations secretly. More abstractly, some sets of selections from a ``menu of options'' might be pareto-inefficient, allowing a Prover to understate their number of operations done. This could be countered by demanding (approximately) pareto-efficient selections, which is incentive-compatible.}
    \end{itemize}
  \end{itemize}
\item
  \textbf{Challenges specific to option C:}

  \begin{itemize}
  \item
    \textbf{Determining workloads' expected physical signatures:}

    \begin{itemize}
    \item
      To determine what physical signature a workload has on given hardware, the Verifier could run the workloads themselves on sufficiently similar hardware (as a physically secured, trusted cluster), though this approach may be vulnerable to backdoored hardware. Alternatively, the Prover may be able to derive signatures by modeling the workload \citep{erdil2024datamovementlimitsfrontier, erdil2025inferenceeconomicslanguagemodels}.
    \end{itemize}
  \item
    \textbf{Spoofed physical signatures:}

    \begin{itemize}
    \item
      Perhaps an AI workload could be condensed to use fewer chip-hours while preserving its physical signature. This would undermine the use of physical signatures for compute accounting. It could thus be valuable to assess the viability of such attacks, and, if needed, to identify constraints or additional measurements that would defeat these attacks.
    \end{itemize}
  \end{itemize}
\item
  \textbf{Challenges for multiple options:}

  \begin{itemize}
  \item
    \textbf{Counting AI chip-hours:} For declared AI compute clusters (undeclared ones being addressed separately through Subgoal 2.B), AI chip-hours could be measured, for example, with electricity meters on the Power Distribution Unit (PDU) connected to each server, or potentially more or less fine-grained units, with additional means such as tamper-evident packaging and video cameras (\hyperref[supplementary-verification-mechanisms]{\ul{Section 4.4}}) to detect tampering or circumvention. Chip-hours are a fairly coarse metric with various physical correlates, and they are familiar as the industry's standard billing metric, so many methods may work.
  \item
    \textbf{Counting declared model operations:} To verify the \emph{number of declared operations} done in given AI data centers in some period, one could (via confidentiality-preserving technology: \hyperref[the-framework]{\ul{Section 3.2}}) (i) apply analytical methods to declared models, data, and code \citep{narayanan2021efficientlargescalelanguagemodel, chowdhery2022palmscalinglanguagemodeling}, and/or re-execute portions of programs with (ii) software profilers or (iii) hardware performance counters \citep{epoch2022estimatingtrainingcompute, Heim2024Cloud}.
  \item
    \textbf{Avoiding false alarms from under-estimating compliant compute use:} In the above options (A) and (B), the Verifier makes a conservatively high estimate of the Prover's MFU, so that a Prover cannot obtain unaccounted-for compute by achieving higher-than-assumed MFU. These conservatively high estimates are, respectively, assuming a highly efficient MFU, and assuming that the MFU is close to the measured HFU. However, these assumptions might not hold for honest Provers, potentially leading honest Provers to be unable to account for their estimated compute use. To mitigate this, for (A), the Verifier could estimate an optimal MFU achievable in practice rather than using a large over-estimate,\footnote{This is closely related to the efficiency problem of achieving high MFU, so there is presumably substantial applicable work and expertise on this problem at AI labs.} and they could technically assist the Prover in reaching the efficient assumed MFU (potentially with compensation for the cost savings).\footnote{This has the political downsides (and upsides) of being technology transfer, but that could be mitigated by financial compensation and by the fact that the efficiency gains are quite limited in principle; at least for AI training, MFU around 50\% of the theoretical upper bound of 100\% is already commonly achieved \citep{Clark2024ComputeThresholds}. Meanwhile, for AI inference, HFU=MFU by default, making option (B) more easily applicable.} For (B), the verification protocol could require the Prover to refrain from surpassing the assumed HFU/MFU ratio (at the limit, require HFU=MFU, though this would carry substantial efficiency costs for training).
  \item
    \textbf{Fine-grained measurement of AI hardware:}

    \begin{itemize}
    \item
      For measuring AI chips' power draw (and potentially other properties), several questions arise:

      \begin{itemize}
      \item
        \emph{What objects to measure:} Measurements could be made at various levels of granularity, from an entire AI data center to individual AI chips. More fine-grained measurements will tend to offer lower error bars,\footnote{If one only measures the power draw of AI chips, then one ``only'' needs to determine the AI chips' energy efficiency to verify the operations done. As more other equipment is brought into the bounds of power measurement (e.g., CPUs, NVLink chips, cooling infrastructure), the energy consumption of this other equipment needs to be determined and subtracted from the total energy consumption to derive the AI chips' energy consumption, introducing error.} but require more sophisticated and numerous equipment.\footnote{For example, measuring an AI data center's whole power draw may be doable with a few standard electricity meters, while precisely measuring individual AI chips' power consumption may require altering a server's design to introduce interposers that sit between AI chips and power supply units---one for each of thousands of AI chips.}
      \item
        \emph{What metrics to measure:} These could be determined based on the analysis, experimentation, and example metrics discussed above and below, considering tradeoffs between precision and cost.
      \item
        \emph{When to measure:} Measurement could be (i) periodic and sufficiently frequent that it would be impractical to lower performance just before measurement, or it could be (ii) random and sufficiently frequent to allow for small error bars.
      \item
        \emph{How to do anti-tamper measurement:} Through tamper-evident or tamper-proof packaging.\footnote{Measurement devices could be made to be (or installed in containers that are) tamper-evident (and then inspected for evidence of tampering) or tamper-proof, the latter option being much more technically challenging \citep{aarne_secure_chips_2024, Kulp2024}. Tamper-evident cameras (\hyperref[supplementary-verification-mechanisms]{\ul{Section 4.4}}) would also help.} \footnote{Inspections may also be needed to ensure that e.g., a data center or server rack does not have hidden power cables, nor unmetered backup generators. Measured power draw could be sanity checked based on observations or other measurements of cooling infrastructure, electrical infrastructure, and heat emissions.}
      \item
        \emph{How to measure with security for the Prover:} Through Prover-inspected equipment being examined only on site, under supervision.\footnote{While remote transmission could raise security issues, the Verifier could take the same approach as the IAEA: having inspectors check device logs on site, under the supervision of the facility owner \citep{baker2023nucleararmscontrolverification}. Also drawing from the IAEA, the data center owner could be allowed to inspect a random sample of measurement devices, to ensure they do not have unauthorized functions.}
      \end{itemize}
    \end{itemize}
  \item
    \textbf{Full-stack secure implementation, including hardware:} The Verifier would need uncompromised software and hardware to securely run their software checks (\hyperref[a.1-full-stack-security-for-technical-verification-mechanisms-implementation]{\ul{Appendix A.1}}).
  \item
    \textbf{Confidentiality-preserving technology:} A secure, confidentiality-preserving technology such as a trusted cluster or Confidential Computing would be needed for confidential data analyses (\hyperref[the-framework]{\ul{Section 3.2}}). (Confidential Computing, though, would be out of scope for off-chip verification.)
  \item
    \textbf{Duplicate model operations:} These are addressed by Subgoal 1.A (\hyperref[a.4-partial-workload-re-execution-with-constraints]{\ul{Appendix A.4}}).
  \end{itemize}
\end{itemize}

\phantomsection\subsubsection{A.7 Verifying Non-AI Workloads in AI Data Centers}\label{a.7-verifying-non-ai-workloads-in-ai-data-centers}

Data center GPUs can be used for a variety of workloads beyond AI, raising the question of how a Verifier could distinguish these non-AI workloads from AI workloads. At a high level, options include:

\begin{enumerate}
\def\labelenumi{\arabic{enumi}.}
\item
  The Prover could:

  \begin{enumerate}
  \def\labelenumii{\alph{enumii}.}
  \item
    Execute the workload on non-AI-specialized chips instead (e.g., CPUs or ASICs)
  \item
    Complete the intended task with neural networks (thus allowing the workload to be verified as an AI workload)
  \end{enumerate}
\item
  Alternatively, the Verifier, in collaboration with the Prover, could:

  \begin{enumerate}
  \def\labelenumii{\alph{enumii}.}
  \item
    Use a verification protocol developed for the non-AI workload, perhaps one analogous to proof-of-learning \citep{jia2021proofoflearningdefinitionspractice} (\hyperref[a.4-partial-workload-re-execution-with-constraints]{\ul{Appendix A.4}}) or, where feasible, analytic methods (\hyperref[supplementary-verification-mechanisms]{\ul{Section 4.4}}).
  \item
    Classify the workload as non-AI (e.g., based on manual or automatic analysis of the code or of data movement patterns, or perhaps the cluster architecture)
  \item
    Use non-confidentiality-preserving methods (for non-sensitive workloads)
  \end{enumerate}
\end{enumerate}

\phantomsection\subsubsection{A.8 Whistleblower Programs}\label{a.8-whistleblower-programs}

\textbf{Background:} Programs and laws that encourage employees to blow the whistle on violations are commonplace \citep{nwc_whistlelaws}, contributing to approximately \$2 billion or more in SEC fines in 2023.\footnote{The SEC reports awarding \$600 million in fines to whistleblowers as part of its whistleblower reward program in 2023, rewards being 10-30\% the value of fines the whistleblower tip contributed to \citep{SEC2023Report}.} In the AI industry, large-scale AI projects tend to involve hundreds of employees (\hyperref[tab:the_number_of_contributors]{\ul{Table 9}})---hundreds of individuals who might be able to report any large-scale violations to a Verifier. In addition to AI developers' own employees, other organizations throughout the AI supply chain have employees who can blow the whistle on some violations, especially undeclared AI data centers. Employees could blow the whistle on a Prover's (i) non-compliant AI activities, (ii) falsified declarations, or (iii) attempts to circumvent another verification mechanism (\hyperref[tab:types_of_employees_who_would_have_information]{\ul{Table 14}}). Formal whistleblower programs could promote appropriate forms of whistleblowing by providing (would-be) whistleblowers with information they can check, disclosure protocols, and incentives (including intrinsic motivation, social norms, protection, and financial rewards). Provers may view formal whistleblower programs as legitimate, so Provers may be willing to take verifiable actions that facilitate whistleblowing (in contrast to espionage), such as allowing employees to privately talk with a Verifier.

\textbf{Challenges and mitigations:}

\begin{itemize}
\item
  \textbf{Secure and confidential communication with potential whistleblowers.} A Prover might try to not only retaliate against whistleblowers, but also entirely block or alter their messages. Standard approaches to secure internet communication (e.g., TLS, VPNs, and Tor) are not designed to secure the communications of parties who may be under video surveillance, or whose computers may be backdoored. Instead, a more secure option is for such employees to make in-person visits to a building physically secured by a Verifier. To prevent the Prover from detecting or blocking whistleblowers' visits to these locations, the verification protocol could require the Prover to periodically send various relevant employees to visit the Verifier-secured building (e.g., as brief visits to an office near the Prover's offices).\footnote{Further anonymity protections could include: ensuring all employees have equal-length visits to the Verifier rather than keeping whistleblowers longer, and strong internal processes for protecting confidentiality.} \footnote{Sending \emph{all} employees on such visits may be especially impractical for larger companies in the AI supply chain; TSMC and AWS, for instance, have around 100,000 employees each \citep{TSMStockAnalysis, gardizy2024googlecloud}. The number of visits could be greatly reduced by limiting visits to primarily the most relevant employees, though this could mean reduced coverage. Still, for cases where violations are known to non-closely-watched employees, more established communication lines like Tor or other anonymous internet communication may also be viable.} \footnote{This raises the problem of potential falsehoods about who the relevant employees are. The Verifier could verify the identities of relevant employees through methods including open-source intelligence (e.g., publication record, social media), checking whether an individual has the knowledge expected of their role, and checking whether the claimed number of employees is approximately standard for the role. Regardless, if a Prover sends forward a group that is incomplete or has impostors, the group members could then blow the whistle on the deception.}
\item
  \textbf{Ensuring whistleblowers have enough information to report signs of violations.} Even if an employee is not aware of a violation, they may have knowledge inconsistent with a Prover's declarations (\hyperref[tab:types_of_employees_who_would_have_information]{\ul{Table 14}}). To learn if this is the case, the Verifier could, within employee interviews, ask the employee to check the Prover's relevant declarations, or to share information that should match the relevant declarations, preferably via a confidentiality-preserving technology. A ``low-tech'' option for confidentiality preservation could involve a carefully overseen personal computer.\footnote{This could work as follows. First, the Prover uploads some information, such as training code and information on compute allocation across projects, to a personal computer. (Other information, like model weights, would typically require a computing cluster rather than a personal computer.) Then, a Prover representative physically oversees the use of this computer to ensure this information is not used for any purposes beyond authorized ones. These authorized uses would be: (i) authorized Prover employees may view claims related to their work, and (ii) the Verifier may extract a cryptographic hash to check that the Prover's claims here match their claims elsewhere.

    To implement (i), a Prover employee could review (selected, relevant-to-the-employee) declarations, under the supervision of a representative of both the Prover and the Verifier---with both of these representatives having the computer's monitor and keyboard out of view. With an appropriate room layout and joint physical security, this arrangement could provide several assurances: (i) the Verifier is assured that the employee reviews the code, (ii) the Prover is assured that the Verifier does not view private declarations, and (iii) the Verifier is assured that the Prover does not view the employee's screen (to make employee intimidation harder).} However, perhaps a Prover can have sufficient compartmentalization and employee loyalty for all accomplices to lie.
\item
  \textbf{Further incentivizing whistleblowers:} Beyond the anonymity protections discussed above, a whistleblower system could likely be strengthened by (potentially mandated) measures that make employees more morally, socially, or financially motivated to blow the whistle on violations. These could include trainings, certifications, knowledge tests, hiring practices, safety culture,\footnote{As one example promoter of a pro-whistleblower culture, industry and political leaders could make statements stressing the social value of compliance and the legitimacy of whistleblowers.} financial rewards \citep{SECWhistleblower, irs2025whistleblower, cftc2025whistleblower},\footnote{Financial rewards for successful whistleblowers could incentivize employees to make fake allegations against their employers. Still, this problem could be mitigated through selective application of rewards, such as rewarding whistleblowers only when their allegations can be independently confirmed.} and asylum or refugee status. Still, as above, it is unclear if these incentives would overcome a major Prover initiative for compartmentalization and employee loyalty.
\item
  \textbf{Avoiding excess disclosure:} Employees could be allowed to disclose only a very small amount of information to the Verifier, as discussed in above footnotes. Further, the Prover and Verifier could jointly state agreed-on, reasonable bounds of protected whistleblowing (including high-level descriptions of potential violations and information to investigate further, but excluding digital transfers of Prover models, data, or code outside of a confidentiality-preserving technology). Parties could also agree on what questions or information a Verifier may share with an employee, so that the Prover could learn from their employees if the Verifier is inappropriately pressuring them to disclose IP.
\end{itemize}

\clearpage

\begin{longtable}[]{
  >{\raggedright\arraybackslash}p{(\linewidth - 2\tabcolsep) * \real{0.2452}}
  >{\raggedright\arraybackslash}p{(\linewidth - 2\tabcolsep) * \real{0.7548}}}
\toprule\noalign{}
\begin{minipage}[t]{\linewidth}\raggedright
Type of violation
\end{minipage} & \begin{minipage}[t]{\linewidth}\raggedright
Some employees who would have information about the violation
\end{minipage} \\
\endhead
\toprule\noalign{}
\rowcolor{black!5!white}
\begin{minipage}[t]{\linewidth}\raggedright
Non-compliant declarations of large-scale AI development or deployment (intended to be detected by Subgoals 1.A or 1.B)
\end{minipage} & \begin{minipage}[t]{\linewidth}\raggedright
• \textbf{AI researchers and engineers} who contribute to the declared activity and thus could notice if model origins, output origins, or evaluations/properties are falsely declared, or if workloads are designed to spoof verification

• \textbf{Officials} who deliberate on, order, or coordinate the violation (e.g., senior executives, top advisory bodies, and lower-level managers, in government and colluding companies)

• \textbf{Spoofers:} researchers and engineers who circumvent on- and off-chip verification mechanisms, if these are present\footnote{Violations that involve software spoofs would likely require AI researchers and engineers, potentially including ones with expertise in low-level implementation. If a Prover sought to circumvent on- or off-chip verification mechanisms at the hardware level, they would likely need hardware security researchers. They may also need accomplices in data centers to do physical tampering, or accomplices earlier in the AI chip supply chain to plant hardware backdoors (\hyperref[personnel-based-verification-layers]{\ul{Section 4.3}}).

  It may be feasible for a Verifier to identify most leading hardware security researchers and give them an opportunity to whistleblow; one hardware security academic we interviewed expressed that there are maybe a handful of groups that do very good offensive (hardware) security research in the United States, and that they know each other (Interview \#2, 2024). This interviewee also estimated that there are 100-200 hardware security researchers in the United States, if one counts all the graduate students and researchers. Another academic we interviewed estimated that there are ``at least'' 200 hardware security researchers in Europe, only counting ``serious researchers'' rather than ones who tried hardware security one time (Interview \#11, 2024). This academic also explained that there are significant hardware security research groups in East Asia, Europe, and the United States.}
\vspace{0.5em}
\end{minipage} \\
\begin{minipage}[t]{\linewidth}\raggedright
Undeclared, large-scale uses of known AI compute clusters (intended to be detected by Subgoal 2.A)
\end{minipage} & \begin{minipage}[t]{\linewidth}\raggedright
• \textbf{AI researchers and engineers} who contribute to the undeclared activity, or to another compliant activity that is altered to hide the non-compliant activity

• \textbf{Officials} who deliberate on, order, or coordinate the violation

• \textbf{Spoofers:} Researchers and engineers who circumvent on- and off-chip verification mechanisms, if these are present

• \textbf{Compute cluster} oversight or management staff
\vspace{0.5em}
\end{minipage} \\
\rowcolor{black!5!white}
\begin{minipage}[t]{\linewidth}\raggedright
Undeclared, large-scale AI compute (intended to be detected by Subgoal 2.B)
\end{minipage} & \begin{minipage}[t]{\linewidth}\raggedright
• \textbf{AI researchers and engineers} who contribute to the undeclared activity, or to another compliant activity that is altered to hide the non-compliant activity

• \textbf{Officials} who deliberate on, order, or coordinate the violation

• \textbf{Spoofers:} Staff who circumvent other verification mechanisms, if present (e.g., by diverting chips and breaking compliance locks)

• \textbf{Data center} construction and operations staff (e.g., maintenance and security staff)

• \textbf{Supplier staff} who supply e.g., AI chips, energy, and other data center equipment

• \textbf{Compute cluster} design and setup staff

• \textbf{Administrative} (e.g., relevant finance, procurement, legal) staff
\vspace{0.5em}
\end{minipage} \\

\bottomrule\noalign{}
\caption{Types of employees who would have information about different types of violations. Personnel-based verification could leverage these employees for verification.} \label{tab:types_of_employees_who_would_have_information}
\endlastfoot
\end{longtable}

\phantomsection\subsubsection{A.9 Deterministic Replication of Neural Network Inference}\label{a.9-deterministic-replication-of-neural-network-inference}

We identify six potential obstacles to replicable neural network inference, which hinder verification. (That is, six potential obstacles in addition to fixing the model version and inputs, which are of course needed.) These obstacles include both non-determinism within a given technology stack, and variation across technology stacks. Fortunately, each of these obstacles can be eliminated via relatively straightforward methods, suggesting exactly replicable inference is feasible.

{
  \setlength{\LTleft}{-33pt}
  \setlength{\LTright}{-26pt}
  \centering
    \Needspace*{8\baselineskip}%
    \begin{longtable}[]{
      >{\raggedright\arraybackslash}p{(\linewidth - 2\tabcolsep) * \real{0.5000} * \real{1.15}}
      >{\raggedright\arraybackslash}p{(\linewidth - 2\tabcolsep) * \real{0.5000} * \real{1.15}}}
    \toprule\noalign{}
    \begin{minipage}[t]{\linewidth}\raggedright
    \textbf{Replicability challenge} in neural network inference
    \end{minipage} & \begin{minipage}[t]{\linewidth}\raggedright
    A \textbf{method} to eliminate the non-determinism
    \end{minipage} \\
    \endhead
    \toprule\noalign{}
    \rowcolor{black!5!white}
    \begin{minipage}[t]{\linewidth}\raggedright
    \textbf{Intentional randomness}: Inference code might explicitly call (pseudo)random number generators, e.g., for sampling with non-zero temperature.
    \end{minipage} & \begin{minipage}[t]{\linewidth}\raggedright
    \textbf{Fix all random seeds}, i.e., have the Prover record them and then share them with the verifier for replication, and avoid unreplicable (``true'') random number generators.
    \vspace{0.2em}
    \end{minipage} \\
    \begin{minipage}[t]{\linewidth}\raggedright
    \textbf{Mixture-of-experts with batch-level routing}: A mixture-of-experts (MoE) model might select which expert an input is routed to in a manner that depends on other inputs in a batch, for efficiency.\footnote{For example, if one expert's instance is already ``busy'' processing many inputs, additional inputs might be routed to other available experts for speed, even if they would be best handled by the ``busy'' expert.} Then, repeatedly running a model on the same input among different batches may lead to different outputs \citep{puigcerver2024sparsesoftmixturesexperts, 152334H2023nondeterminism}.
    \vspace{0.2em}
    \end{minipage} & \begin{minipage}[t]{\linewidth}\raggedright
    \textbf{Fix batches}, i.e., replicate inference at the level of entire batches rather than individual inputs.
    \end{minipage} \\
    \rowcolor{black!5!white}
    \begin{minipage}[t]{\linewidth}\raggedright
    \textbf{Non-replicable rounding errors:} Due to rounding errors, floating-point addition and multiplication (and fixed-point multiplication) are not associative; e.g., the order in which terms are added (the ``accumulation order'') can affect results \citep{Goldberg1991FloatingPoint}. As a result, when accumulation order is not deterministic, floating point arithmetic has inconsistent results. A non-deterministic accumulation order (whether within or across devices) can occur due to e.g., algorithmic choices \citep{twosigma2017tensorflow} and device differences \citep{nagarajan2019deterministicimplementationsreproducibilitydeep}. Aside from accumulation order, the use of different numerical precisions for activations or intermediate computations could lead to different rounding errors.
    \end{minipage} & \begin{minipage}[t]{\linewidth}\raggedright
    The simplest solution is to \textbf{do inference with integer operations of a consistent precision}, by converting (``quantizing'') floating-point model weights to integers, and also using consistent precisions for intermediate computations. Quantization is common for efficiency \citep{jacob2017quantizationtrainingneuralnetworks}, though if one trains in FP8, then quantizing offers no advantage on H100 GPUs \citep{NvidiaH100}. Integer arithmetic is associative (as it is exact), avoiding the problem of floating-point arithmetic (with the potential exception of overflows, discussed below). Another option is using ``\textbf{fully deterministic {[}computing{]} infrastructure},'' as Google DeepMind did for training Gemini 1 \citep{geminiteam2024geminifamilyhighlycapable}. This infrastructure leveraged TPUs’ long-standing determinism \citep{jouppi2017indatacenterperformanceanalysistensor}. In contrast, with GPUs, as a then-Google employee explained, ``the threading model means you get better performance with atomics {[}and thus non-deterministic operation order{]}'' \citep{Bradbury2021Tweet}.
    \vspace{0.2em}
    \end{minipage} \\
    \begin{minipage}[t]{\linewidth}\raggedright
    \textbf{Non-replicable overflow:} Some overflow behaviors, such as saturation, lead to non-associative arithmetic.\footnote{To see this, consider an example of addition of integers in the range -128 to 127, assuming the overflow behavior is saturation (``maxing out''). Then (127 + 50) + (-50) would be computed as 127 + (-50) = 77, while 127 + (50 + (-50)) would have the different result 127.} As a result, as above, non-deterministic accumulation order could have non-deterministic results. Similarly, tech stacks which differ in their overflow behavior could have different results.
    \vspace{0.2em}
    \end{minipage} & \begin{minipage}[t]{\linewidth}\raggedright
    \textbf{Use overflow behavior that is associative}, such as wrapping, i.e., modular arithmetic.
    \end{minipage} \\
    \rowcolor{black!5!white}
    \begin{minipage}[t]{\linewidth}\raggedright
    \textbf{Inconsistent rounding procedures:} The Prover and Verifier might use different rounding procedures, beyond different accumulation orders. Rounding arises even with integer operations, as some computations (e.g., layer normalization) involve division.
    \vspace{0.2em}
    \end{minipage} & \begin{minipage}[t]{\linewidth}\raggedright
    The Prover and Verifier could \textbf{use the same rounding procedures}. In a conventional transformer with integer operations, consistent rounding could be achieved via minor edits to the few library functions that use division.
    \end{minipage} \\
    \begin{minipage}[t]{\linewidth}\raggedright
    \textbf{Hardware errors:} Finally, errors at the hardware level (e.g. from manufacturing defects or cosmic rays) could unpredictably change outputs.
    \end{minipage} & \begin{minipage}[t]{\linewidth}\raggedright
    Hardware errors are very rare in practice---e.g., two studies find ``soft'' memory hardware errors on the order of up to one or thirty per \emph{billion hours} of a device's computation \citep{Slayman2011SoftErrors, Sridharan2015MemoryErrors}---so they can be addressed via \textbf{occasionally re-doing computations} that result in high discrepancies, and if needed monitoring hardware health and using error correction code (ECC) memory.
    \vspace{0.5em}
    \end{minipage} \\
    
    \bottomrule\noalign{}
    \caption{Some replicability challenges for neural network inference and methods to address them.} \label{tab:some_replicability_challenges}
    \endlastfoot
    \end{longtable}
}

\phantomsection\subsubsection{A.10 Storing Sensitive Data for Verification}\label{a.10-storing-sensitive-data-for-verification}

Some verification mechanisms, such as partial workload re-execution (\hyperref[a.4-partial-workload-re-execution-with-constraints]{\ul{Appendix A.4}}), require AI training data or usage prompts to be stored for verification. However, the data in question may be highly sensitive, e.g., it could be individuals' medical data or states' classified intelligence. The mere storage of such data by the relevant AI company could pose risks of leaks and might be in tension with current data protection regulations or consumer desires. Indeed, some prominent AI companies offer opt-in ``zero data retention'' options \citep{OpenAIDataControls, AnthropicDataRetention} There are practical reasons to store training data,\footnote{E.g., facilitating multiple training runs or epochs on the same data and facilitating debugging.} but these may not apply as strongly to usage data.

\textbf{Options for improving confidentiality in data storage.} There are various options for addressing confidentiality concerns around data storage, and these options could be combined. Here, we only consider addressing concerns regarding data \emph{storage} by a party demonstrating their compliance; concerns regarding data \emph{examination} by the Verifier are meant to be addressed by confidentiality-preserving technologies like Confidential Computing (\hyperref[on-chip-verification-layer]{\ul{Section 4.1}}) and trusted clusters (\hyperref[prerequisites-off-chip-devices]{\ul{Section 4.2.1.1}}).

\begin{enumerate}
\def\labelenumi{\arabic{enumi}.}
\item
  \emph{Data storage exemption for small data owners:} Actors who own small amounts of data used for AI inference could be allowed to verify the compliance of the data by simply asserting it to the Verifier, without further checks. With an appropriate definition of ``small,'' a large-scale violation would require collusion among an impractically high number of data owners, which could be revealed by associated personnel-based verification mechanisms (\hyperref[personnel-based-verification-layers]{\ul{Section 4.3}}). (Know-your-customer checks for a small sample of data owners would also be needed to ensure the ``small users'' really are small users.) However, the resulting conclusion would likely be imprecise, as small data users might not notice if their data is processed by a somewhat smaller model than claimed, which could undermine compute accounting.
\item
  \emph{Option for data owners to store their own data:} The data owners could keep their own logs of the data and submit it when asked. This may be especially feasible for larger organizations, so this option may be a good complement to a data storage exemption for small data owners.
\item
  \emph{Short data storage period:} The data storage period could be short (perhaps days), especially when no discrepancies are found.
\item
  \emph{Secure and local data storage:} Data could be stored securely, e.g., encrypted in a physically secure data center, and/or locally in the data owner's jurisdiction.
\item
  \emph{Storing only a verifiably small sample of data:} As a more technically complex option, the Prover may be able to do verifiably random sampling of their own data and then only store the small sample of their data. However, verifiably random sampling likely requires exact replication (\hyperref[a.9-deterministic-replication-of-neural-network-inference]{\ul{Appendix A.9}}) and solutions to further complications.\footnote{As a technical implementation in the context of large-scale AI deployment, the Prover could store only data with a hash in a certain range (the hash being e.g., the Merkle root of a (prompt, logits, model) tuple). However, this scheme could be foiled by a Prover e.g., making small perturbations to the logits or to a continuous input (e.g., an image input) until the Prover finds values that produce a desired hash; that would enable an arbitrary set of (prompt, logits, model) tuples to look like a small, random sample of tuples. To counter this, exact replication (\hyperref[a.9-deterministic-replication-of-neural-network-inference]{\ul{Appendix A.9}}) or perhaps precisely modeled non-determinism would catch sufficiently large perturbations. Still, there may be input perturbations that cause no change to the output (this is necessarily true when the input space is larger than the output space, and perhaps such perturbations can be made easier to find by an adversarially trained model); ruling out such perturbations may require further constraints on the Prover and testing. (On the other hand, small input perturbations may be infeasible for text inputs represented as discrete tokens.)} A more feasible variant may be for the Verifier to blindly do random sampling of data for the Prover to store for later analysis, by sampling over hashes of data provided by the Prover.\footnote{However, ``live'' sampling might allow a Prover to manipulate the sample by changing their inference in response to which queries are sampled (e.g., perhaps generating duplicates until the first duplicate in a set is sampled).}
\end{enumerate}

\textbf{Overhead costs of storing usage data are tiny.} Latency and costs of data storage could also matter, but back-of-the-envelope calculations suggest latency and cost overheads would be on the order of 0.01\% or lower compared to just doing inference on the data,\footnote{An H100 GPU (conservatively assuming 100\% utilization, running a 70B-parameter dense model, with 2-byte inputs plus 2-byte outputs) can process data at up to the FLOP per second * data per forward pass / FLOP per forward pass, or $1.98\text{e}15 \cdot \frac{2 + 2}{2 \cdot 70\text{e}9} \approx 57\text{ KB/s}$, while PCIe Gen5 can transfer data at 128 GB/s \citep{NvidiaH100}, and some PCIe Gen4 SSDs read and write data at around 7 GB/s \citep{Burek2024}. This back-of-the-envelope calculation suggests the latency overhead from copying data to an SSD would be on the order of 0.001\% (assuming the lower SSD rate of 7 GB/s).} \footnote{PCIe Gen4 SSDs have storage capacities such as 2 TB for \$90 \citep{Burek2024}. If they were to store data for a month and be amortized over a year, the cost would be $\frac{90}{2\text{e}12 \cdot 12} = 3.75\text{e-}12\text{ \$/byte}$, while the cost of \emph{processing} a byte with a large AI model (70B-parameter dense model, with 2-byte data, with an H100 GPU at a conservatively assumed 100\% utilization and for \$2/hr) \citep{GPUUtils2023H100Pricing, Clark2024ComputeThresholds} would be $70\text{e}9 \cdot 2 \cdot \frac{2}{1.98\text{e}15 \cdot 60 \cdot 60} \cdot \frac{1}{2} = 1.96\text{e-}8\text{ \$/byte}$, implying a storage cost overhead on the order of 0.01\%.} since processing data with a large AI model is so much more slow and expensive than simply storing the data.

\phantomsection\subsection{B. Broader Regime Design}\label{b.-broader-regime-design}

\phantomsection\subsubsection{B.1 Compute Accounting vs. Other Kinds of Accounting}\label{b.1-compute-accounting-vs.-other-kinds-of-accounting}

AI compute is relatively specialized and large in its physical footprint, making it more suitable to being accounted for than other resources used in AI development and deployment (\hyperref[tab:prominent_resources]{\ul{Table 16}}).

\begin{longtable}[]{
  >{\raggedright\arraybackslash}p{(\linewidth - 4\tabcolsep) * \real{0.3333}}
  >{\raggedright\arraybackslash}p{(\linewidth - 4\tabcolsep) * \real{0.3333}}
  >{\raggedright\arraybackslash}p{(\linewidth - 4\tabcolsep) * \real{0.3333}}}
\toprule\noalign{}
\begin{minipage}[t]{\linewidth}\raggedright
\textbf{Resource} used for AI development and deployment
\end{minipage} & \begin{minipage}[t]{\linewidth}\raggedright
\textbf{Specialization:} To what extent is the resource narrowly used for large-scale AI?
\vspace{0.5em}
\end{minipage} & \begin{minipage}[t]{\linewidth}\raggedright
\textbf{Physical footprint:} How large is the physical footprint of this resource?
\end{minipage} \\
\endhead
\toprule\noalign{}
\begin{minipage}[t]{\linewidth}\raggedright
\textbf{Compute}: high-end AI chips, typically in data centers
\end{minipage} & \begin{minipage}[t]{\linewidth}\raggedright
\cellcolor{green!10!white}
Relatively specialized
\end{minipage} & \begin{minipage}[t]{\linewidth}\raggedright
\cellcolor{green!10!white}
Relatively large\footnote{``Large scale AI training and deployment is highly resource intensive, often requiring thousands of specialized chips in a high performance cluster hosted in a large data center consuming large amounts of power'' \citep{sastry2024computingpowergovernanceartificial}.}
\end{minipage} \\
\begin{minipage}[t]{\linewidth}\raggedright
\textbf{Data}: AI training datasets
\end{minipage} & \begin{minipage}[t]{\linewidth}\raggedright
\cellcolor{yellow!10!white}
Somewhat\footnote{Similar datasets, such as large text corpuses or data from human feedback, are often used for smaller-scale AI development and deployment.}
\end{minipage} & \begin{minipage}[t]{\linewidth}\raggedright
\cellcolor{red!10!white}
Small
\end{minipage} \\
\begin{minipage}[t]{\linewidth}\raggedright
\textbf{Algorithms}: e.g., model architectures and optimization algorithms for large-scale AI
\end{minipage} & \begin{minipage}[t]{\linewidth}\raggedright
\cellcolor{green!10!white}
Relatively specialized\footnote{Similar algorithms can be used for AI development and deployment of varying scales, though leading AI labs increasingly use proprietary algorithms primarily for large-scale AI.}
\end{minipage} & \begin{minipage}[t]{\linewidth}\raggedright
\cellcolor{red!10!white}
Small
\end{minipage} \\
\begin{minipage}[t]{\linewidth}\raggedright
\textbf{Human capital}: frontier AI researchers and engineers
\end{minipage} & \begin{minipage}[t]{\linewidth}\raggedright
\cellcolor{yellow!10!white}
Somewhat\footnote{AI technical staff of course do not spend all of their time or energy on their AI work, and many people can gain AI expertise.}
\end{minipage} & \begin{minipage}[t]{\linewidth}\raggedright
\cellcolor{yellow!10!white}
Medium
\end{minipage} \\
\begin{minipage}[t]{\linewidth}\raggedright
\textbf{Electricity}: power for data centers
\end{minipage} & \begin{minipage}[t]{\linewidth}\raggedright
\cellcolor{yellow!10!white}
Somewhat, in some cases\footnote{Though electric power of course has many uses, the energy density of typical AI data centers is unusually high. Relatedly, large AI data centers often have dedicated electrical substations \citep{pilz2023computescalebroadinvestigation} or adjacent power plants \citep{DOE2025NuclearDC}. However, AI compute could be decentralized to reduce its energy density.}
\end{minipage} & \begin{minipage}[t]{\linewidth}\raggedright
\cellcolor{green!10!white}
Relatively large\footnote{Generating large amounts of electricity requires power plants, though transporting electricity does not require as much infrastructure, potentially making it challenging to trace power plants to all their uses.}
\end{minipage} \\
\begin{minipage}[t]{\linewidth}\raggedright
\textbf{Water}: water for the cooling systems of data centers
\end{minipage} & \begin{minipage}[t]{\linewidth}\raggedright
\cellcolor{red!10!white}
No
\end{minipage} & \begin{minipage}[t]{\linewidth}\raggedright
\cellcolor{yellow!10!white}
Varies\footnote{Some data centers use large amounts of water for cooling, while others do not. ``On average, evaporative cooling systems use approximately 1,800 to 2,900 liters of water per megawatt per hour,'' but an alternative cooling method---the use of ``chillers''---``obviates the need for water, {[}though{]} it demands substantial electrical power'' \citep{pilz2023computescalebroadinvestigation}.}
\end{minipage} \\

\bottomrule\noalign{}
\caption{Prominent resources used for AI development and deployment \citep{Buchanan2020AITriad, pilz2023computescalebroadinvestigation}, by two properties that make verification less intrusive: specialization and physical footprint. More specialized resources can be overseen with less need to oversee unrelated activities, which could pose major confidentiality and logistical challenges.} \label{tab:prominent_resources}
\endlastfoot
\end{longtable}

\phantomsection\subsubsection{B.2 Verification of Narrower Rules}\label{b.2-verification-of-narrower-rules}

Our verification framework (\hyperref[verification-framework]{\ul{Section 3}}) is a general framework for verifying rules on AI models, data, and code created or used in large-scale AI development \emph{and} deployment, including negative rules (e.g., rules that \emph{none} of an actor's large-scale AI developments or deployments pose unmanageable risks). However, for verifying compliance with some hypothetical, narrower (and perhaps, more blunt) rules, not all verification subgoals are needed:

\begin{itemize}
\item
  \textbf{Compute ownership caps (i.e., ``compute caps''):} To verify a compute cap---i.e., that the AI compute clusters some actor owns have no more than some maximum combined computing power---a Verifier can just ask for declarations of compute ownership and then only complete Subgoal 2.B: verifying that there are no undeclared, large-scale AI compute clusters. Since the rule being verified here would not be sensitive to usage, there is no need to verify claims about usage.
\item
  \textbf{Positive rules:} To verify rules that just require executing some workloads (rather than \emph{not} executing some workloads), e.g., a rule that requires AI companies to actually run inference consistently with what they claim (rather than running a cheaper model), a Verifier just needs Subgoal 1.A: verifying correctness.
\item
  \textbf{Negative rules about large-scale AI development:} To just verify that an actor refrained from certain kinds of AI development (without verifying claims about AI \emph{deployment}), it is not necessary to verify the correctness of claimed uses of AI compute for purposes other than AI development. Instead, it can suffice to otherwise verify the total amount of AI compute used for AI development vs. other purposes---such as via verified workload classification by cloud providers \citep{Heim2024Cloud}, or via some chips being locked into ``fixed sets'' that are impractical for training \citep{Kulp2024}---and then to just carry out verification with regards to the compute that \emph{was} used for AI development.
\item
  \textbf{Negative rules about large-scale AI deployment:} The case is the same as the above, except switching ``development'' with ``deployment.''
\end{itemize}

\phantomsection\subsubsection{B.3 Acting on Ambiguous Findings}\label{b.3-acting-on-ambiguous-findings}

A Verifier might have ambiguous findings---evidence that is inconclusive about the Prover's compliance.\footnote{For example, whistleblowers, interviews, and intelligence agencies might uncover merely suggestive evidence, and some technical tests may reveal anomalies that have legitimate explanations.} What should the Verifier do then? We highlight some increasingly escalatory options.

\textbf{Requests for clarification.} Simple requests for clarification could resolve some issues. Studies of arms control have highlighted permanent consultative commissions as helpful in this regard.\footnote{One study writes: ``Perhaps the most important form of co-operation is a continuing process of consultation among the parties, institutionalized in a consultative commission made up of highly qualified experts. The purpose of such a commission must be entirely on the side of preserving agreements and building confidence by dealing promptly and objectively with any ambiguities, misunderstandings or technical violations which arise'' \citep{krass1985verification}.}

\textbf{Focused investigation.} Ambiguous findings could trigger a focused investigation of the specific organizations, facilities, activities, or employees whose compliance is ambiguous, at greater cost than would be practical for general verification. The Verifier could apply their verification methods with increased intensity to the suspicious area.\footnote{This could involve temporarily increasing the rate of random sampling, e.g., for selecting interviewees, AI data centers, AI chips, or workload portions to inspect. A focused inspection could also involve dedicating more analysts, intelligence agents, satellites, or computational power to the suspect area, including by running new technical tests. The IAEA has a similar practice; their common response to some ambiguous findings is ``to re-establish the physical inventory'' of a nuclear facility, ``which is time consuming and costly'' \citep{iaea_sip_guide_2014}} The associated costs also incentivize the Prover to carry out their role carefully, to reduce the incidence of focused investigations. These intensified efforts would either reveal more clear evidence of non-compliance or fail to do so; in either case, the ambiguity would be at least partly resolved.

\textbf{More intrusive verification.} If needed, the Verifier could escalate an investigation to include more intrusive verification methods than would normally be allowed. For example, the Verifier could (with the Prover's cooperation) increase the amount of information that compliance tests output.\footnote{Minimally, compliance tests would output only a yes/no determination of compliance, to minimize their potential for leaking sensitive information. The amount of information they output could be increased to gain more clarity at the expense of potentially leaking more information; there is a large gap between revealing only 1 bit and revealing terabytes, which may be needed to encode large model weights or full datasets. Still, intermediate sizes of outputs (e.g., kilobytes) would have downsides: potentially leaking algorithms or small amounts of training or usage data.} Additionally, the Prover and Verifier could authorize an expanding set of humans (not just automated programs) to directly inspect the Prover's declarations, beginning with the Prover's most relevant employees (who may already have this access via whistleblower programs) (\hyperref[a.8-whistleblower-programs]{\ul{Appendix A.8}}) and potentially escalating to the Verifier directly inspecting Prover code, which would violate confidentiality-preservation.

\textbf{Precautionary pauses.} As an emergency measure, in extreme cases the Verifier could demand some or all of the Prover's AI compute clusters be turned off while a suspected violation is investigated, or take other actions to mitigate imminent risks. Such a pause would assure the Verifier that a violation using declared AI compute clusters is not ongoing, but it could come at a high economic cost.

\textbf{Partial penalties or threats.} The enforcing parties could apply a penalty in part or with some probability, to deter a strategy of using multiple ambiguous violations to carry out a significant violation.

\phantomsection\subsection{C. Methodology Details}\label{c.-methodology-details}

\phantomsection\subsubsection{C.1 Methodology for Analysis}\label{c.1-methodology-for-analysis}

Our analysis consisted of the following steps.

\textbf{1. Developing a framework of verification subgoals.} To develop and assess a framework that breaks down verification into a series of subgoals (\hyperref[fig:framework_of_verification_subgoals]{\ul{Figure 1}}), we took as a starting point a framework used by the International Atomic Energy Agency (IAEA), identified through our literature review. The IAEA divides verification into (i) verifying that declarations are ``correct'' and (ii) verifying that declarations are ``complete'' \citep{rosenthal_iaea_safeguards}. These ultimately corresponded to Subgoal 1.A and Subgoal 2 in our framework. We then identified ways this framework fell short of the following criteria, and modified the framework until it met the criteria:

\begin{itemize}
\item
  \emph{Deductive validity:} If all subgoals are completed perfectly, this justifies a series of claims from which the desired confirmation of compliance (\hyperref[context-for-this-framework]{\ul{Section 3.1}}; \hyperref[verification-scope-and-research-methodology]{\ul{Section 2}}) deductively follows.\footnote{In practice, presumably any subgoal will not be completed perfectly, so the conclusion of compliance would follow probabilistically (e.g., one might conclude there is a 90\% probability that an actor is compliant).} Without this, verification subgoals would not suffice for verifying compliance.
\item
  \emph{Flexibility:} The framework can represent existing verification proposals identified in our literature review, a few additional potential verification systems we considered,\footnote{These other potential verification systems were: verification done only via whistleblowers, verification done only by intelligence agencies, and verification done using operations accounting but not on-chip mechanisms.} and variants. This makes the framework broadly applicable for analysis.
\item
  \emph{Avoiding excess complexity:} The framework's subgoals all contribute to its validity or flexibility, to avoid wasting analysts' time or government resources.\footnote{We also considered an additional aspect of avoiding excess complexity, which may be clearest by example. Suppose the last two subgoals in our verification framework, 2.A and 2.B, were merged into a single subgoal. Then the mechanisms for this merged subgoal would be every possible pairing of a Subgoal 2.A mechanism with a Subgoal 2.B mechanism. This would be an unnecessarily long and complicated list of mechanisms. We sought to avoid such complexity where feasible by factoring a verification subgoal into more or different subgoals.}
\end{itemize}

\textbf{2. Identifying verification mechanisms.} From the sources described above, we identified candidate verification options by compiling verification mechanisms that:

\begin{itemize}
\item
  Have been explicitly proposed, in our reviewed literature, for verifying rules on AI;
\item
  Have been used for verification in other contexts (e.g., domestic regulation, arms control) per our reviewed literature;
\item
  Have been proposed for related purposes (e.g., enforcing rules on AI);
\item
  Would leverage a known regularity, `fingerprint,' or resource requirement of AI activities; \emph{or}
\item
  Are variants of the above, varied to mitigate specific weaknesses or to complete a different verification subgoal than originally considered.
\end{itemize}

\textbf{3. Assessing, red teaming, and enhancing verification mechanisms; and identifying open problems.} For each identified mechanism, we iterated between assessing the mechanism and enhancing it. In the assessment stage, we evaluated various properties of each mechanism: what subgoals in our verification framework the mechanism could complete or support, its \emph{probability of detecting} a violation quickly\footnote{For most verification mechanisms we consider, parties can make the mechanism detect any violation more quickly by increasing the frequency at which declarations are reported and checked for compliance. For example, declarations could hypothetically be submitted and verified every few months, or every few days. Thus, parties can typically make a verification mechanism X times more timely by accepting approximately X times greater costs. However, past some point, further speed may be of limited value; for example, if requests for clarification and political deliberations would take days, perhaps it would not be a priority to speed up verification by a few hours. Still, some mechanisms such as tamper-proof, compliance-locked AI chips or server enclosures (\hyperref[mechanisms]{\ul{Section 4.1.1}}; \hyperref[mechanisms-1]{\ul{Section 4.2.1}}) would hypothetically guarantee that a violation is not done at all, making the question of detection speed unnecessary.} if done by a highly motivated major government, the frequency of \emph{false alarms}, the \emph{confidentiality} and \emph{security} for the Prover, the \emph{setup speed} in terms of time required for R\&D and implementation, and the financial or computational \emph{cost}. We focus on these properties because history and incentives suggest they will be important for the acceptability of a verification regime \citep{krass1985verification, coe_arms_control_2019, baker2023nucleararmscontrolverification, Nevo2024}. After assessing each property and identifying challenges for it,\footnote{To assess these properties, we considered, respectively: how a Prover might seek to circumvent a verification mechanism; how legitimate variation or randomness might trigger a verification mechanism, what information and exploitable control points would be transferred to the Verifier; the amount of R\&D and physical infrastructure development needed; and the costs associated with all major components of the mechanism (often estimated with a back-of-the-envelope calculation or prior empirical results).} we considered how the mechanism could be enhanced to address these challenges (e.g., through a different implementation or additional compliance tests), and then we repeated the assessment on the enhanced version of the mechanism, up to the point where further assessment or enhancement appeared to require a substantial research project of its own.

We included a candidate verification option in our overview of options (i.e., in \hyperref[fig:verification_layers_consist_of_distinct_mechanisms_for_each_verification_subgoal]{\ul{Figure 2}} and \hyperref[verification-mechanisms-and-layers]{\ul{Section 4}}) if our subsequent analysis suggested that the verification option:

\begin{enumerate}
\def\labelenumi{\arabic{enumi}.}
\item
  At least plausibly meets all desired criteria (e.g., robustness, confidentiality protection, and cost---discussed further below) for completing at least one verification subgoal, while being meaningfully distinct from other included verification options in terms of its tradeoffs or assumptions; \emph{or}
\item
  Likely strengthens another verification mechanism that meets the above condition (1), while being meaningfully distinct from other included verification options.
\end{enumerate}

\textbf{4. Identifying and analyzing verification layers.} We generated verification layers by starting with the list of plausibly robust verification mechanisms (i.e., those meeting criterion (1) above). Then, we searched for ways to partition this list of mechanisms into verification layers, i.e., identify (as much as possible) disjoint subsets of these mechanisms such that each subset has similar tradeoffs and assumptions and contains one mechanism applicable to each verification subgoal. Finally, we analyzed the verification layers by using the fact that each layer's challenges are the combined challenges of the mechanisms that make up the layer, already examined per the above analysis.

\phantomsection\subsubsection{C.2 Methodology for Expert Interviews and Interview Protocol}\label{c.2-methodology-for-expert-interviews-and-interview-protocol}

\textbf{Interviewee selection:} To identify expert interviewees, we considered: our professional networks, including in academia, industry, and nonprofits; authors of papers from our literature review; and other experts recommended by our initial interviewees. Within these groups, we looked for individuals with expertise in highly relevant fields, determined based on our literature review: computer security, computing hardware, machine learning, and AI policy (especially frontier AI verification).

Our interviews complied with policies on human subjects research, including by maintaining interviewee confidentiality per our data safeguarding plan, to err on the side of caution.

\textbf{Interview protocol:} Our interviews, which were semi-structured, were guided by our interview protocol, which is reproduced below in a lightly condensed form.

This is not a comprehensive or formal protocol. We will prioritize or vary questions based on an interviewee's areas of expertise, we will phrase questions based on an interviewee's existing familiarity, and we will ask follow-up questions based on an interviewee's initial responses.

Prior to asking the following questions, we will briefly reiterate the goals of the study and the fact that participation and all questions are optional. We may show interviewees a few draft reference figures summarizing some of the report's content, so they can more easily identify room for improvement.

Questions for experts with broad relevant expertise (listed in the default planned order):

\begin{itemize}
\item
  Does the verification framework in this figure seem like a useful, valid, and clear way to break down verification?
\item
  What are some promising verification mechanisms for completing each verification step?
\item
  Are there promising verification mechanisms that we're missing?
\item
  Have we miscategorized the potential functions or reliability of any of the verification mechanisms, as summarized in this figure? Are we missing major challenges faced by any?
\item
  Are there other considerations or challenges you'd suggest we keep in mind?
\item
  With regard to our section identifying directions for future work, what types of research or future work would be especially productive here? What would lay the groundwork for rigorous red-teaming by world-leading organizations?
\item
  What specific research projects would most advance the field?
\item
  How could this report be more useful?
\item
  Overall, how promising does each category of verification mechanisms seem?
\item
  Who else should we talk to, or what else should we read, to inform this study?
\end{itemize}

Example questions for experts with highly specialized expertise:

\begin{itemize}
\item
  How precisely could an AI accelerator's utilization be estimated by physical measurements?
\item
  How might information from data center suppliers be used to detect covert AI data centers?
\item
  To what extent might AI data center maintenance be automated?
\item
  Could most AI hardware's secure boot functionality be made infeasible to deactivate through a firmware update?
\item
  How likely is it that Confidential Computing on H100 GPUs is securely implemented?
\item
  What would be the performance penalty of entirely running a large AI training or inference workload with Confidential Computing?
\end{itemize}

\phantomsection\subsubsection{C.3 Additional R\&D Problems for Verification}\label{c.3-additional-rd-problems-for-verification}

This report highlights selected R\&D problems for verification (\hyperref[tab:rd_challenges_duplicate]{\ul{Table 10}}; \hyperref[open-problems-in-verification]{\ul{Section 5}}). In this appendix, we list additional verification R\&D problems that tentatively did \emph{not} meet our criteria\footnote{Recall our criteria: ``R\&D problems that are relatively challenging, have been subject to relatively little work compared to their difficulty, could play major roles in verification, and would not create highly abusable tech''} (\hyperref[open-problems-in-verification]{\ul{Section 5}}) for inclusion in the selected problems, but may still be valuable to work on:

\begin{itemize}
\item
  \textbf{Thresholds for AI chips and AI compute clusters:} For AI chip monitoring to be implemented (e.g., by chain-of-custody verification, network taps, or analog sensors), one must determine which chips and clusters are in scope \citep{reuel2025openproblemstechnicalai}, ideally including all compute that would enable serious violations while excluding all the rest. Some cases are unambiguously in-scope (e.g., large data centers of leading GPUs), but there are many edge cases.\footnote{Edge cases might include: older AI chips (e.g., A100s), high-end consumer GPUs (e.g., RTX 5090s), CPUs with powerful AI modules, and smaller data centers.} Still, there is already substantial relevant research, including sophisticated modeling \citep{erdil2024datamovementlimitsfrontier, erdil2025inferenceeconomicslanguagemodels}, data collection \citep{EpochMachineLearningHardware2024}, and definitions of AI chips for export controls \citep{patel_wafer_2023, Burga2025H20Problem}.
\item
  \textbf{Evaluations, mitigations, and rule specifications:} As we discuss (\hyperref[open-problems-in-verification]{\ul{Section 5}}), effective rule specifications, which may include technical evaluations and risk mitigations, are under-developed and crucial. Still, these challenges are already subject to extensive research, including by industry and governments \citep{frontier_model_forum_publications, shenk2024evaluating, uk_ai_security_institute_work}, more so than the R\&D problems we highlight (\hyperref[tab:rd_challenges_duplicate]{\ul{Table 10}}).
\item
  \textbf{Counting AI chip-hours:} This does not appear to require significant R\&D (\hyperref[a.6-compute-accounting-via-analog-sensors]{\ul{Appendix A.6}}).
\item
  \textbf{Replicability:} Non-deterministic workloads could pose challenges for verification. Still, the sources of non-determinism we identified appear sufficiently controllable without significant R\&D (\hyperref[a.9-deterministic-replication-of-neural-network-inference]{\ul{Appendix A.9}}). Remaining implementation difficulties may be resolved over the course of pilot programs.
\item
  \textbf{Estimating AI chips' theoretical performance:} This does not appear to require significant R\&D (\hyperref[a.1-full-stack-security-for-technical-verification-mechanisms-implementation]{\ul{Appendix A.1}}).
\item
  \textbf{Abuse-resistant enforcement:} This does not appear to require significant R\&D, as there are multiple existing proposals and some tradeoffs may be inevitable (\hyperref[on-chip-verification-layer]{\ul{Section 4.1}}).
\item
  \textbf{Counting declared model operations:} This does not appear to require significant R\&D, as there is significant relevant existing work (\hyperref[a.6-compute-accounting-via-analog-sensors]{\ul{Appendix A.6}}).
\item
  \textbf{Using information from data center suppliers:} This does not appear to require significant R\&D and would only be valuable as a supplemental mechanism (\hyperref[supplementary-verification-mechanisms]{\ul{Section 4.4}}).
\item
  \textbf{Adversarial robustness of design information verification:} This does not appear to require significant R\&D and would only be valuable as a supplemental mechanism (\hyperref[supplementary-verification-mechanisms]{\ul{Section 4.4}}).
\item
  \textbf{Sensitive data storage:} This does not appear to require significant R\&D (\hyperref[a.10-storing-sensitive-data-for-verification]{\ul{Appendix A.10}}).
\item
  \textbf{Robust, high-level workload classification:} Distinguishing between training and inference workloads is too coarse-grained to verify most of the rules in our scope (\hyperref[rules-on-ai-models-data-and-code]{\ul{Section 2.1}}), especially as synthetic data generation (inference) may be an increasingly large component of training. Still, it could be valuable (\hyperref[supplementary-verification-mechanisms]{\ul{Section 4.4}}).
\end{itemize}

Theoretically, one could also pursue R\&D on lie detection for interviews of personnel, but we do not recommend this because of the potential for such technology to be abused.

\phantomsection\subsubsection{C.4 Scope: Additional Notes}\label{c.4-scope-additional-notes}

Further detail on our definitions of ``large-scale'' compute use:

\begin{itemize}
\item
  To see why a lower threshold could intrude on consumer hardware, note that 100 high-end gaming GPUs---widely sold consumer products \citep{garreffa_rtx_4090_sales_2022}---have the same computational power as 33 contemporaneous AI chips (H100 GPUs) \citep{NVIDIA2023AdaArchitecture, NvidiaH100}.\footnote{This is counting theoretical INT8 operations per second without sparsity.} Thus, to practically verify how the computing power of \emph{tens} of AI chips is used, one would have to practically verify that no actor secretly assembles 100 high-end gaming GPUs.
\item
  We do not give a more precise definition, as that could misleadingly suggest that a specific quantity is known to be especially significant. To the contrary, a smaller cluster can typically execute the same workloads as a slightly larger cluster if given slightly more time. Additionally, small differences in training or inference compute (all else equal) tend to correspond to small differences in model performance \citep{owen2024predictable, openai2024gpt4technicalreport, OpenAI2024Reasoning}.
\item
  The practicality of verifying AI compute use above some threshold would likely be affected, not just by how high that threshold is in absolute terms, but also by how high that threshold is relative to the total amount of AI compute. A small fraction of all AI compute could fall within the margin of error of verification mechanisms, especially for mechanisms such as analog sensors that may require more approximate analysis.
\end{itemize}

Further detail on the AI models, data, and code we consider:

\begin{itemize}
\item
  We do not assume verified models are \emph{necessarily} publicly available via public interfaces or published weights, which would enable more direct testing. Instead, we consider verifying models that could range from open-source models to AI companies' internal, advanced models.
\end{itemize}

\phantomsection\subsubsection{C.5 Related Work}\label{c.5-related-work}

We draw on a range of related technical, historical, and strategic work, including:

\phantomsection\paragraph{C.5.1 Work on Frontier AI Verification}\label{c.5.1-work-on-frontier-ai-verification}

\textbf{Overviews of AI verification:} Previous papers overview options for AI verification \citep{brundage2020trustworthyaidevelopmentmechanisms, wasil2024verificationmethodsinternationalai, Scher2024Verification, harack_verification_2025}. These works explore a range of verification mechanisms. This report aims to provide an overview with improved clarity and detail (\hyperref[contributions]{\ul{Section 1.1}}).

\textbf{Verifying rules on large-scale AI}: Beyond overview papers, some research has advanced specific verification proposals. \cite{shavit2023doescatchchinchillaverifying} makes a high-level proposal for verifying rules on large-scale AI development through compute monitoring. \cite{choi2023toolsverifyingneuralmodels}, building on work by \cite{jia2021proofoflearningdefinitionspractice}, propose and experimentally demonstrate a protocol for verifying ``training transcripts'' of neural networks. \cite{trager_civilian_ai_governance_2023} propose an international governance framework for civilian AI involving certification of state jurisdictions, though the focus on civilian AI means this approach is not designed for robustness to major state circumvention efforts. \cite{petrie2025technicaloptionsflexiblehardwareenabled} propose Flexible Hardware-Enabled Guarantees (FlexHEGs), a family of verification mechanisms involving network taps and tamper-resistant enclosures.

\textbf{Confidentiality-preserving technologies and secure machine learning:} Researchers have advanced cryptographic methods for confidential audits or implementations of AI workloads, including zero-knowledge proofs (ZK-SNARKs) \citep{10.1145/3627703.3650088, waiwitlikhit2024trustlessauditsrevealingdata, sun2024zkllmzeroknowledgeproofs} and Confidential Computing \citep{CCC2021TechnicalAnalysis}. Confidential Computing may be highly applicable to verifying international agreements on AI (\hyperref[on-chip-verification-layer]{\ul{Section 4.1}}). On the other hand, ZK-SNARKs, despite rapid progress in relative terms, as of early 2025 still have enormous compute overhead costs\footnote{A back-of-the-envelope estimate suggests a 500,000x compute overhead. Sun, et al. \citep{sun2024zkllmzeroknowledgeproofs} report a prover time of 803 seconds per inference verification for Llama-13B, using an A100 with 40GB of memory. Since this implementation ``fully leverag{[}es{]} parallel computing resources'' and uses the majority of available memory, the throughput of this method would be around 1 token per 803 seconds. In contrast, one benchmark of running ordinary (non-ZK) inference on the same model with the same GPU reports a throughput of 668 tokens per second \citep{schmid_llama2_sagemaker_benchmark_2023}: approximately 500,000 times greater.} and do not preserve the confidentiality of model architectures.\footnote{Prior work in ZK-SNARKs \citep{10.1145/3627703.3650088, waiwitlikhit2024trustlessauditsrevealingdata, sun2024zkllmzeroknowledgeproofs} all does not preserve the privacy of AI model architectures, which AI developers often treat as highly sensitive IP.}

\phantomsection\paragraph{C.5.2 Broader Related Work}\label{c.5.2-broader-related-work}

\textbf{Arms control verification:} A large literature explores verification of historical arms control treaties, including their institutional design \citep{Dai2002InformationSystems}, negotiation \citep{coe_arms_control_2019, carlson_iaea_safeguards_2020}, implementation \citep{c7bd1558-c885-3aa5-a326-301078b3df37, krass1985verification}, technical methods \citep{krass1985verification, ota_verification_technologies_1990, rosenthal_iaea_safeguards}, and applicability to AI \citep{baker2023nucleararmscontrolverification}. Much of this work focuses on nuclear arms control agreements, where verification has been especially developed. The Conventional Armed Forces in Europe Treaty had verification measures similar to those of nuclear arms control \citep{nti_cfe_2016}. The Chemical Weapons Convention authorizes ``challenge inspections,'' but none have been conducted \citep{aca_chemical_weapons_norm_2021}. The Biological Weapons Convention lacks formal verification, and it has been notoriously violated \citep{nti_bwc_2024}.

\textbf{Verification frameworks:} The International Atomic Energy Agency (IAEA) verifies declarations' correctness and completeness \citep{rosenthal_iaea_safeguards}, as do many financial audits \citep{cambridge_financial_audit}. Our verification framework draws on these frameworks (\hyperref[c.1-methodology-for-analysis]{\ul{Appendix C.1}}).

\textbf{Governance of AI compute}: \cite{sastry2024computingpowergovernanceartificial} argue that computing power presents promising opportunities for AI governance, including compute verification.

\textbf{On-chip / hardware-enabled governance mechanisms}: \cite{aarne_secure_chips_2024} and \cite{Kulp2024} propose hardware-enabled governance mechanisms as potential tools to advance U.S. export control priorities, including via verifying compliant use. Aarne, et al. also discuss potential applications for verification of international agreements on AI.

\textbf{Compute accounting and measurement}: \cite{Heim2024Cloud} discuss compute accounting as a governance mechanism, particularly in the context of enforcing domestic regulations. \cite{epoch2022estimatingtrainingcompute}, among others, propose and implement methods for estimating the compute used to train a model based on hyperparameters or hardware use.

\clearpage

\phantomsection\section{Glossary}\label{glossary}

\begin{figure}[ht]
  \noindent
  \makebox[\textwidth][c]{%
    \includegraphics[width=6.5in,height=3.26389in]{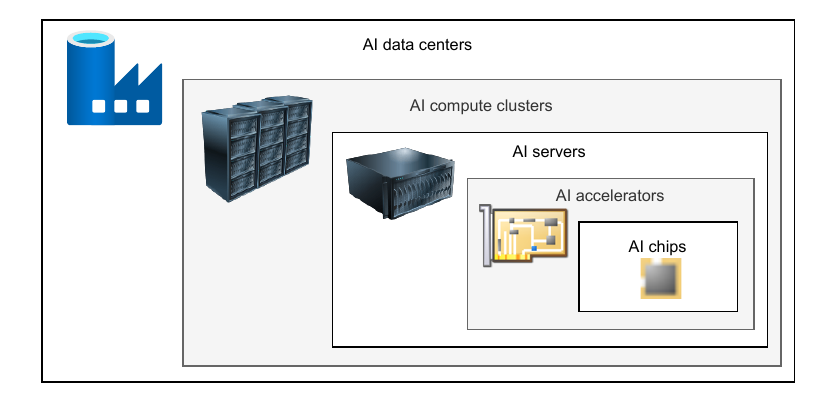}%
  }
  \caption{How AI computing hardware is organized. AI data centers are facilities that host AI compute clusters and supporting infrastructure. AI compute clusters are made up of racks of AI servers and support equipment. These servers are enclosed units of AI accelerators, supporting chips, and other devices. AI accelerators are circuit boards with AI chips, memory, and other components. AI chips are computer chips that efficiently execute computational operations for AI development or deployment. The figure's images are generic images for illustrative purposes.}
  \label{fig:how_ai_computing_hardware_is_organized}
\end{figure}

\emph{AI accelerator:} A circuit board containing an AI chip, along with memory and other components (\hyperref[fig:how_ai_computing_hardware_is_organized]{\ul{Figure 9}}).

\emph{AI chip:} A computer chip that efficiently executes computational operations for AI development or deployment, such as NVIDIA H100s (\hyperref[fig:how_ai_computing_hardware_is_organized]{\ul{Figure 9}}).

\emph{AI compute cluster:} An interconnected collection of AI servers and support equipment (\hyperref[fig:how_ai_computing_hardware_is_organized]{\ul{Figure 9}}).

\emph{AI data center:} A facility that hosts AI compute clusters and supporting infrastructure (\hyperref[fig:how_ai_computing_hardware_is_organized]{\ul{Figure 9}}).

\emph{AI server:} An enclosed unit of AI accelerators, together with supporting chips and other supporting equipment (\hyperref[fig:how_ai_computing_hardware_is_organized]{\ul{Figure 9}}).

\emph{Artificial intelligence (AI):} Computer programs capable of sophisticated cognitive tasks, and related technologies, particularly in the deep learning paradigm.

\emph{Prover:} An entity, such as a company or government, that makes a claim which they aim to convince a Verifier of.

\emph{Verification:} The act or process of checking whether a claim is true, such as checking whether a party has complied with a requirement as claimed.

\emph{Verifier:} An entity, such as a government agency or international agency, that aims to verify a claim.

\emph{Workload:} A computer program, such as a program for AI development or deployment, particularly on a compute cluster.

\clearpage

\phantomsection\section{Abbreviations}\label{abbreviations}

AI Artificial intelligence

AISI AI Security Institute or (Center for) AI Standards and Innovation

ASIC Application-Specific Integrated Circuit

CC Confidential Computing

CPU Central Processing Unit

CS Computer science

DARPA Defense Advanced Research Projects Agency

ECC Error correction code

FLOP Floating point operation(s)

FlexHEG Flexible Hardware-Enabled Guarantee

GPU Graphics processing unit

HDL Hardware Description Language

HFU Hardware FLOP (floating point operations) utilization

HSM Hardware security module

IAEA International Atomic Energy Agency

IC Intelligence community

IP Intellectual property

IT Information technology

MFU Model FLOP (floating point operations) utilization

ML Machine learning

NIST National Institute of Standards and Technology

NSF National Science Foundation

OP Operation(s)

OS Open-source or operating system

OSINT Open-source intelligence

PDU Power Distribution Unit

ROM Read-only-memory

R\&D Research and development

SEC Securities and Exchange Commission

TLS Transport Layer Security

TPU Tensor Processing Unit

UK United Kingdom

UN United Nations

US United States

VPN Virtual private network

\clearpage

\bibliographystyle{plainnat} 
\bibliography{references.bib}

\end{document}